\documentclass[pre,amsmath,twocolumn]{revtex4}
\usepackage{bm}
 \usepackage{amssymb}
 \usepackage{graphicx}
 \usepackage{subfigure}
 
\begin{document}

 \title{Constraint relaxation leads to jamming}

\author{Eial Teomy$^{1}$ and Yair Shokef$^{1,2}$}

\affiliation{%
$^{1}$School of Mechanical Engineering, Tel Aviv University, Tel Aviv 69978, Israel}
\affiliation{$^{2}$Sackler Center for Computational Molecular and Materials Science, Tel Aviv University, Tel Aviv 69978, Israel}
\begin{abstract}
Adding transitions to an equilibrium system increases the activity. Naively, one would expect this to hold also in out of equilibrium systems. This surprising effect is caused by adding heretofore forbidden transitions into less and less active states. We demonstrate, using relatively simple models, how adding transitions to an out of equilibrium system may in fact reduce the activity and even cause it to vanish. We investigate six related kinetically-constrained lattice gas models, some of which behave as naively expected while others exhibit this non-intuitive behavior. We introduce a semi-mean-field approximation describing the models, which agrees qualitatively with our numerical simulation.
\end{abstract}

  \maketitle

\section{Introduction}

For an equilibrium system, adding more transitions to its state space increases the activity. Naively, one would think that this intuitive result holds also for systems that are out of equilibrium. However, unlike in equilibrium systems, the added transitions between states do not have to obey detailed balance, and may lead to an absorbing state, and thus decrease the activity. Conceptually, this is similar to several phenomena seen in many-particle out-of-equilibrium systems, such as the faster-is-slower effect \cite{Helbing2000,Parisi2005,Garcimartin2014,Sticco2017,Chen2018}, slower-is-faster effect \cite{Gershenson2015,Tachet2016}, motility induced phase separation \cite{Fily2012,Redner2013,Cates2015,Cugliandolo2017,Digregorio2018,Whitelam2018,Klamser2019,Merrigan2020}, in which the total activity decreases as the activity of the individual particles increases.

\begin{figure}
\includegraphics[width=\columnwidth]{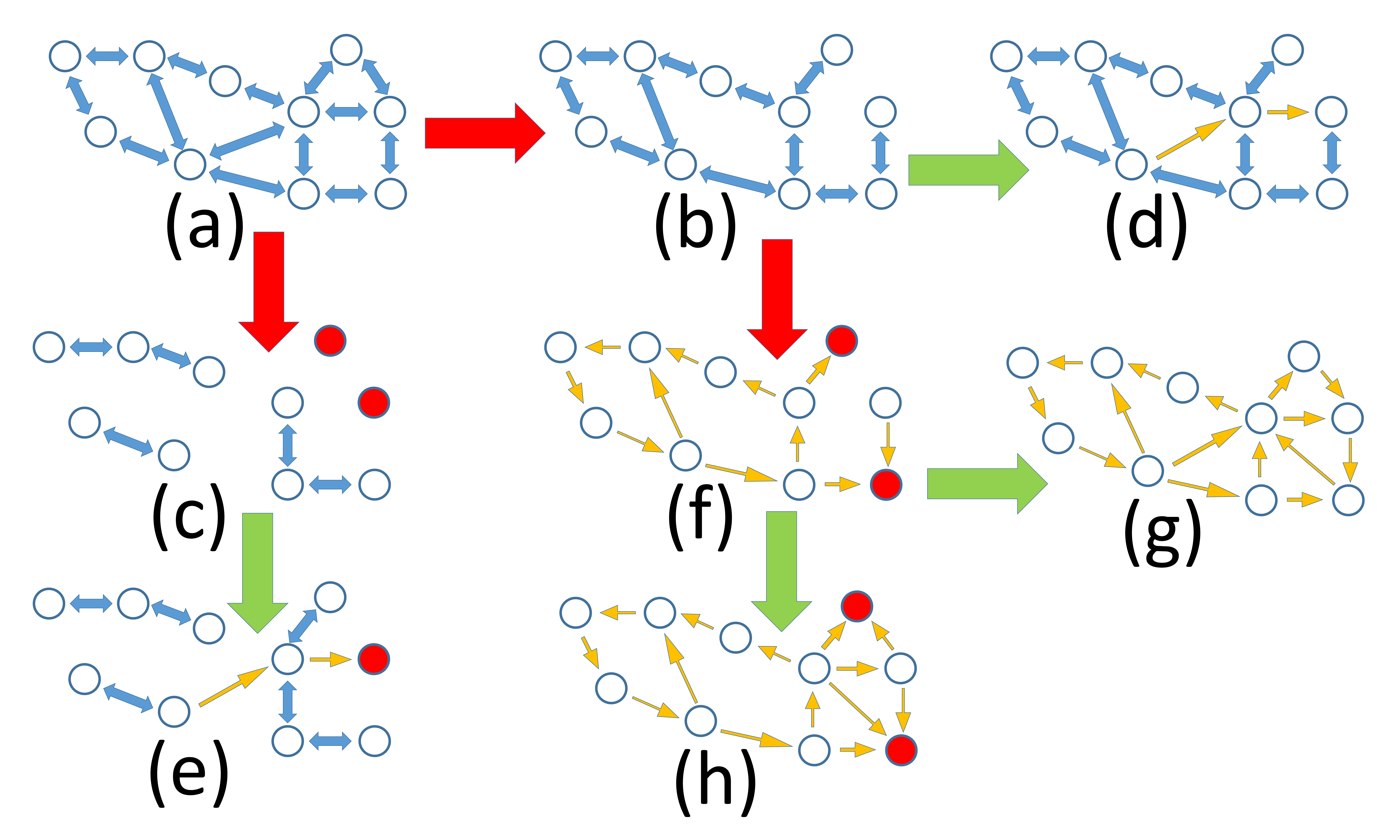}
\caption{A sketch of discrete dynamical systems and the transitions between their states: (a-b) Ergodic equilibrium systems, (c) a system which obeys detailed balance but is not ergodic, (d-h) non-equilibrium systems violating detailed balance. The big red arrows between systems represent a removal of transitions, and the big green arrows represent an addition of transitions. In each system, absorbing states are denoted by a red circle, two way transitions by a blue double-sided arrow, and one way transitions by an orange arrow.}
\label{sketch}
\end{figure}

Consider for example an ergodic system in equilibrium, depicted in Fig. \ref{sketch}a. A concrete example of such a system is the symmetric exclusion process \cite{Spitzer1970}, which is a lattice gas in which each particle may hop to a neighboring site if the target site is vacant. Removing transitions and their reciprocal transition keeps the system in equilibrium (Fig. \ref{sketch}b), but may cause it to become non-ergodic (Fig. \ref{sketch}c). A class of models which demonstrate this is kinetically constrained models (KCMs) \cite{Jackle1994,Ritort2003,Toninelli2006,Toninelli2007,Jeng2008,Biroli2008,Jeng2010,Garrahan2010,Ghosh2014,Ohta2014,Segall2016}, in which a particle may hop to a neighboring site if both the target site is vacant, and the neighborhood of the particle satisfies some model-dependent rule, both before and after the move. These models obey detailed balance, since by construction if a transition is allowed, its reverse is also allowed at the same rate. Essentially, adding this kinetic constraint removes some of the bonds from the transition graph of Fig. \ref{sketch}a and transforms the system into the one schematically illustrated in Fig. \ref{sketch}b. At high enough particle density, these models become non-ergodic, as schematically depicted in Fig. \ref{sketch}c. 

When one-way transitions are added to the system, it is driven out of equilibrium, as shown for instance in going from Fig. \ref{sketch}b to Fig. \ref{sketch}d. In KCMs, this corresponds to allowing some of the moves which are prohibited by the kinetic constraint, but not their reverse moves. If the original KCM is ergodic, these additional one-way transitions increase the activity in the system, as is the case in going from Fig. \ref{sketch}b to Fig. \ref{sketch}d. However, if the original KCM is non-ergodic, these additional one-way transitions may create a path into absorbing states, and decrease the long time activity in the system, as is the case when going from Fig \ref{sketch}c to Fig. \ref{sketch}e. Another way to add transitions is connecting the system to external reservoirs \cite{Sellitto98,Sellitto2002,Sellitto2002b,Goncalves2009,Teomy2017,Arita2018}.

Another way to drive dynamical systems out of equilibrium is to make all transitions one-way only, as shown in going from Fig. \ref{sketch}b to Fig. \ref{sketch}f. This transformation might turn the long time activity to zero, but not necessarily. In lattice gases, this can be achieved by allowing the particles to move only in one direction \cite{Spitzer1970,Derrida1993}. Adding one-way transitions to this system, may increase the activity (Fig. \ref{sketch}g) or decrease it (Fig. \ref{sketch}h). In extreme cases, either adding or removing transitions may jam the system, i.e. cause the long time activity to become zero, or unjam it, i.e. increase the long time activity from zero to a finite value. In this paper we investigate such extreme behavior and provide concrete examples for this non-intuitive result. A less extreme method is to break detailed balance by biasing the particles to move in a certain direction \cite{Sellitto2000,Levin2001,Fielding2002,Arenzon2003,Fernandes2003,Sellitto2008,Shokef2010,Turci2012}.

We consider six related modified KCMs, one of which is the equilibrium Kob-Andersen (KA) model \cite{Kob1993}, and the others are out of equilibrium variants of it, which add or remove one-way transitions. By investigating these models numerically, we demonstrate how in some cases adding transitions increases the activity in the system, while in other cases it counter-intuitively decreases the activity and even jams the system. We also derive a semi-mean-field (SMF) analytical approximation for the activity which qualitatively captures the behavior observed numerically. Although the models we consider here are relatively simple, our results can be generalized to other, more complicated systems driven out of equilibrium by adding transitions. The models are described in Section \ref{sec_models} and their activity is investigated in Section \ref{sec_activity}. Section \ref{sec_conclusion} concludes the paper. The technical derivations of our results are presented in the Appendices.

\section{The models}
\label{sec_models}

\begin{figure}
\includegraphics[width=0.3\columnwidth]{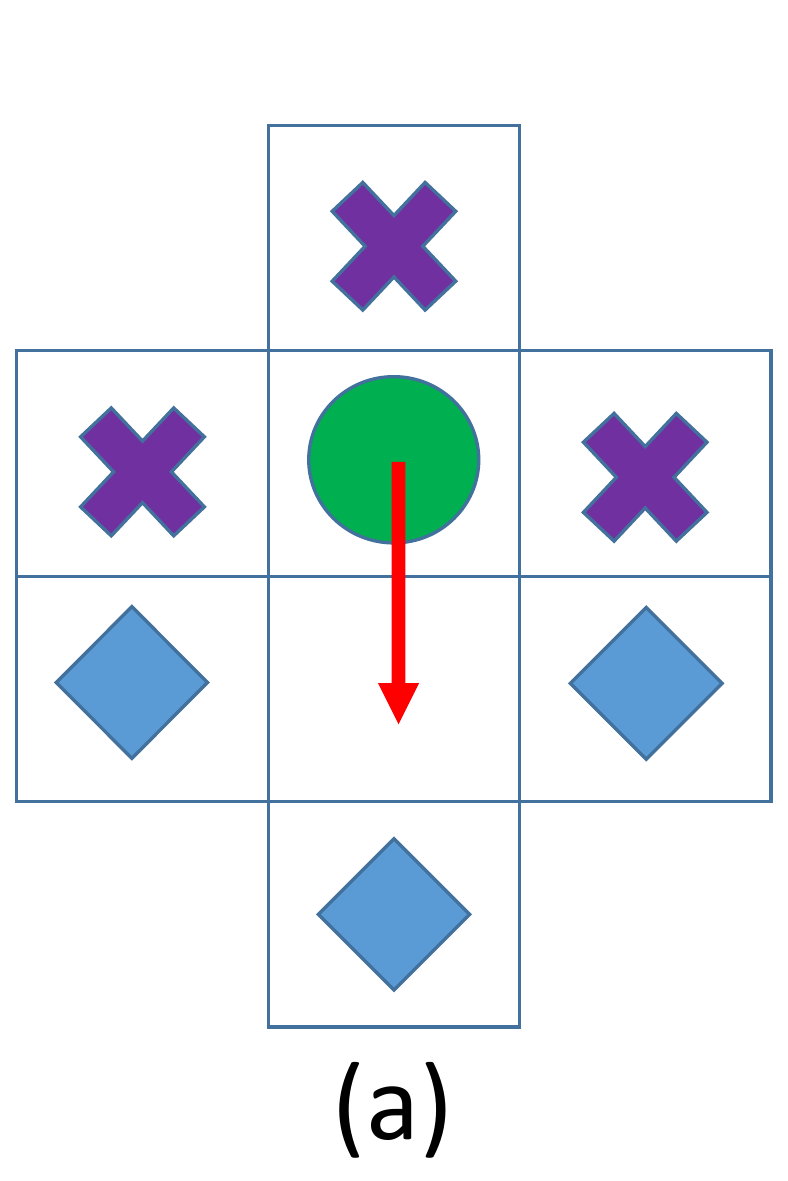}
\includegraphics[width=0.3\columnwidth]{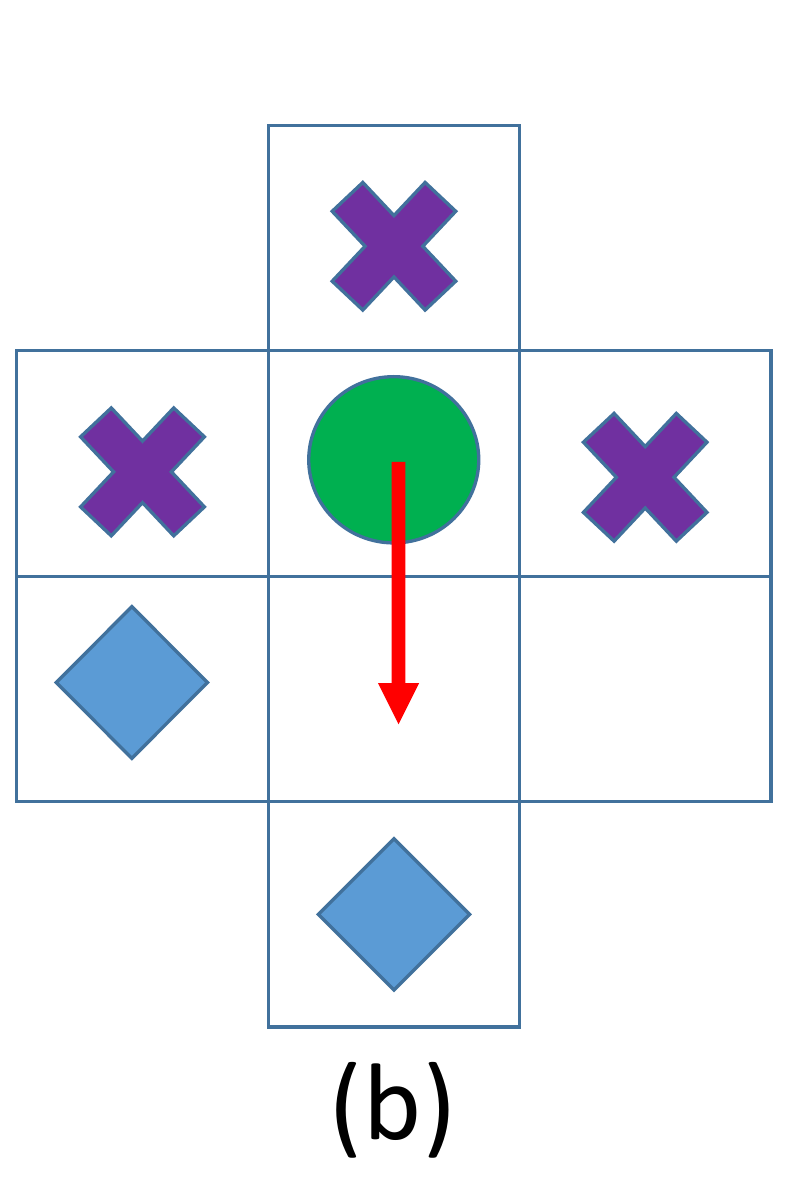}\\
\includegraphics[width=0.3\columnwidth]{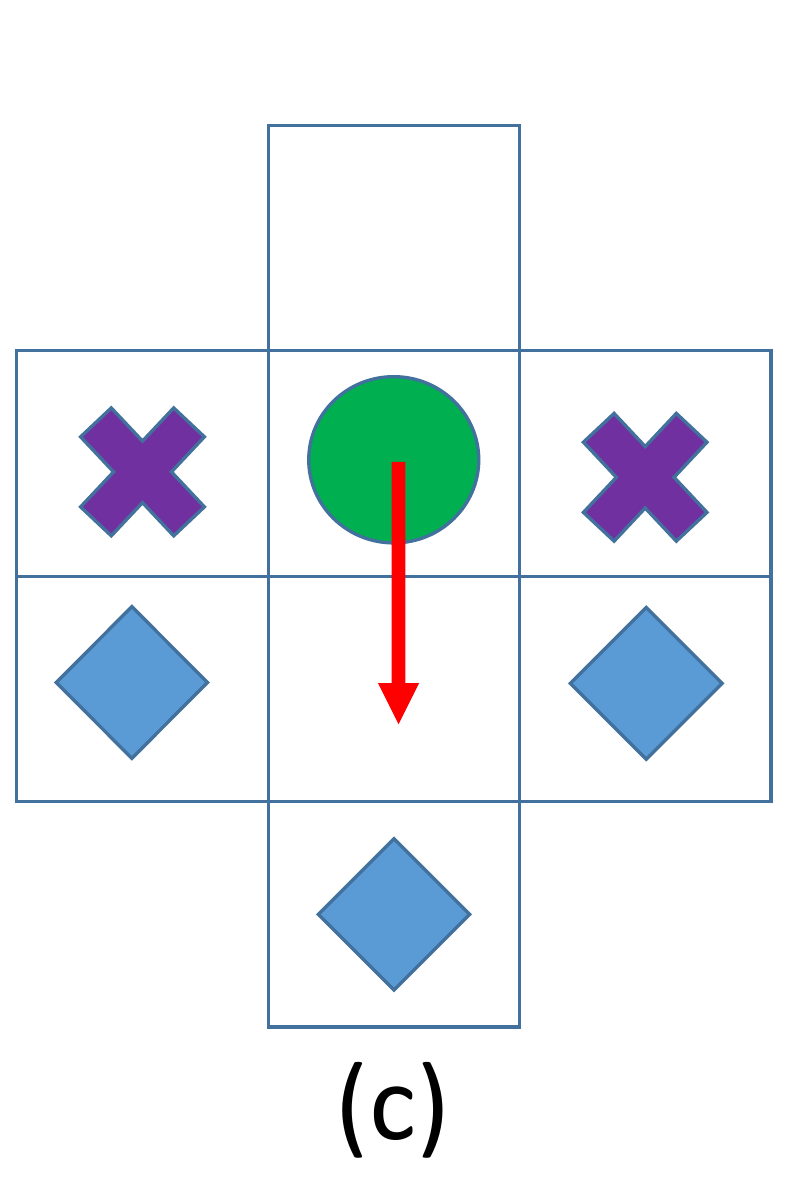}
\includegraphics[width=0.3\columnwidth]{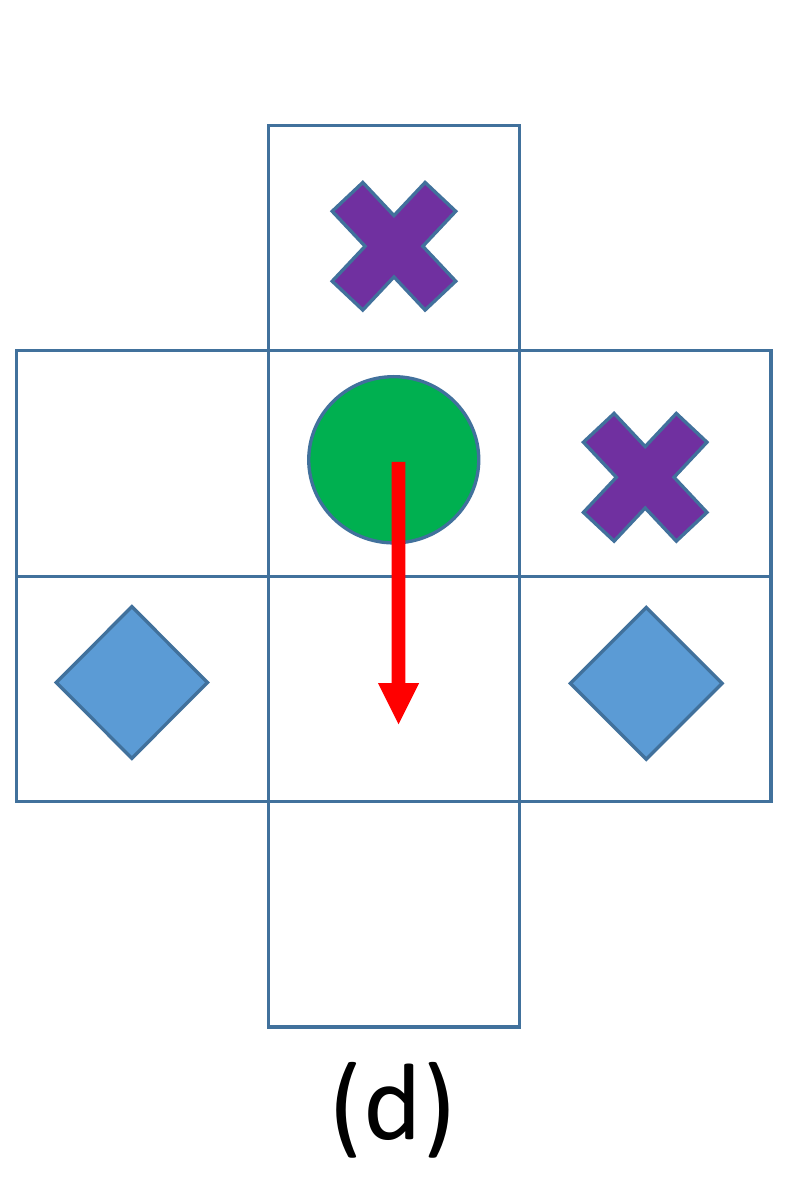}
\caption{An illustration of the kinetic constraints for a particle (green circle) moving to one of its nearest neighbors. (a) The three sites marked by a purple $\times$ are the Before group, and the three sites marked by a blue $\diamond$ are the After group. In the KA and DKA models at least one of the sites in the Before group and at least one of the sites in the After group needs to be vacant in order for the particle to move. In the BKA and DBKA models only the Before group is checked, while in the AKA and DAKA models only the After group is checked. (b) The particle can move in the AKA and DAKA models. (c) The particle can move in the BKA and DBKA models. (d) The particle can move in all six models.}
\label{ka_rules}
\end{figure}

In this paper we consider six related models. The first model, from which all the others are derived, is the KA KCM on a 2D square lattice. In this model, a particle can hop to one of its four neighboring sites if that site is vacant and if both before and after the move at least two of the particle's four neighbors are vacant, see Fig. \ref{ka_rules}. This model obeys detailed balance with respect to a trivial Hamiltonian; for each allowed move, also the reverse move is allowed and at the same rate. In the steady state the occupancy of all states is equal and there are no probability currents between the states of the system. In the infinite size limit the KA model is always ergodic, while in finite systems it jams at some size-dependent density due to finite-size effects \cite{Teomy2012,Teomy2014b,Toninelli2004,Teomy2014}. In a system of size $L\times L$, the critical density in the KA model is given by $\rho^{\rm{KA}}_{c}(L)=1-\lambda(L)/\ln L$, where $\lambda(L)$ depends weakly on $L$, converges to $\pi^{2}/18\approx0.55$ in the $L\rightarrow\infty$ limit, and is approximately $\lambda(L)\approx0.25$ for all system sizes considered in this paper \cite{Holroyd2003,Teomy2014}.

A system is jammed when it contains particles that will never be able to move, while an unjammed system does not. Note that if a particle cannot move at the current configuration, but will be able to move if some other particles move, then the system is not jammed. For example, the particles marked by an empty circle in Fig. \ref{rattlers} will never be able to move no matter how the three particles marked with a green circle and a purple $\times$ move, and therefore the system depicted there is jammed. We define the activity as the number of moves per unit time, and thus a jammed system may still be active if some of the particles in it can move.

We now define two variants of the KA model, namely the After-KA (AKA) and the Before-KA (BKA) models. In the AKA (BKA) model, a particle can hop to an adjacent vacant site if after (before) the hop at least two of its four neighbors are vacant. As opposed to the KA model, the AKA (BKA) model allows a particle to move regardless of the occupancy of the neighbors before (after) the move. Hence these two models both allow all the moves of the KA model as well as additional moves. These two models are out of equilibrium, since some transitions, namely some of those that are prohibited in the KA model, are allowed here but their inverse transitions and not.

The last three models we consider are the driven variants of the three aforementioned models, which we call the DKA, DAKA and DBKA models. In these driven models all particles can move only along one of the four directions, which we designate as down. For such a move to occur the same kinetic constraints are required to hold as in the KA, AKA, and BKA models, respectively. The DKA model was recently investigated numerically \cite{Bolshak2019} and it was found that the steady state current vanishes beyond a certain non-trivial critical density. Similar results were found for a variant of the KA model in which the particles can move in all four directions, but are biased in a particular direction \cite{Sellitto2008,Turci2012}.

\section{Activity}
\label{sec_activity}
\subsection{Definition and Mean-Field Approximation}
\begin{figure}
\includegraphics[width=\columnwidth]{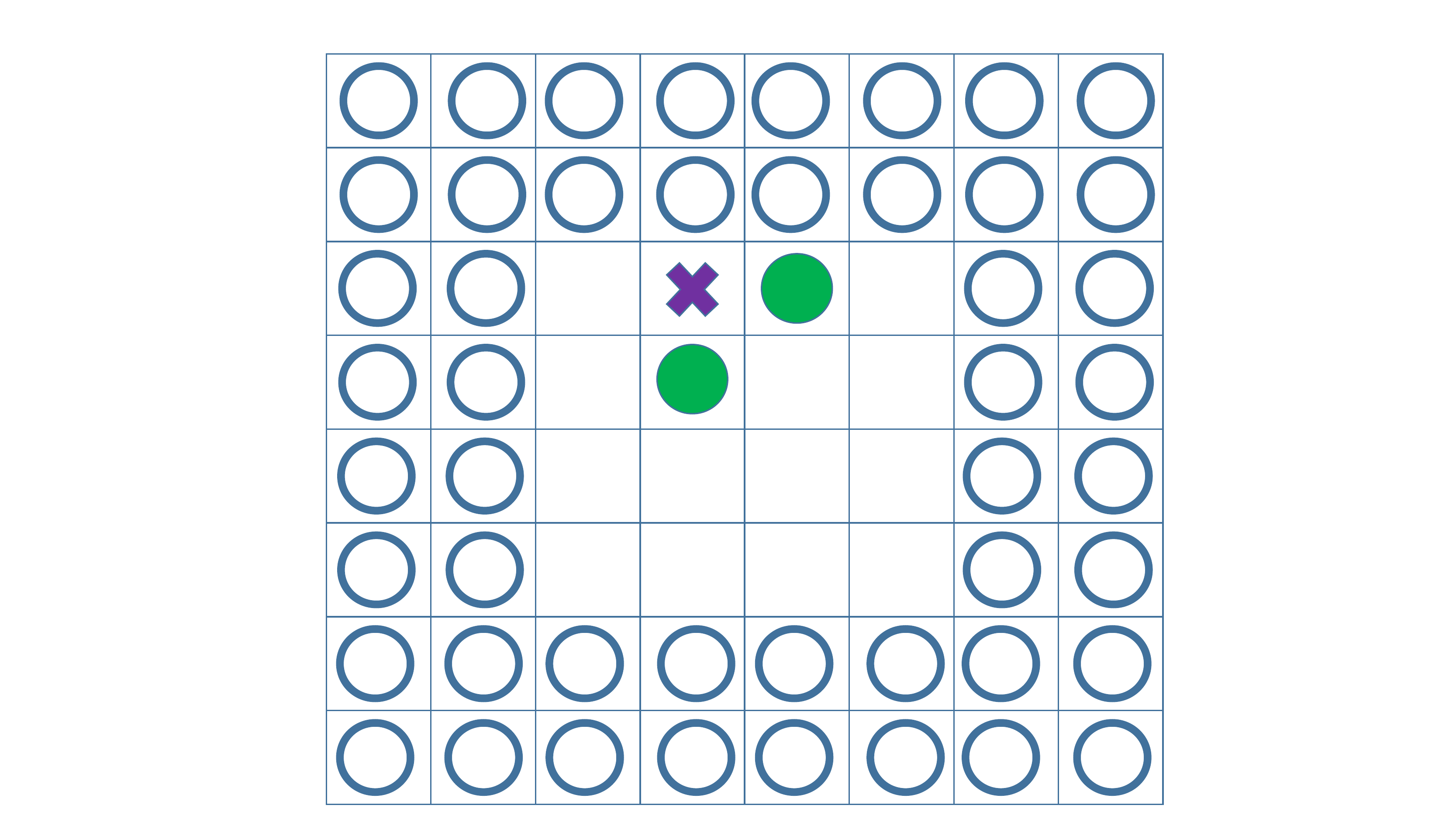}
\caption{An illustration of rattlers trapped inside a cage in the KA or BKA models. The particles on the edges (empty circles) are permanently frozen since at least three of their nearest neighbors are occupied by other permanently frozen particles. Although the $\times$ particle cannot currently move, it will be able to after the particles marked with a solid green circle move.}
\label{rattlers}
\end{figure}

In this section we investigate the activity in the system after it had reached the steady state. We define the activity, $K$, as the number of moves per unit time per lattice site. In the driven models it is equal to the current, and it may be written as $K=\rho P_{\rm{F}}$, where $\rho$ is the fraction of occupied sites and $P_{\rm{F}}$ is the probability that a given particle can move downwards. In the undriven models the activity is equal to
\begin{align}
K=\frac{\rho}{4}\sum^{4}_{n=1}nP_{\rm{F},n} ,
\end{align}
where $P_{\rm{F},n}$ is the fraction of particles that can move in $n$ of the four directions. Note that while an undriven system may be jammed, i.e. that a finite fraction of the particles are permanently frozen and will never be able to move whatever the future dynamics of the system may be, there could still be rattlers, which are particles able to move back and forth inside a confined space, and thus the activity does not vanish in those cases. See Fig. \ref{rattlers} for an illustration of such a case.

We start by considering a mean-field (MF) approximation of the activity, in which we ignore all correlations between occupancies of neighboring sites. This approximation is exact in the KA model, for which there are no correlations \cite{Kob1993}. The MF approximation for the activity in the various models is
\begin{align}
&K^{\rm{KA}}_{\rm{MF}}=K^{\rm{DKA}}_{\rm{MF}}=\rho\left(1-\rho\right)\left(1-\rho^{3}\right)^{2} ,\nonumber\\
&K^{\rm{AKA}}_{\rm{MF}}=K^{\rm{BKA}}_{\rm{MF}}=K^{\rm{DAKA}}_{\rm{MF}}=K^{\rm{DBKA}}_{\rm{MF}}=\nonumber\\
&=\rho\left(1-\rho\right)\left(1-\rho^{3}\right) .
\end{align}
For the KA and DKA models, the terms on the right hand side correspond respectively to the probabilities that a site is occupied, that its neighbor in the chosen direction of motion is vacant, and that at least one of the three sites both in the Before and the After group is vacant. In the BKA and DBKA (AKA and DAKA) models, the last term correspond to the probability that at least one of the three sites in the Before (After) group is vacant. Note that under the MF approximation there is no difference between the driven and undriven models, and furthermore the AKA and BKA have the exact same MF behavior. Also note that the MF activity in the KA model is lower than the MF activity for the AKA and BKA models, due to the extra constraint in the KA model. The MF activity is finite for all densities and vanishes only either when $\rho=0$ and there are no particles that can move and contribute to the activity, or when $\rho=1$ and the system is fully occupied such that there are no vacant sites that particles can move into.

However, as we will show below, for each of the five non-equilibrium models there is a finite, non-trivial critical density above which the activity in the steady state vanishes. We now derive a SMF approximation for the activity, which considers some of the correlations in the system, and then we will compare it to simulation results. Our SMF approximation predicts a finite, non-trivial value for the critical density at which the activity vanishes and thus qualitatively captures the simulation results. However, the SMF approximation does not capture the numerical values of the critical densities in the different models.

\subsection{Semi-Mean Field Approximation}
We describe here a sketch of the SMF approximation for the driven models, with the full details given in the appendices. It is straightforward, yet more lengthy to follow the same steps and obtain the SMF approximation also for the undriven models.
In the SMF approximation for the driven models, at any moment in time we divide all particles into three groups: free (F), jammed (J), and blocked (B). The particles in the free group are those that can move. The particles in the jammed group are those that have a vacancy in the site below them, but cannot move in their next step solely due to the kinetic constraint. The particles in the blocked group are those whose neighboring site in the direction of the flow is occupied and therefore cannot move regardless of the kinetic constraint. We denote the fractions of particles in the free, jammed, and blocked groups by $P_{\rm{F}}$, $P_{\rm{J}}$, and $P_{\rm{B}}$, respectively, where by construction
\begin{align}
P_{\rm{F}}+P_{\rm{J}}+P_{\rm{B}}=1 .\label{p3e1}
\end{align}

Now we write a master equation for the rates for each particle to change its type
\begin{align}
&\frac{\partial P_{\alpha}}{\partial t}=\sum_{\beta\neq\alpha}r_{\beta,\alpha}P_{\beta}-\sum_{\beta\neq\alpha}r_{\alpha,\beta}P_{\alpha} ,\label{eveq}
\end{align}
where $\alpha,\beta={\rm F,J,B}$ and $r_{\alpha,\beta}$ is the rate in which a particle of type $\alpha$ changes into a particle of type $\beta$. The rates themselves depend on $P_{\rm{F}}$ such that Eq. (\ref{eveq}) represents a set of three coupled nonlinear equations, which may be reduced to two equations using Eq. (\ref{p3e1}). In order to find an approximation for the rates, we assume that each site not accounted for in the type of the state before the transition is occupied with probability $\rho$, and that within each group the probability to be in each of the microscopic states is proportional to its uncorrelated probability. 

\begin{figure}[t!]
\includegraphics[width=0.3\columnwidth]{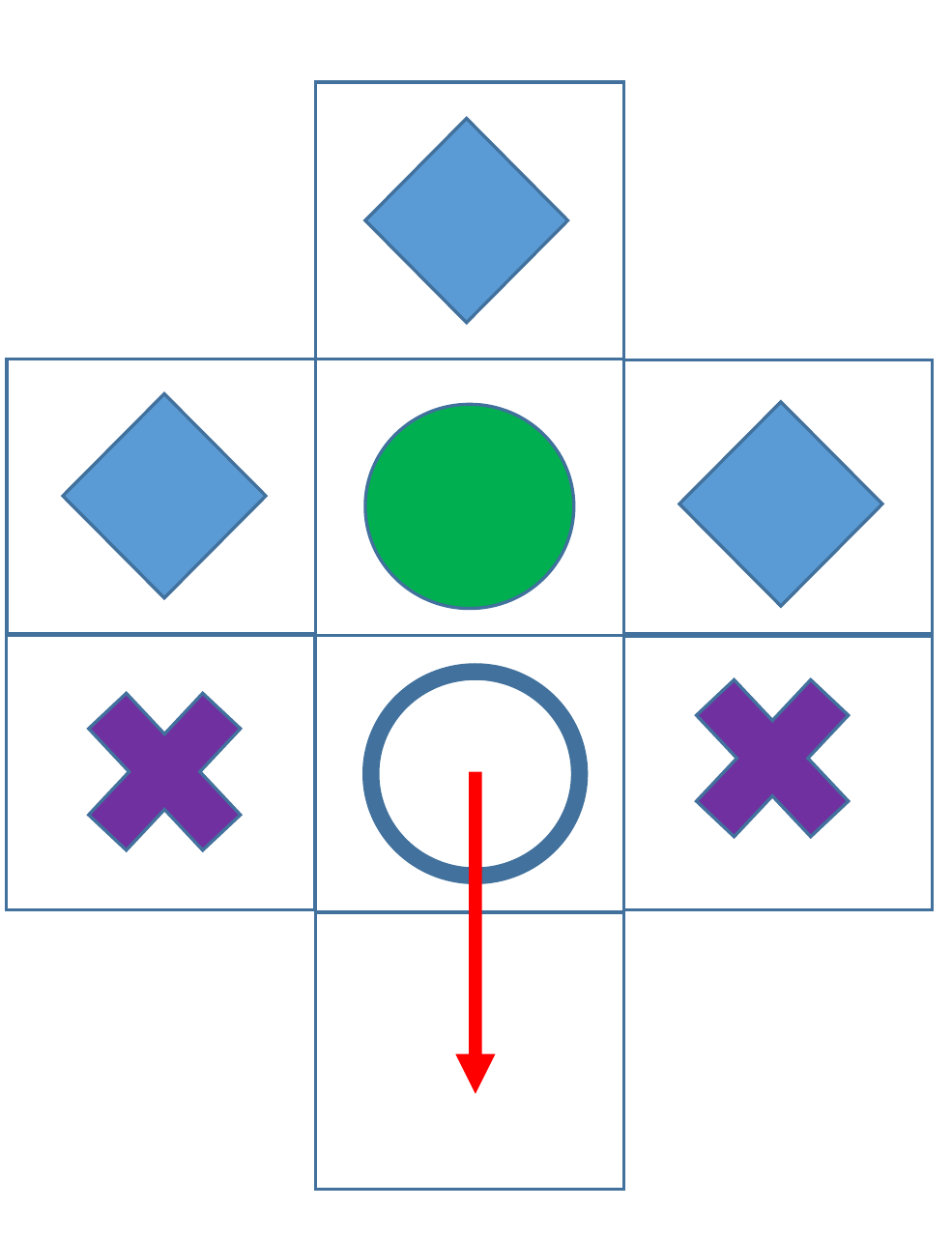}
\caption{An illustration of the transition from a blocked state to a free state in the DBKA model. The main particle whose state changes in marked by a full green circle, and the blocking particle by an empty circle. At least one of the $\times$ sites needs to be vacant in order for the blocking particle to move, and at least one of the $\diamond$ sites needs to be vacant in order for the main particle to be free after the blocking particle moves.}
\label{ratebf}
\end{figure}

For example, consider the rate $r_{{\rm B},{\rm F}}$ in the DBKA model illustrated in Fig. \ref{ratebf}. The configuration before the transition consists of the blocked particle and the blocking particle below it. In order for the blocked particle to change its type to a free particle, two independent conditions should be satisfied. First, the blocking particle needs to move. The blocking particle can move only if it is free itself. The kinetic constraint for the blocking particle to be free is that at least one of the three adjacent sites, except the site below it, is vacant. Since the site above it is occupied by the blocked particle, we approximate the probability that the blocked particle is free given that the site above it is occupied by the uncorrelated fraction of free sites with one of the three neighbors occupied, i.e. $(1-\rho^2)/(1-\rho^3)$. Note that if the blocking particle is free, the site below it must be vacant, and therefore the probability that it is vacant is already included in the probability that the blocking particle is free. The second condition for the blocked particle to change its type to a free particle, is that at least one of its three other neighbors is vacant, the probability of which we approximate by $1-\rho^{3}$. Altogether, the rate $r_{\rm{B},\rm{F}}$ in the DBKA model is given by
\begin{align}
r_{\rm{B},\rm{F}}=P_{\rm{F}}\frac{1-\rho^{2}}{1-\rho^{3}}\left(1-\rho^{3}\right)=\left(1-\rho^{2}\right)P_{\rm{F}} .
\end{align}
The other rates are generated in a similar fashion for the three driven models, as detailed in Appendix \ref{app_driven}.

For all six models, the rates $r_{\rm{B},\alpha}$ and $r_{\rm{J},\alpha}$ are proportional to $P_{\rm{F}}$ since they involve the movement of a particle besides the blocked or jammed main particle, and the rates $r_{\rm{F},\alpha}$ are linear in $P_{\rm{F}}$, since they contain terms which correspond to the movement of the main particle and to movement of other particles. Therefore, we may write the rates as
\begin{align}
&r_{\rm{B},\alpha}=\omega_{\rm{B},\alpha}P_{\rm{F}} ,\nonumber\\
&r_{\rm{J},\alpha}=\omega_{\rm{J},\alpha}P_{\rm{F}} ,\nonumber\\
&r_{\rm{F},\alpha}=\Omega_{\rm{F},\alpha}+\omega_{\rm{F},\alpha}P_{\rm{F}} ,\label{omega_def}
\end{align}
with $\omega_{\alpha,\beta}$ and $\Omega_{\alpha,\beta}$ depending only on the density, and obviously different for the six different models.

We now look for stationary solutions of Eq. (\ref{eveq}) under the condition $0\leq P_{\rm{F}},P_{\rm{J}},P_{\rm{B}}\leq 1$. The solution $P_{\rm{F}}=0$ is always a stationary solution. We find that if there is another stationary solution with $P_{\rm{F}}>0$, then it is unique and given by
\begin{widetext}
\begin{align}
P_{\rm{F}}=\frac{\left(\Omega_{\rm{F},\rm{B}} + \omega_{\rm{J},\rm{B}}\right) \left(\omega_{\rm{B},\rm{F}} - \omega_{\rm{J},\rm{F}}\right) - \left(\omega_{\rm{B},\rm{F}} + \omega_{\rm{B},\rm{J}} + \omega_{\rm{J},\rm{B}}\right)\left(\Omega_{\rm{F},\rm{B}} + \Omega_{\rm{F},\rm{J}} - \omega_{\rm{J},\rm{F}}\right)}{\left(\omega_{\rm{F},\rm{B}}+\omega_{\rm{F},\rm{J}}\right)\left(\omega_{\rm{B},\rm{J}}+\omega_{\rm{J},\rm{B}}\right)+\left(\omega_{\rm{B},\rm{J}}+\omega_{\rm{F},\rm{B}}\right)\omega_{\rm{J},\rm{F}}+\omega_{\rm{B},\rm{F}}\left(\omega_{\rm{F},\rm{J}}+\omega_{\rm{J},\rm{B}}+\omega_{\rm{J},\rm{F}}\right)} .\label{pfss}
\end{align}
\end{widetext}
In Appendix \ref{app_stability} we derive Eq. (\ref{pfss}) and show that if this solution exists, it is also stable. In Appendix \ref{app_stabpf0} we investigate the stability of the $P_{\rm{F}}=0$ state under the SMF approximation, and find that for large enough $P_{\rm{B}}$, the solution is stable. 

For the three driven models, as well as for the BKA model, we find that within the SMF approximation there is some finite, model-dependent critical density $0<\rho_{c}<1$ such that for densities higher than the critical density $\rho>\rho_{c}$, solving Eq. (\ref{pfss}) yields a negative $P_{\rm{F}}$ and thus it does not exist, while for $\rho<\rho_{c}$ we find that $P_{\rm{F}}>0$. Therefore, the critical density is defined as the solution to Eq. (\ref{pfss}) with $P_{\rm{F}}=0$. The critical densities we get from this SMF approximation are $\rho^{{\rm SMF}}_{{\rm DKA}}=0.792$, $\rho^{{\rm SM}}_{{\rm DAKA}}=0.933$, $\rho^{{\rm SMF}}_{{\rm DBKA}}=0.679$ and $\rho^{{\rm SMF}}_{{\rm BKA}}=0.858$. 

In the driven models, the SMF approximation involves three different states. In the undriven models, we need to account for whether in each of the four directions the particle is free to move, blocked or jammed, which gives a total of $3^{4}=81$ states, which reduce to $20$ by rotational and inversion symmetry. In the BKA model, however, the number of states is reduced to six, since a particle is jammed in a certain direction only if it is blocked in the other three directions. We therefore present in Appendix \ref{app_bka} also the derivation of the SMF activity in the BKA model. We leave the derivation of the AKA and KA models, which are straightforward but cumbersome, to future publications. However, we expect that the SMF approximation in the KA model would yield the exact same result as the MF approximation for that model since by construction it has no correlations. Thus, it would only be interesting to develop the SMF approximation for the AKA model.

\subsection{Numerical Results}
\begin{figure}
\includegraphics[width=0.6\columnwidth]{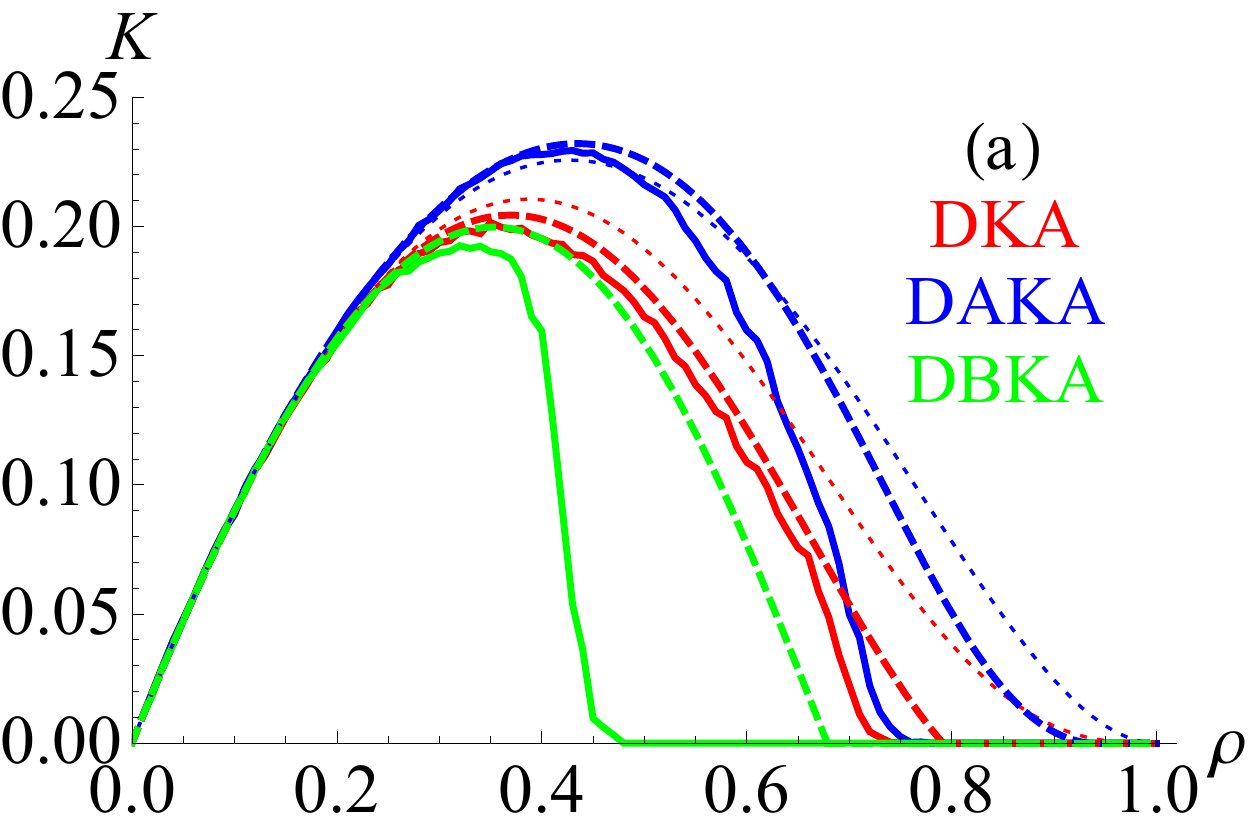}\\
\includegraphics[width=0.6\columnwidth]{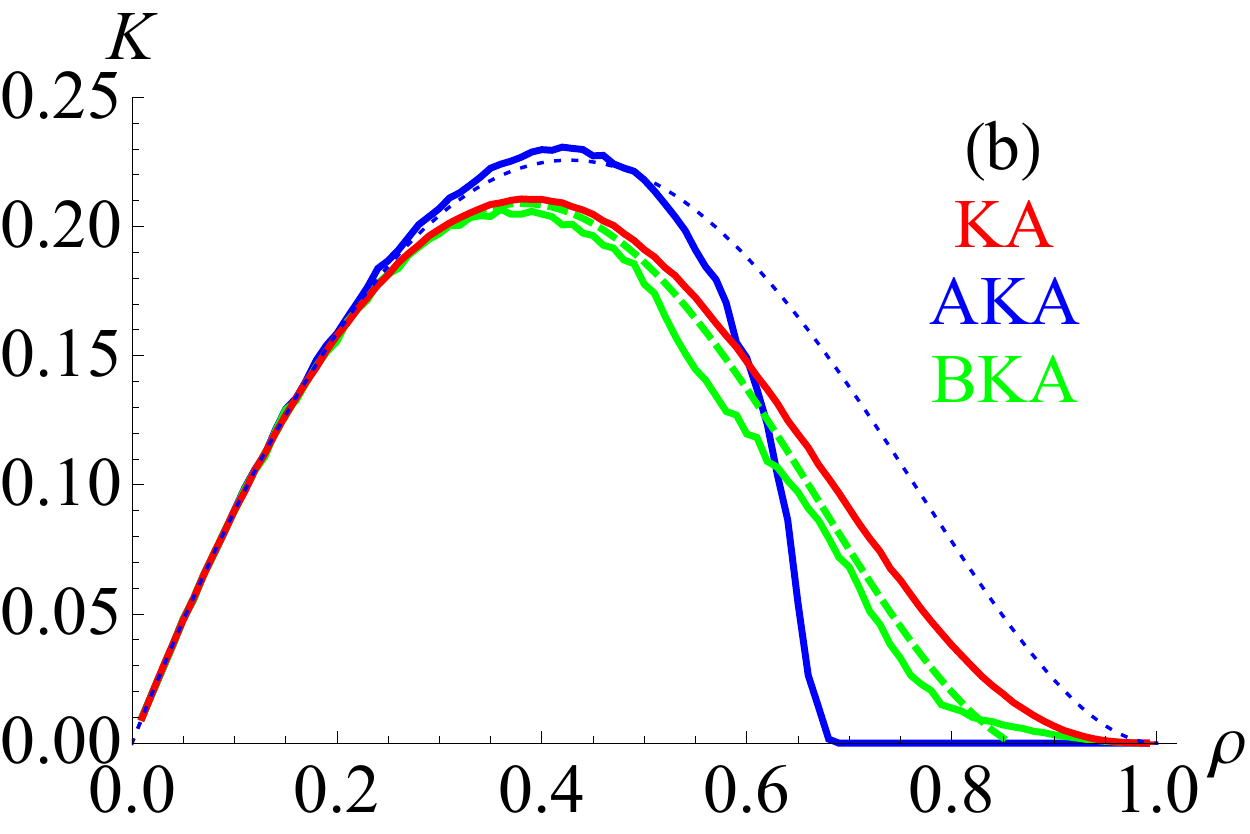}
\caption{Activity as a function of density for the three driven models (a) and the three undriven models (b), both from simulations results (continuous lines) and under the MF (dotted) and SMF (dashed) approximations. Note that the MF approximation is identical for the AKA/BKA/DAKA/DBKA models.}
\label{activity_plot}
\end{figure}

\begin{figure}[t]
\includegraphics[width=0.4\columnwidth]{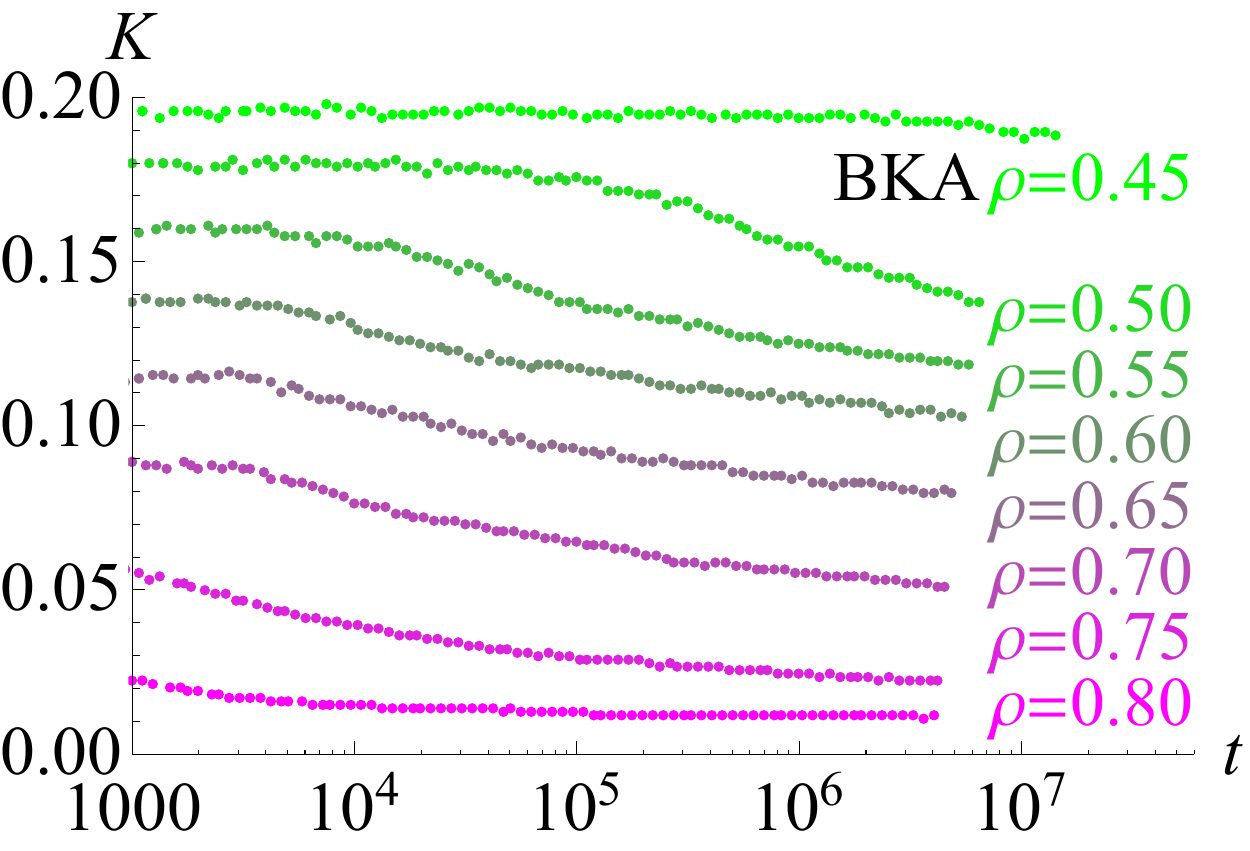}
\includegraphics[width=0.4\columnwidth]{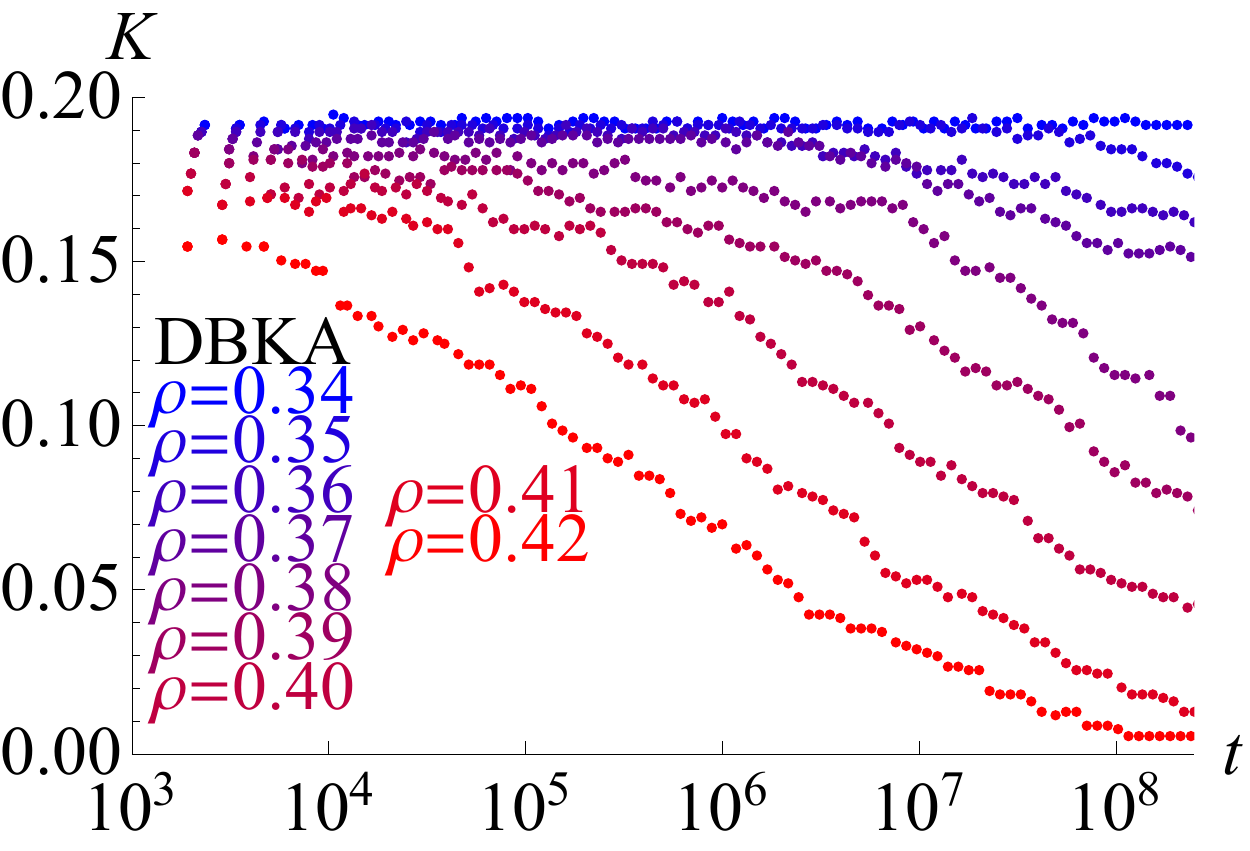}
\caption{Activity $K$ as a function of time for the BKA and DBKA models for different densities.}
\label{decay}
\end{figure}

We simulated the six models on a $30\times30$ lattice with periodic boundary conditions. For each density we averaged over $100$ realizations, which start from different random initial conditions. We also performed simulations on larger systems up to $100\times100$ (not shown), and found very small deviations due to finite-size effects \cite{Teomy2012,Teomy2014b,Toninelli2004,Teomy2014}. Figure \ref{activity_plot} compares the steady state activity evaluated from the simulations, the MF approximation and the SMF approximation. The SMF approximation over-estimates the activity in the simulations for all six models. While the KA, DKA, AKA and DAKA models converge to the steady state rather rapidly, the BKA and DBKA models converge very slowly for an intermediate range of densities ($0.37<\rho<0.43$ for the DBKA model and $0.50<\rho<0.81$ for the BKA model), as shown in Fig. \ref{decay}.

\begin{figure}[h!]
\includegraphics[width=0.6\columnwidth]{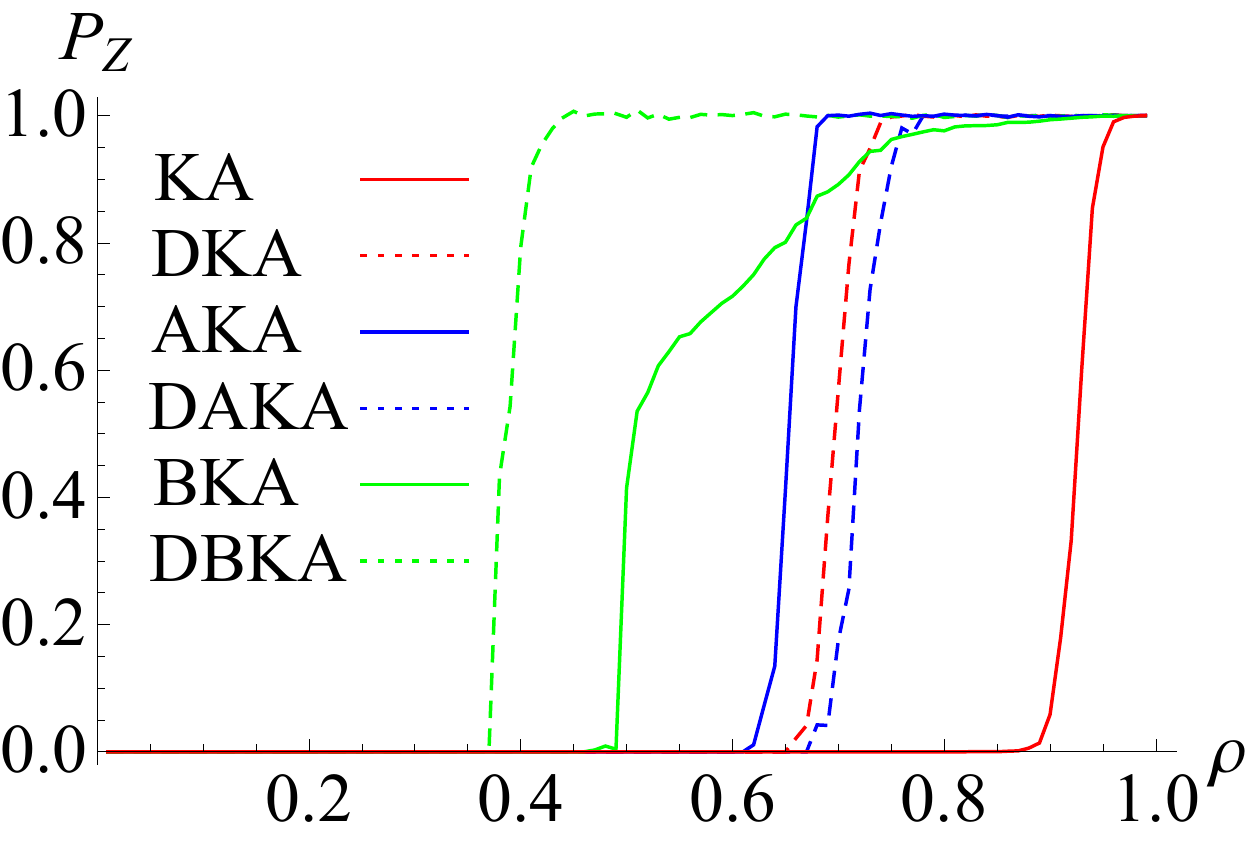}
\caption{Lower bound on the fraction of frozen particles, $P_{{\rm Z}}$, as a function of density $\rho$ after a very long time. Note that in the BKA model in the density range $0.5<\rho<0.81$, the system did not yet reach the steady state.}
\label{frozen_fig}
\end{figure}

We also measure in the simulations a lower bound on the fraction of frozen particles, $P_{{\rm Z}}$, i.e. those that will never be able to move. For a given configuration, we do this by an iterative culling procedure \cite{Jeng2008,Teomy2015}. This procedure starts by removing all mobile particles. In this new configuration, some particles which could not move before can now move, and we remove them too. We continue this procedure until all the remaining particle, if any, cannot be removed. This procedure gives a lower bound, since any particle which remains after this process is necessarily a frozen particle, but it is possible that some frozen particles have been removed \cite{Teomy2015}. Note that this procedure is not done during the dynamics, but on a snapshot of the system. Except for the BKA model which did not reach the steady state at intermediate densities ($0.5<\rho<0.81$), we find from Fig. \ref{frozen_fig} that $P_{{\rm Z}}$ jumps from $0$ to $1$ at some critical density. This critical density is the same one at which the activity vanishes, since zero activity implies $P_{{\rm Z}}=1$.

Note that the critical density of the KA model in a system of size $30\times30$ is $\rho^{\rm{KA}}_{c}(30)\approx0.93$ \cite{Teomy2014}, much higher than the numerically obtained critical densities of the other five models. This is in contrast to the MF approximation, which predicts $\rho^{\rm{KA}}_{c}=1$. However, as stated above, the finite value of the critical density in the KA model is strictly a well understood finite size effect, which does not affect significantly the critical densities of the other five models.

\begin{figure*}
\includegraphics[width=2\columnwidth]{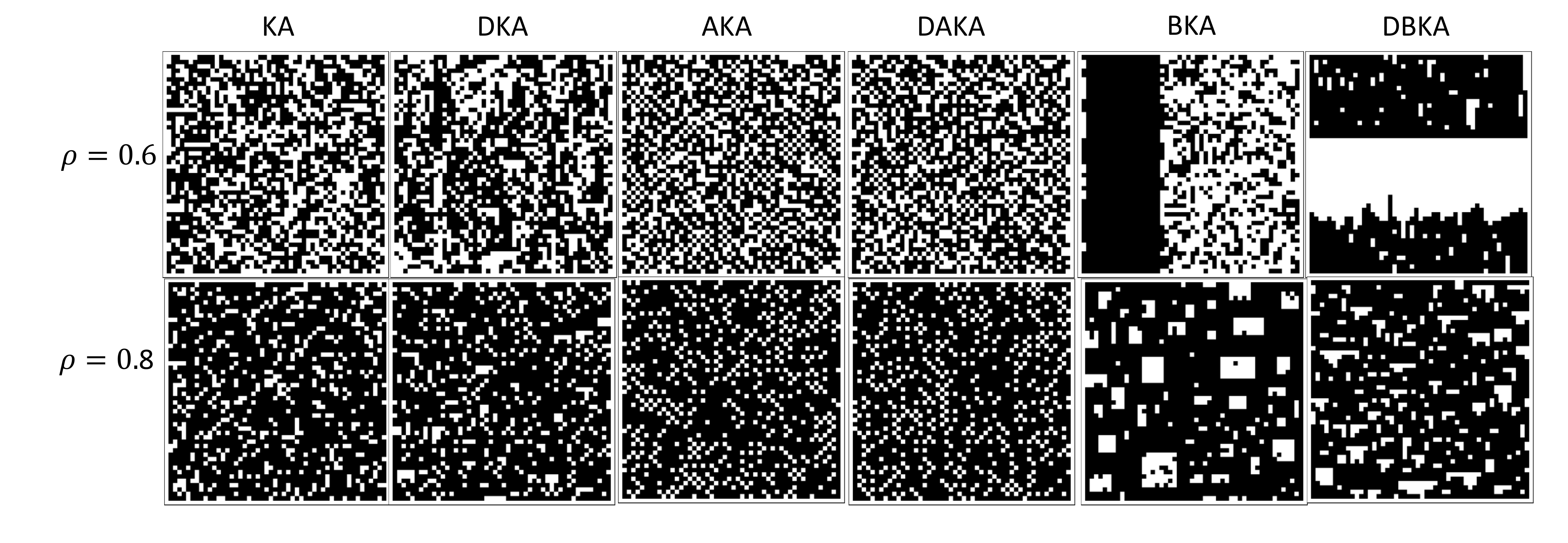}
\caption{A snapshot of $50\times 50$ systems after a long time at density $\rho=0.6$ (top) and $\rho=0.8$ (bottom), for six models.}
\label{snapshot_big}
\end{figure*}
In order to better understand the behavior of the system, we present in Fig. \ref{snapshot_big} typical configurations of a $50\times50$ system after a very long time. In the KA, DKA, AKA, DAKA and DBKA models the system has reached the steady state, while in the BKA it has not.

In the KA model, the system is always in equilibrium and there are no correlations between the occupancies of the sites. In the DKA model, the system is jammed at $\rho=0.8$ and there are scattered structures of vacancies. At $\rho=0.6$ the system is not jammed, but these structures can still be seen. These structures have been investigated in \cite{Bolshak2019,Turci2012}. The AKA and DAKA models appear very similar. At $\rho=0.8$ the system is jammed, and the vacancies tend to be arranged in a checkerboard pattern. At $\rho=0.6$ the system is not jammed, but there are jammed regions with a checkerboard pattern. These checkerboard patterns are the sparsest locally jammed structures in the AKA and DAKA models, and due to their symmetry may be extended indefinitely. Hence, once a checkerboard pattern appears it is unstable only at its boundary. However, above the critical density the accumulation of particles at its boundary does not allow the pattern to break, but rather causes it to grow.

\begin{figure}
\includegraphics[width=0.6\columnwidth]{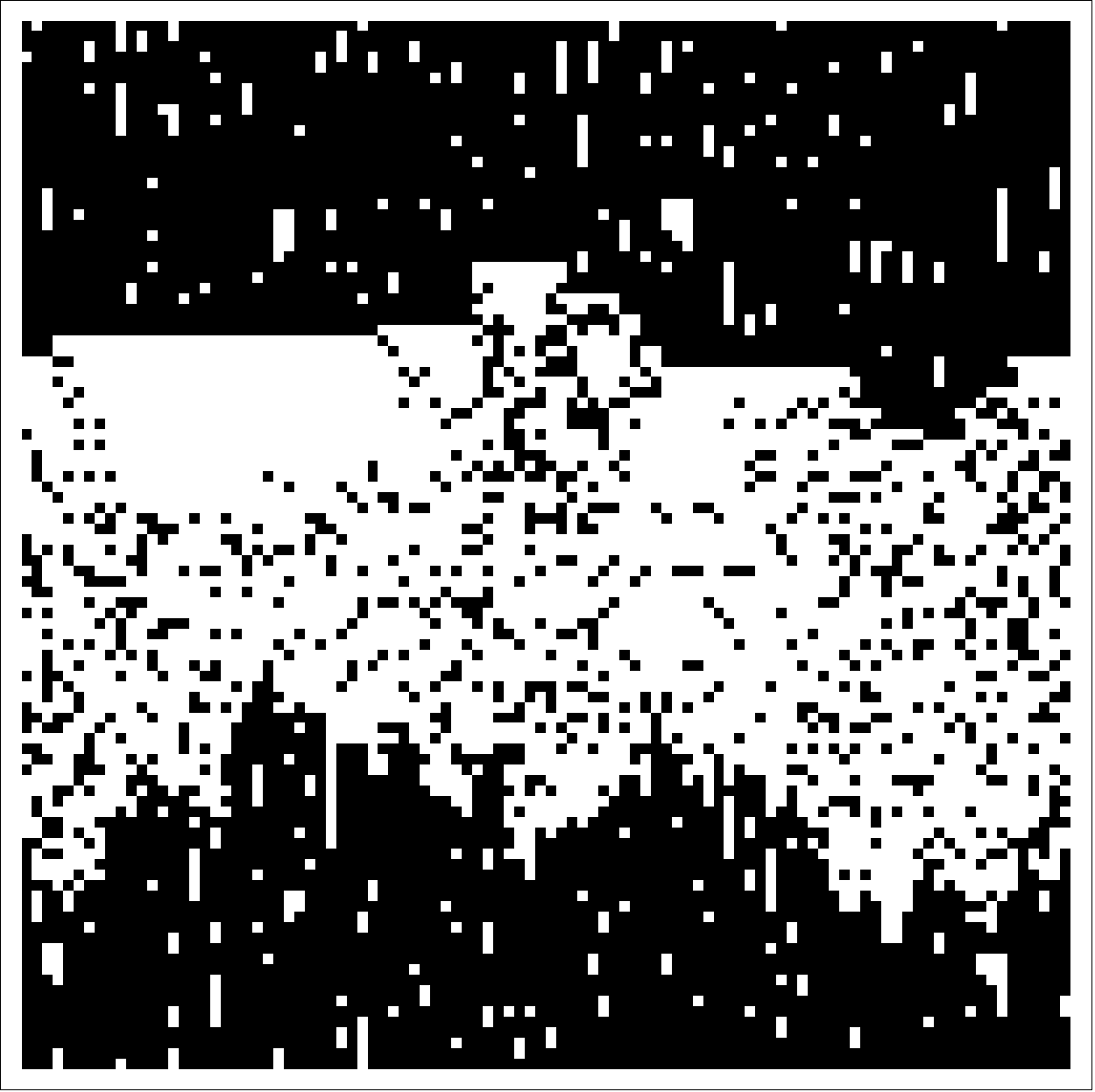}
\caption{A snapshot of a $100\times100$ system in the DBKA model at $\rho=0.6$}
\label{snow}
\end{figure}

The behavior of the BKA model and the DBKA model is more interesting. Before investigating the configurations, we note that any particle that is part of two consecutive full rows is permanently frozen, since it is blocked in three directions and jammed in the other direction. In the KA, DKA, AKA and DAKA models such a configuration cannot be generated dynamically, since a particle is prohibited from completing the second row. However, in the BKA and DBKA models such a configuration can be generated dynamically. At high densities in the DBKA model the system is jammed, and there are structures of vacancies reminiscent of those in the DKA model. At lower densities a front develops, which after some time settles into two full consecutive rows. At that point, these two rows cannot move, and thus the system becomes jammed after all the remaining particles drop onto these rows. The two consecutive rows always form in the direction normal to the driving. A snapshot of a $100\times100$ system in the DBKA model at $\rho=0.6$ before the onset of jamming is shown in Fig. \ref{snow}. The spontaneous formation of these jammed structures is the cause for the slow relaxation in the DBKA model, since it generally takes a very long time for this event to occur.

In the BKA model at low densities two consecutive full rows or columns can be generated dynamically, which then behave as an unmovable wall inside the system. This wall can grow thicker as other particles form full rows or columns adjacent to it. However, outside the wall, the remaining particles are still active. These walls spontaneously form in either of the two axes. As the density increases the system becomes divided into rectangles with rattlers, which may be thought of as enclosures between two pairs of parallel walls. If there are enough rattlers inside a rectangle, they can decrease the size of the rectangle by forming a full row or column adjacent to the rectangle's edge. If there are not enough rattlers to form a full row or column, they will continue rattling inside the rectangle forever. We hypothesize that in the infinite size limit, the density of infinitely rattling particles goes to zero. The slow relaxation in the BKA model is due to the rare events of rattlers forming a full row or column and decreasing the size of the rectangle. Since the formation of the wall is a collective effect of $O(L)$ particles, the relaxation time scale increases with system size.

\subsection{Temporal behavior of the activity}
The SMF approximation can also be used to describe the temporal behavior of the activity.
Starting form a random, uncorrelated initial condition, the SMF approximation can give insight into the short time dynamics, namely the temporal derivative of the activity at time $t=0$, before correlations start developing in the system. Since at $t=0$ the sites are uncorrelated, the activity at time $t=0$ is equal to the MF approximation.

\begin{figure}
\includegraphics[width=0.6\columnwidth]{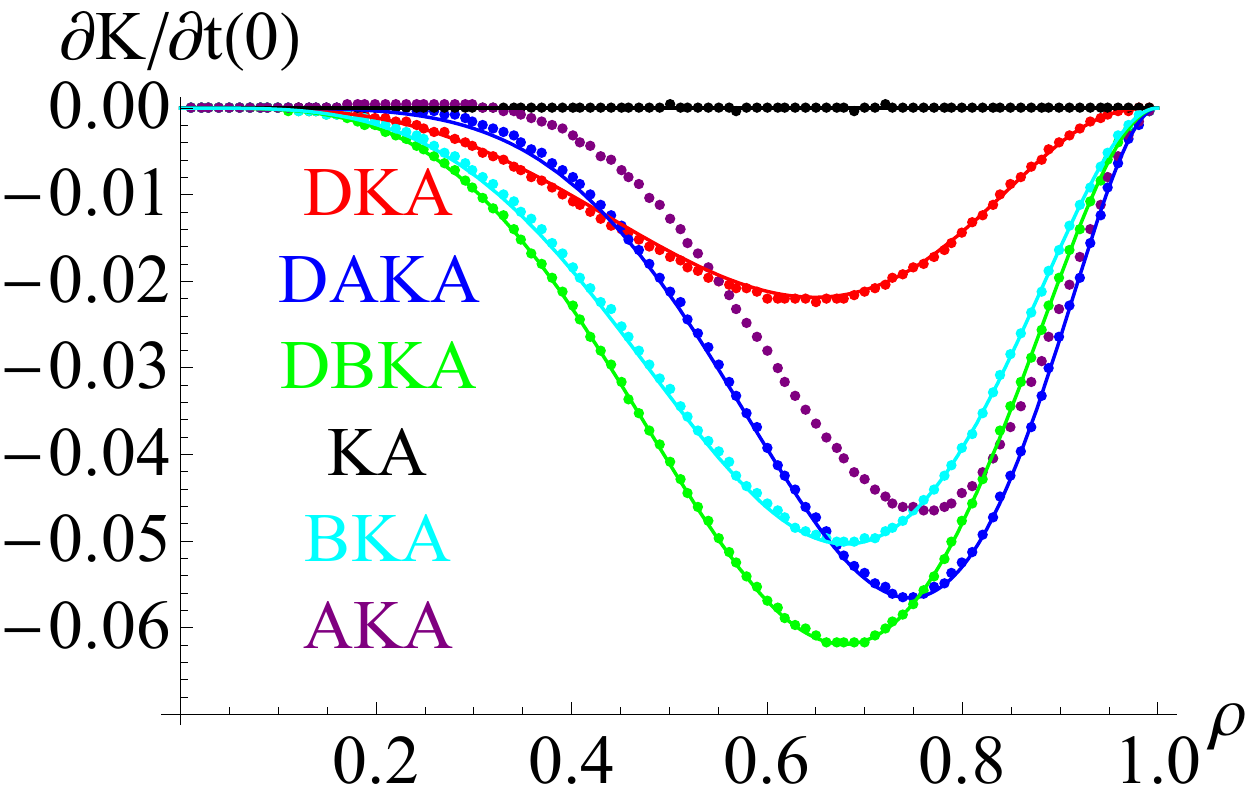}
\caption{The rate of change of the activity, $\partial K/\partial t$ at $t=0$ from the simulations (symbols) and the SMF approximation (continuous lines). The numerical results are averages over $10^5$ runs of $100\times100$ systems.}
\label{dkt_t0}
\end{figure}

The temporal derivative, $\partial K/\partial t$, is given by the SMF approximation with the probabilities $P_{\alpha}$ given by their MF values. Using Eq. (\ref{eveq}), we find that for the driven models
\begin{align}
&\left.\frac{\partial K^{\rm{DKA}}}{\partial t}\right|_{t=0}=\nonumber\\
&-\rho^{4}\left(1-\rho\right)^{3}\left(1+2\rho-\rho^{3}+2\rho^{4}+3\rho^{5}+\rho^{6}+\rho^{7}\right) ,\nonumber\\
&\left.\frac{\partial K^{\rm{DAKA}}}{\partial t}\right|_{t=0}=-\rho^{5}\left(1-\rho\right)^{2}\left(1+3\rho+\rho^{2}\right) ,\nonumber\\
&\left.\frac{\partial K^{\rm{DBKA}}}{\partial t}\right|_{t=0}=-\rho^{4}\left(1-\rho\right)^{2}\left(2+\rho+\rho^{2}-\rho^{3}\right) .
\end{align}
For the BKA model we find that
\begin{align}
\left.\frac{\partial K^{\rm{BKA}}}{\partial t}\right|_{t=0}=-\frac{1}{4}\rho^{4}\left(1-\rho^{2}\right)^{2}\left(7+\rho+6\rho^{2}-4\rho^{3}\right) .
\end{align}
In these four models, $\partial K/\partial t$ at $t=0$ is negative for all densities. Figure \ref{dkt_t0} shows the excellent agreement between the simulations and the analytical results. In the KA model, $\partial K^{{\rm KA}}/{\partial t}=0$ at all times and for all densities, since correlations never develop there.

\begin{figure}
\includegraphics[width=0.6\columnwidth]{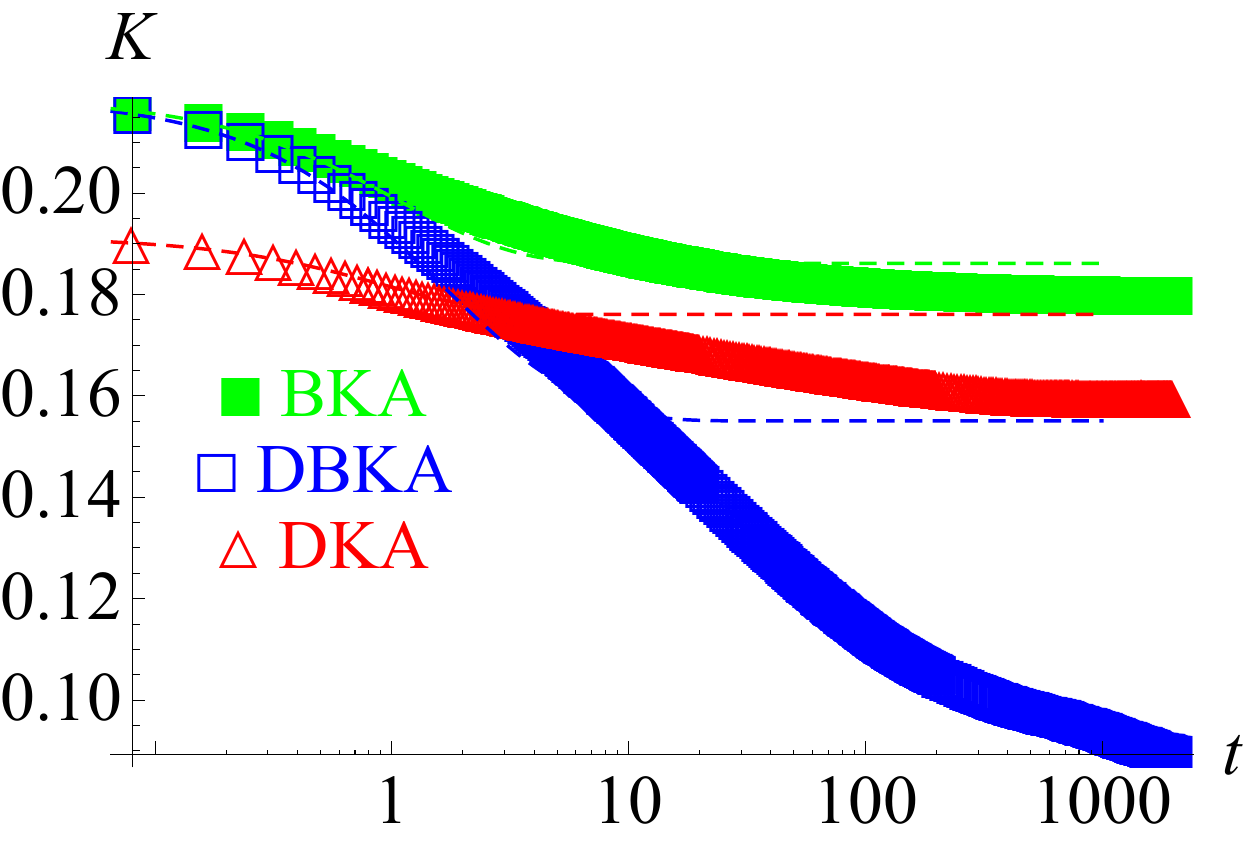}
\caption{The temporal behavior of the activity in the BKA, DBKA and DKA models for density $\rho=0.5$. The dashed lines are the SMF approximation. The numerical results are averages over $10^5$ runs of $100\times100$ systems.}
\label{ktime_other}
\end{figure}

\begin{figure}
\includegraphics[width=0.4\columnwidth]{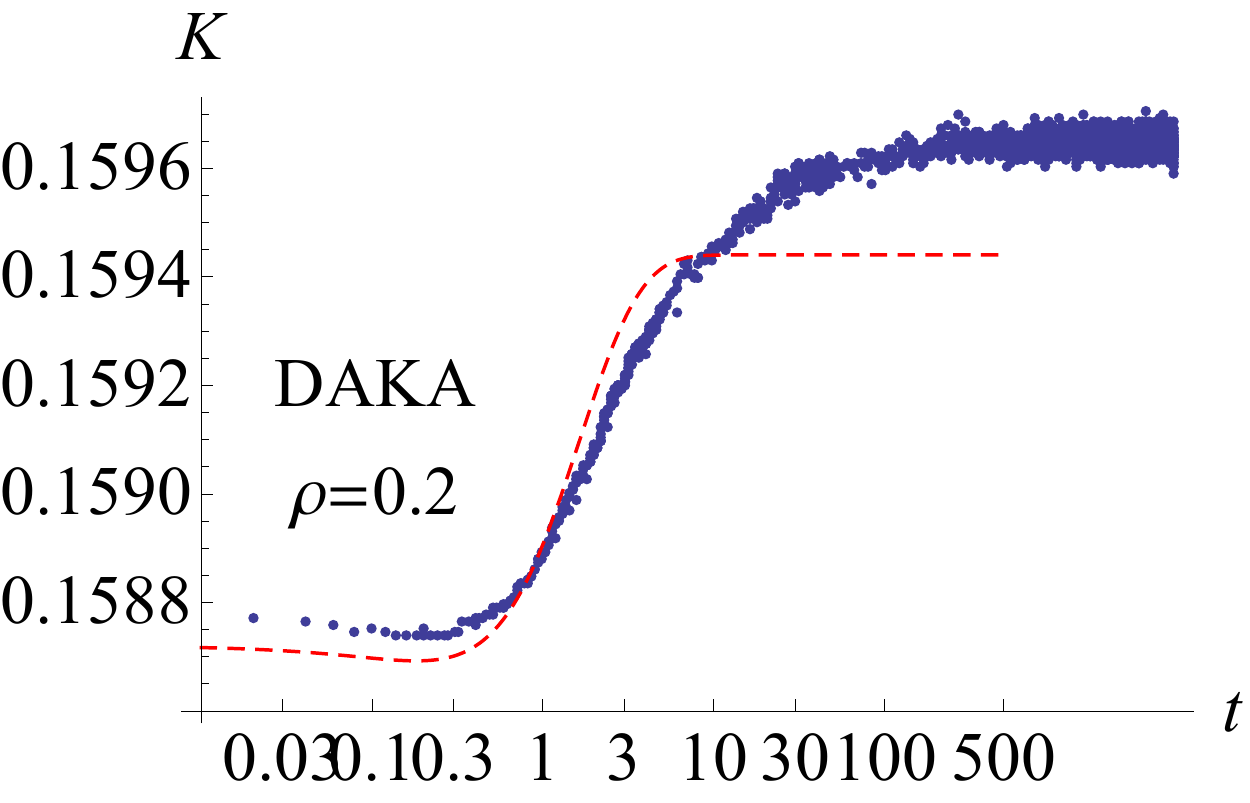}
\includegraphics[width=0.4\columnwidth]{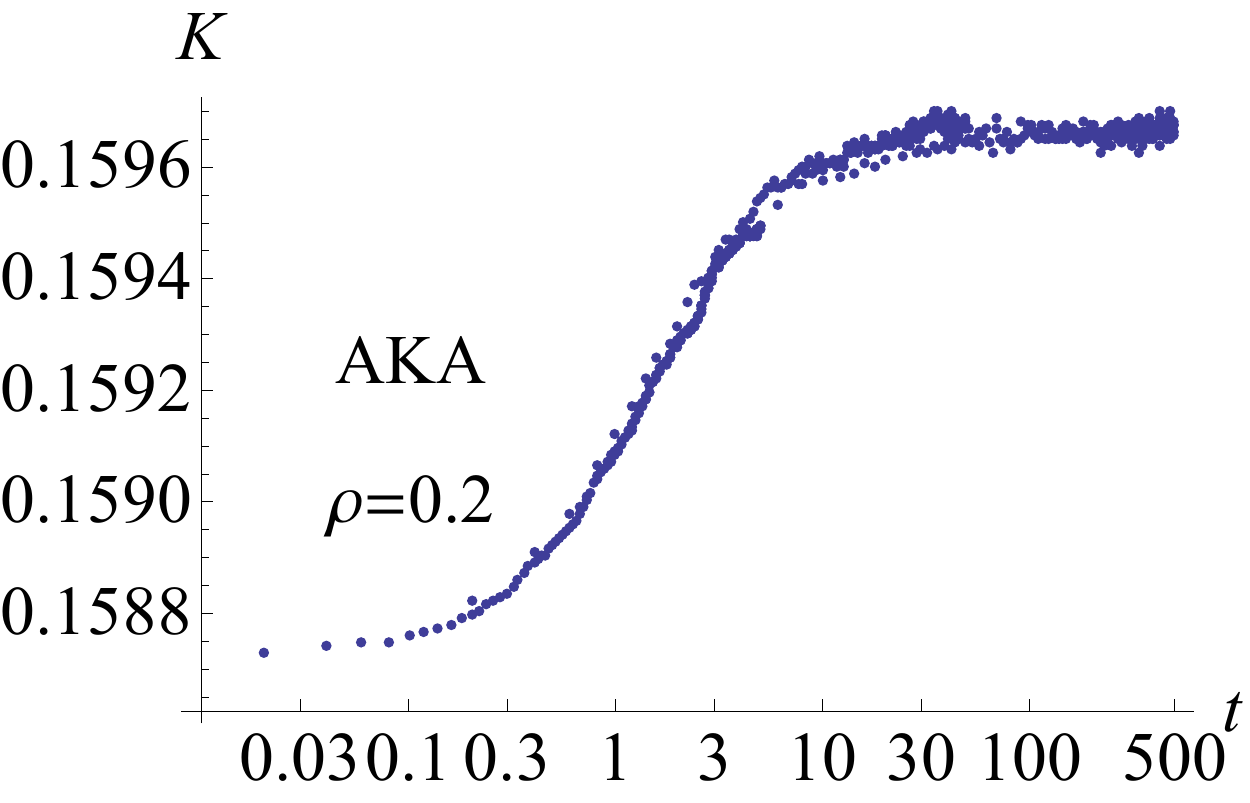}\\
\includegraphics[width=0.4\columnwidth]{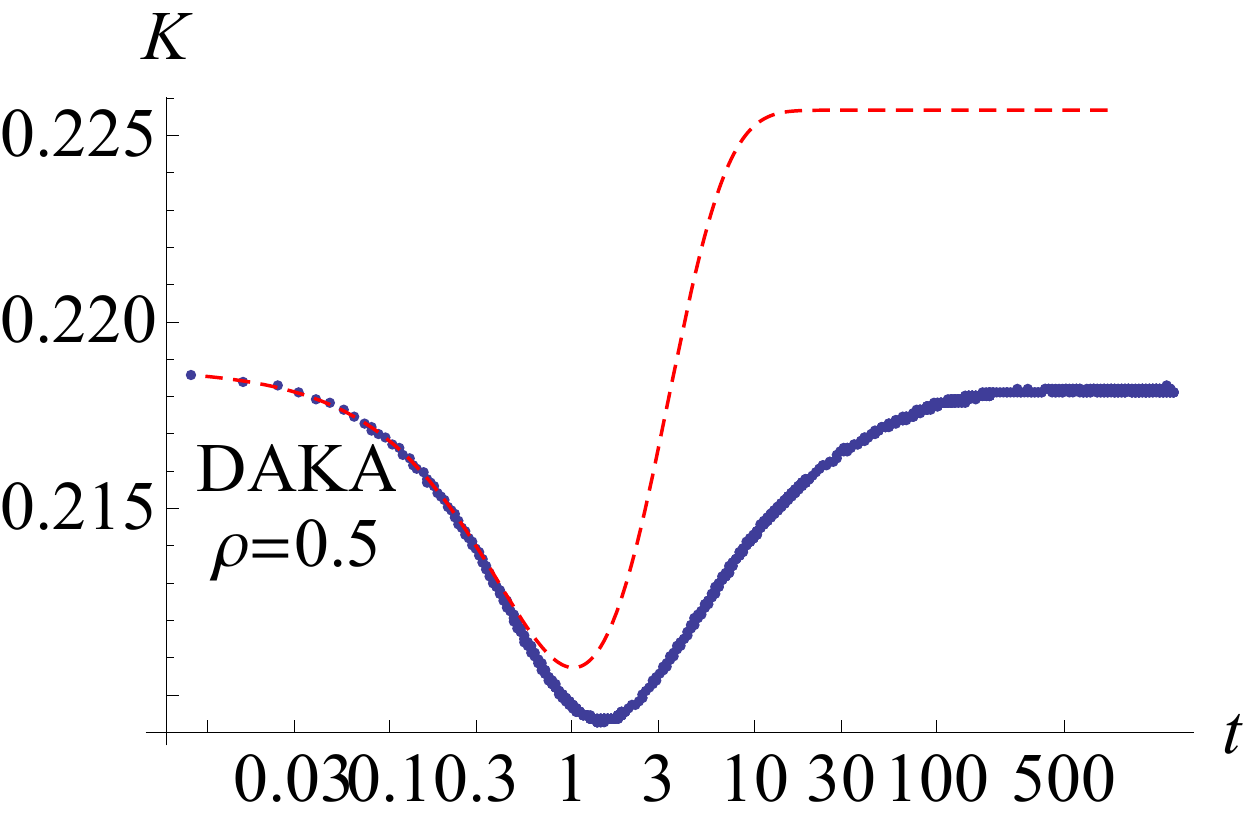}
\includegraphics[width=0.4\columnwidth]{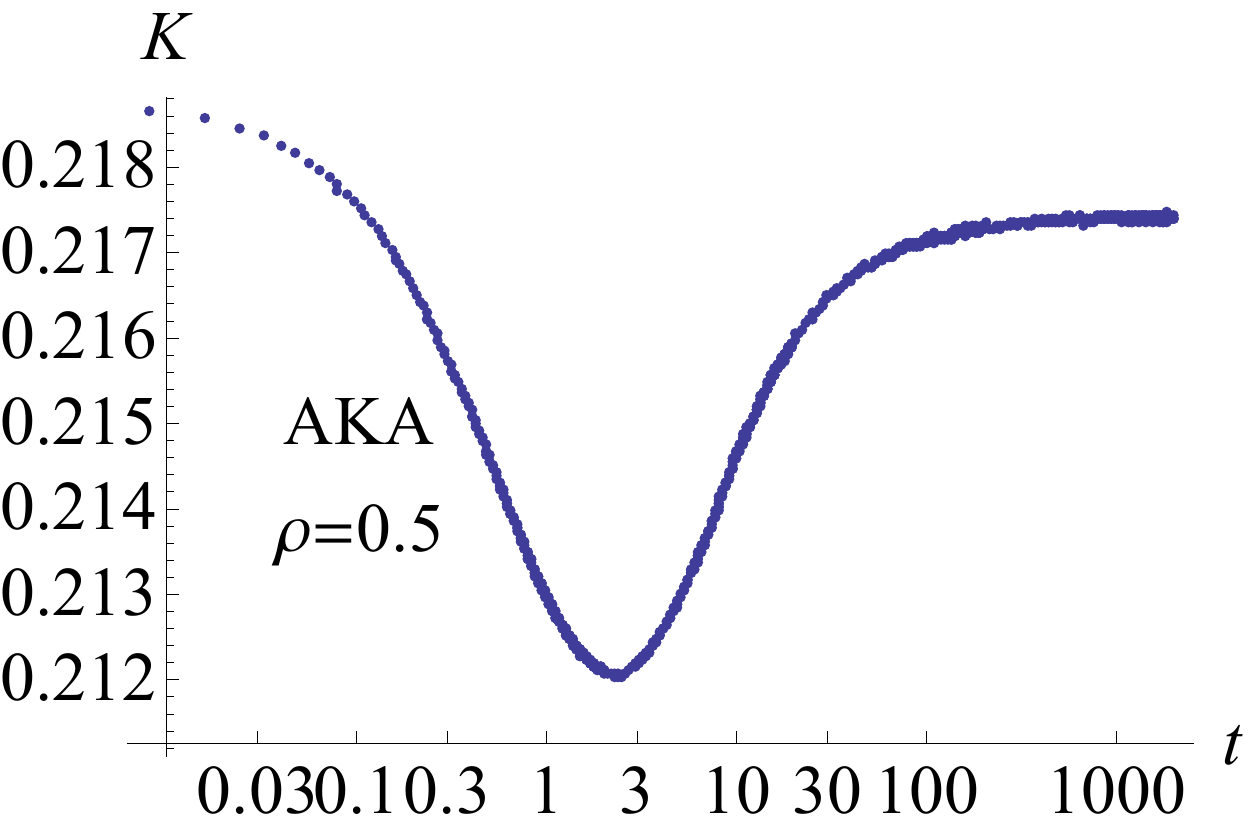}\\
\includegraphics[width=0.4\columnwidth]{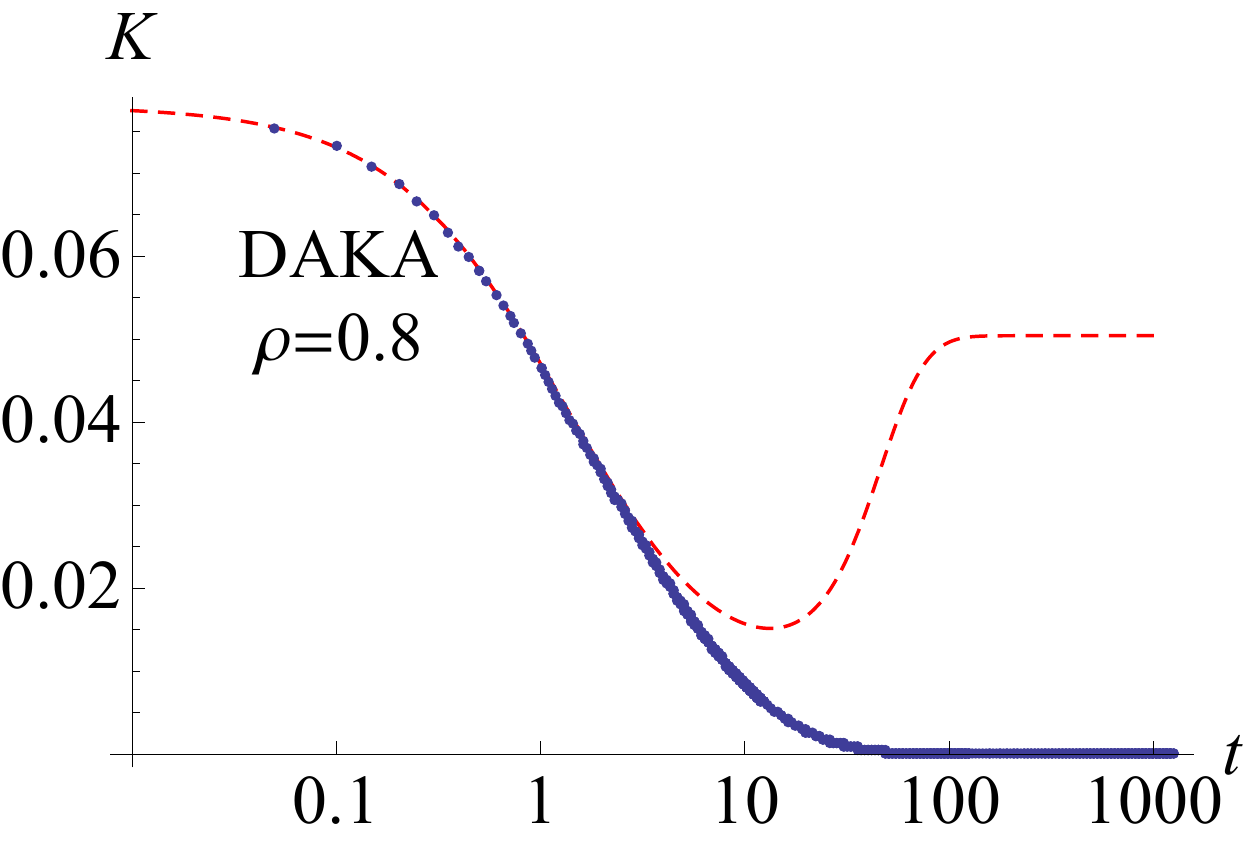}
\includegraphics[width=0.4\columnwidth]{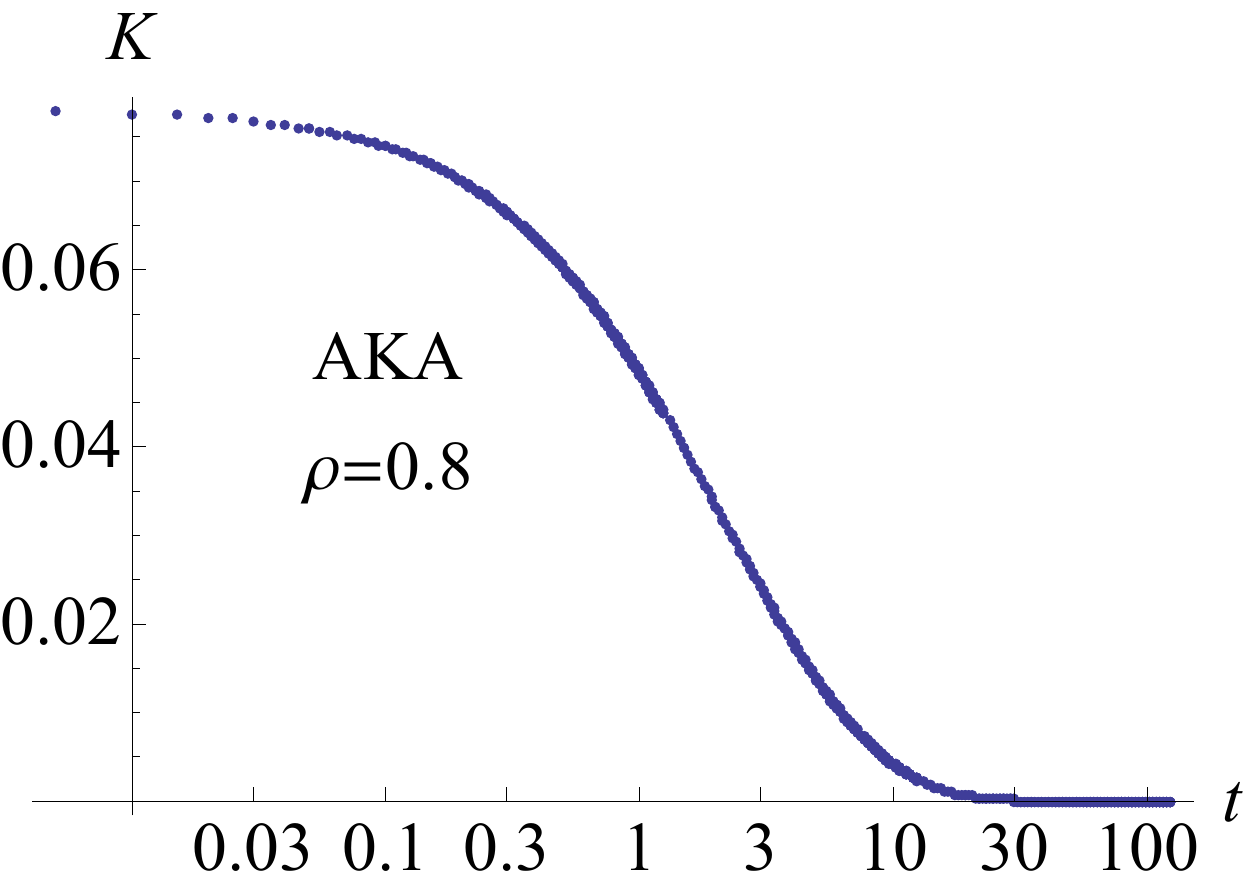}
\caption{The temporal behavior of the activity in the AKA and DAKA models for three different densities. The dotted red line is the SMF approximation. The numerical results are averages over $10^5$ runs of $100\times100$ systems.}
\label{ktime}
\end{figure}

Numerically, we see in Fig. \ref{ktime_other} that in the DKA, BKA and DBKA models, the activity decreases monotonically with time, while in Fig. \ref{ktime} we see that in the AKA and DAKA models, the activity is not monotonic with time for $\rho<\rho_{{\rm m}}$, with $\rho^{\rm{AKA}}_{{\rm m}}\approx0.66$ and $\rho^{\rm{DAKA}}_{{\rm m}}\approx0.64$. As shown in Fig. \ref{ktime}, for $\rho<\rho_{{\rm m}}$ the activity in the AKA and DAKA models first decreases until it reaches a minimum at time $t_{{\rm min}}$, then increases until it reaches the steady state, while for $\rho>\rho_{{\rm m}}$ it is monotonically decreasing. This result is counter-intuitive; one would expect that the activity would either increase or decrease monotonically with time, depending on whether the system becomes less or more restricted. 

The SMF approximation can qualitatively explain this behavior. Consider the evolution equation for $P_{\rm{B}}$ in the three driven models, given explicitly by
\begin{align}
&\frac{\partial P^{\rm{DKA}}_{\rm{B}}}{\partial t}=\frac{1-\rho^{2}}{1-\rho^{3}}\left(\rho-P^{\rm{DKA}}_{\rm{B}}\right)P^{\rm{DKA}}_{\rm{F}} ,\nonumber\\
&\frac{\partial P^{\rm{DAKA}}_{\rm{B}}}{\partial t}=\left[\frac{\rho\left(1-\rho^{2}\right)}{1-\rho^{3}}-P^{\rm{DAKA}}_{\rm{B}}\right]P^{\rm{DAKA}}_{\rm{F}} ,\nonumber\\
&\frac{\partial P^{\rm{DBKA}}_{\rm{B}}}{\partial t}=\left[\rho-P^{\rm{DBKA}}_{\rm{B}}\frac{1-\rho^{2}}{1-\rho^{3}}\right]P^{\rm{DBKA}}_{\rm{F}} ,
\end{align}
with the initial condition $\left.P_{\rm{B}}\right|_{t=0}=\rho$. In the DKA model, $P^{\rm{DKA}}_{\rm{B}}=\rho$ at all times, while in the DAKA (DBKA) model $\frac{\partial P_{\rm{B}}}{\partial t}$ is negative (positive) for all $\rho$. This means that $P^{\rm{DAKA}}_{\rm{B}}$ ($P^{\rm{DBKA}}_{\rm{B}}$) decreases (increases) monotonically with time from its initial value, $\rho$, to its steady state value. 

Now consider the evolution equation for $P_{\rm{F}}$ in the three driven models, which has the form
\begin{align}
\frac{\partial P_{\rm{F}}}{\partial t}=\left(C_{0}+C_{\rm{B}}P_{\rm{B}}+C_{\rm{F}}P_{\rm{F}}\right)P_{\rm{F}} ,\label{pfgen}
\end{align}
where $C_{0},C_{\rm{B}}$ and $C_{\rm{F}}$ depend only on the density. For the DKA and DBKA models, $\partial P_{\rm{F}}/\partial t$ is negative for all values of $P_{\rm{B}}$ and $P_{\rm{F}}$ between their initial values and the steady state values, for all densities, and thus $P_{\rm{F}}$ in these two models monotonically decreases with time, and so does the activity. In the DAKA model, however, there are values of $P_{\rm{B}}$ and $P_{\rm{F}}$ between their initial and steady state values for which $\partial P_{\rm{F}}/\partial t$ is positive. Furthermore, for $P^{\rm{DAKA}}_{\rm{B}}$ equal to its steady state value, and $\rho<\rho^{SMF}_{DAKA}\approx0.933$, we find that $\partial P_{\rm{F}}/\partial t$ is positive for all values of $P^{\rm{DAKA}}_{\rm{F}}$ between its initial and steady state values.

\begin{figure}
\includegraphics[width=0.6\columnwidth]{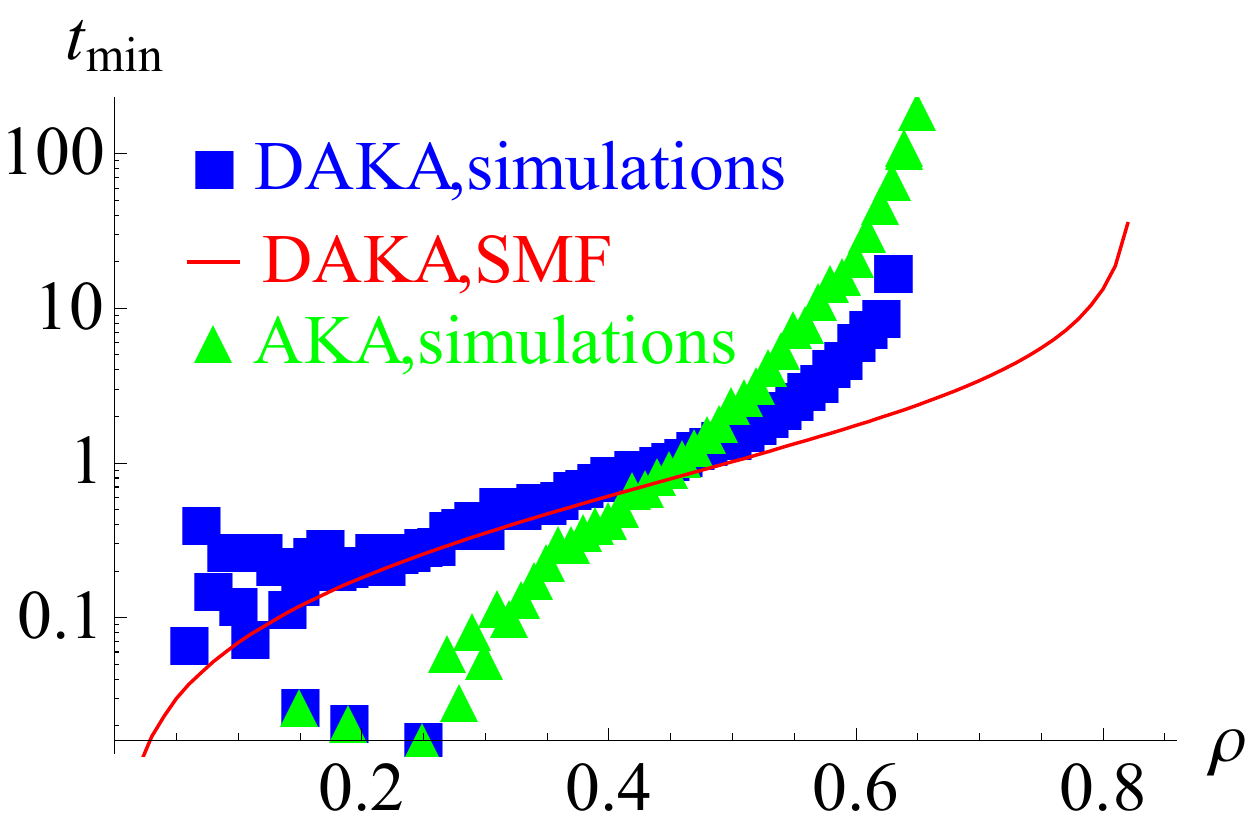}
\caption{The time at which the activity reaches a minimum, $t_{{\rm min}}$, as a function of density $\rho$.}
\label{tmin}
\end{figure}

Figure \ref{tmin} shows the value of $t_{{\rm min}}$ vs. the density. The SMF approximation predicts that $t_{{\rm min}}$ diverges at $\rho^{\rm{DAKA}}_{{\rm m}}\approx 0.827$, while according to the simulations it diverges at $\rho\approx0.64$.

\section{Discussion}
\label{sec_conclusion}
In this paper we showed that counter-intuitively, adding transitions to a dynamical system may decrease its activity. We analyzed this scenario by investigating six related lattice gas models: the equilibrium KA model, and five non-equilibrium variants of it (AKA,BKA,DKA,DAKA,DBKA). In some cases adding transitions increases the activity (DKA$\rightarrow$DAKA, KA$\rightarrow$AKA at small densities), while in other cases it decreases the activity (DKA$\rightarrow$DBKA, KA$\rightarrow$BKA). 

The difference lies in the topology of the phase-space for each model. For example, consider the undriven models KA, AKA and BKA. The phase space of the KA model is composed of a large part which contains states in which none of the particles are permanently frozen, and many small parts each of them contains states in which a specific subset of the particles cannot move, most of them due to the particles being in two (or more) consecutive rows or columns. The AKA and BKA models add transitions between the different disjoined parts. In the BKA model, the permanently frozen walls cannot be broken, but they can form dynamically. Therefore, the added transitions in the BKA model between the different parts of the state space is into a more jammed structure. In the AKA model, the permanently frozen walls can be broken, and so the added transitions between the parts allow the system to escape from these jammed parts. At high enough density, there are other jammed structures which are 2D in nature, not quasi-1D as the walls. The AKA model also allows transitions into these 2D jammed structures, but not out of them, and thus at high enough density it also jams.

It would be interesting to continue investigating the new models we described in this paper: AKA, DAKA, BKA and DBKA. For example, the critical density we found in the simulations is for a system of size $30\times30$, and there are bound to be finite-size effects. Also, the simulations always started from an uncorrelated initial condition, and an interesting question is how does the initial condition affects the dynamics, since the initial conditions affect even models which obey detailed balance \cite{Corberi2009}. Other points which are worth investigating are the correlations and the relaxation time, especially in the BKA and DBKA models.

\section*{Acknowledgments}
We thank Gregory Bolshak, Rakesh Chatterjee, and Erdal O\u{g}uz for fruitful discussions. This research was supported in part by the Israel Science Foundation Grant No. 968/16 and by the National Science Foundation Grant No. NSF PHY-1748958.

\appendix

\section{Derivation of the transition rates in the driven models}
\label{app_driven}
In this Appendix we derive the transition rates for the three driven models. Each possible transition is illustrated in the configuration before the move, and an arrow shows which particle moves where. The particle whose state changes between free, blocked, and jammed is called the \textit{main particle}, and it is marked in the illustrations with a green circle. A site with an empty circle represent an occupied site, and a site without any mark represent a vacant site. Other symbols represent sites which may or may not be occupied, and are specified for each rate. In the derivation of the rates, we use various kinds of conditional probabilities, which are evaluated in Section \ref{app_cond}. The rates for the DAKA, DBKA and DKA model are derived in Sections \ref{app_daka}, \ref{app_dbka} and \ref{app_dka}, respectively. The derived rates $\omega_{\alpha,\beta}$ and $\Omega_{\alpha,\beta}$ for the three driven models, related to the rates $r_{\alpha,\beta}$ via Eq. (\ref{omega_def}), are summarized in Table \ref{table_rates}. Note that $\omega_{{\rm J,B}}=\omega_{{\rm F,B}}=0$ for all the driven models, since a particle can change into a blocked state only if it moves downwards.
\begin{widetext}
\begin{center}
\begin{table}
\begin{tabular}{c|c|c|c}
 & DKA & DAKA & DBKA \\\hline
$\omega_{{\rm B,J}}$ & $\frac{\rho^{3}\left(1-\rho^{2}\right)}{1-\rho^{3}}$ & $\rho^{2}$ & $\frac{\rho^{3}\left(1-\rho^{2}\right)}{1-\rho^{3}}$ \\
$\omega_{{\rm B,F}}$ & $1-\rho^{2}$ & $1-\rho^{2}$ & $1-\rho^{2}$ \\
$\omega_{{\rm J,F}}$ & $\frac{1}{2-\rho^{3}}\left[1+\frac{2\left(1-\rho^{2}\right)}{1-\rho^{3}}+\frac{2\left(1-\rho^{2}\right)^{2}}{\left(1-\rho^{3}\right)^{2}}\right]$ & $1+2\frac{1-\rho^{2}}{1-\rho^{3}}$ & $2\frac{1-\rho^{2}}{1-\rho^{3}}$ \\
$\Omega_{{\rm F,B}}$ & $\frac{\rho\left(1-\rho^{2}\right)}{1-\rho^{3}}$ & $\frac{\rho\left(1-\rho^{2}\right)}{1-\rho^{3}}$ & $\rho$ \\
$\omega_{{\rm F,J}}$ & $\frac{\rho^{3}\left(1-\rho^{2}\right)}{\left(1-\rho^{3}\right)^{2}}\left[1+\frac{2\left(1-\rho\right)\left(2+2\rho-\rho^{3}\right)}{\left(1-\rho^{3}\right)^{2}}\right]$ & $\frac{2\rho^{3}}{\left(1-\rho^{3}\right)^{2}}$ & $\frac{\rho^{3}}{1-\rho^{3}}\left(1+2\frac{1-\rho^{2}}{1-\rho^{3}}\right)$ \\
$\Omega_{{\rm F,J}}$ & $\frac{\rho^{3}\left(1-\rho\right)}{1-\rho^{3}}$ & $\frac{\rho^{3}\left(1-\rho\right)}{1-\rho^{3}}$ & $0$ \\
\end{tabular}
\caption{The rates $\omega_{\alpha,\beta}$ and $\Omega_{\alpha,\beta}$ for the three driven models.}
\label{table_rates}
\end{table}
\end{center}
\end{widetext}

\subsection{Conditional probabilities}
\label{app_cond}
In this section we evaluate several conditional probabilities by using a MF approach. We start with the probability that a particle is free given that one of its constraining sites is occupied, $P_{\rm{F}|\rho}$. The conditional probability is given by
\begin{align}
P_{\rm{F}|\rho}=\frac{P_{\rm{F}\cap\rho}}{\rho} ,
\end{align}
where $P_{\rm{F}\cap\rho}$ is the probability that the particle is free \textit{and} one of its constraining sites is occupied. We approximate $P_{\rm{F}\cap\rho}$ as $P_{\rm{F}}$ multiplied by the mean-field fraction of free configurations with one constraining site occupied, i.e.
\begin{align}
P_{\rm{F}\cap\rho}=\frac{\rho\left(1-\rho^{2}\right)}{1-\rho^{3}}P_{\rm{F}} .
\end{align}
Hence,
\begin{align}
P_{\rm{F}|\rho}=\frac{\left(1-\rho^{2}\right)}{1-\rho^{3}}P_{\rm{F}} .
\end{align}
In a similar fashion, we find that the probability that a particle is free given that one of its constraining sites is vacant, $P_{\rm{F}|v}$, is
\begin{align}
P_{\rm{F}|v}=\frac{P_{\rm{F}\cap v}}{1-\rho}=\frac{1}{1-\rho^{3}}P_{\rm{F}} .
\end{align}

The probability that a particle is free given that the site below is vacant, $P_{\rm{F}|t}$, is
\begin{align}
P_{\rm{F}|t}=\frac{P_{\rm{F}\cap t}}{1-\rho}=\frac{P_{\rm{F}}}{1-\rho} .
\end{align}
the probability that a particle is free given that one of its constraining sites is occupied and the site below it is vacant, $P_{\rm{F}|\rho t}$, is
\begin{align}
P_{\rm{F}|\rho t}=\frac{P_{\rm{F}\cap \rho\cap t}}{\rho\left(1-\rho\right)}=\frac{P_{\rm{F}\cap \rho}}{\rho\left(1-\rho\right)}=\frac{P_{\rm{F}|\rho}}{1-\rho} .
\end{align}
The probability that a particle is free given that one of its constraining sites and the site below it are vacant, $P_{\rm{F}|vt}$, is
\begin{align}
P_{\rm{F}|vt}=\frac{P_{\rm{F}\cap v\cap t}}{\left(1-\rho\right)^{2}}=\frac{P_{\rm{F}\cap v}}{\left(1-\rho\right)^{2}}=\frac{P_{\rm{F}|v}}{1-\rho} .
\end{align}

The probability that a site is occupied given that it is a constraining site of a free particle, $P_{\rho|{\rm F}}$, is given by
\begin{align}
P_{\rho|{\rm F}}=\frac{P_{\rm{F}\cap\rho}}{P_{\rm{F}}}=\frac{\rho\left(1-\rho^{2}\right)}{1-\rho^{3}} ,
\end{align}
and the probability that a site is vacant given that it is a constraining site of a free particle, $P_{v|{\rm F}}$, is given by
\begin{align}
P_{v|{\rm F}}=\frac{P_{\rm{F}\cap v}}{P_{\rm{F}}}=\frac{1-\rho}{1-\rho^{3}} ,
\end{align}
The probability that two sites are occupied given that they are constraining sites in the same group of a free particle, $P_{\rho^{2}|{\rm F}}$, is given by
\begin{align}
P_{\rho^{2}|{\rm F}}=\frac{P_{\rm{F}\cap\rho^{2}}}{P_{\rm{F}}}=\frac{\rho^{2}\left(1-\rho\right)}{1-\rho^{3}} .
\end{align}

In the DKA model, we are also interested in the probabilities that a particle is free given some condition on both groups of constraining particles, $P_{\rm{F}|s_{1},s_{2}}$, and the probabilities that the two groups satisfy a certain condition given that they are constraining groups of a free particle, $P_{s_{1},s_{2}|{\rm F}}$. Since in the MF approximation the occupancies of the two groups are independent of each other, we find that
\begin{align}
&P_{\rm{F}|s_{1},s_{2}}=\frac{P_{\rm{F}|s_{1}}P_{\rm{F}|s_{2}}}{P_{\rm{F}}} ,\nonumber\\
&P_{s_{1},s_{2}|{\rm F}}=P_{s_{1}|{\rm F}}P_{s_{2}|{\rm F}} .
\end{align}
Similarly, the probability that a particle is free given some conditions on both groups of constraining particles and that the site below is vacant, $P_{\rm{F}|s_{1},s_{2},t}$, is given by
\begin{align}
P_{\rm{F}|s_{1},s_{2},t}=\frac{P_{\rm{F}|s_{1},s_{2}}}{1-\rho} .
\end{align}

We are also interested in the conditional probabilities in the DKA model that the two constraining groups are in a certain configuration given that the main particle is jammed. Specifically, the probability that one of the constraining groups is fully occupied and the other is not, $P_{\rho^{3},1-\rho^{3}|{\rm J}}$, and the probability that one of the constraining groups is fully occupied and in the other one specific site is vacant and at least one of the other two is also vacant, $P_{\rho^{3},v\left(1-\rho^{2}\right)}$. These probabilities are given by
\begin{align}
&P_{\rho^{3},1-\rho^{3}|{\rm J}}=\frac{P_{\rm{J}\cap\rho^{3}\cap 1-\rho^{3}}}{P_{\rm{J}}}=\frac{\rho^{3}\left(1-\rho^{3}\right)}{\rho^{6}+2\rho^{3}\left(1-\rho^{3}\right)} ,\nonumber\\
&P_{\rho^{3},v\left(1-\rho^{2}\right)|{\rm J}}=\frac{P_{\rm{J}\cap\rho^{3}\cap v\left(1-\rho^{2}\right)}}{P_{\rm{J}}}=\frac{\rho^{3}\left(1-\rho\right)\left(1-\rho^{2}\right)}{\rho^{6}+2\rho^{3}\left(1-\rho^{3}\right)} .
\end{align}

\subsection{DAKA model}
\label{app_daka}
\subsubsection{$r^{\rm{DAKA}}_{\rm{B},\rm{J}}$ and $r^{\rm{DAKA}}_{\rm{B},\rm{F}}$}
\begin{figure}
\includegraphics[width=0.3\columnwidth]{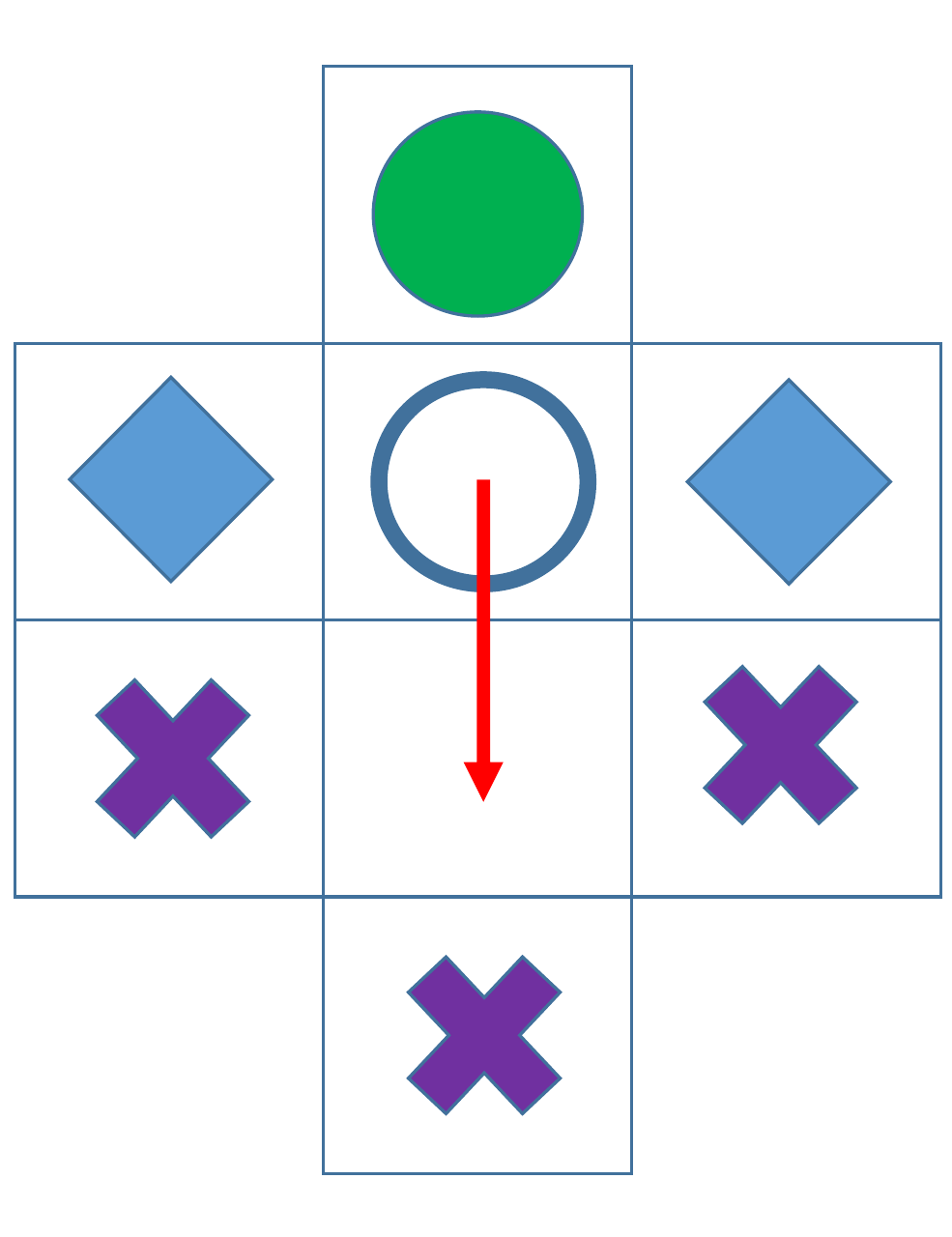}
\caption{An illustration of the transitions ${\rm B\rightarrow J}$ and ${\rm B\rightarrow F}$ in the DAKA model}
\label{rate_fig1}
\end{figure}

Before the move, the blocked configuration consists of the main particle (green circle in Fig. \ref{rate_fig1}) and the blocking particle (empty circle). The state of the main particle can change only if the blocked particle moves. The blocked particle can move with probability $P^{\rm{DAKA}}_{\rm{F}}$, since the occupancy of the three sites comprising its kinetic constraint (purple $\times$) is irrelevant to the state of the main particle either before or after the move. If both sites marked with a blue $\diamond$ are occupied, which occurs with probability $\rho^{2}$, then after the move the main particle will be jammed, otherwise it will be free. Therefore,
\begin{align}
&r^{\rm{DAKA}}_{\rm{B},\rm{J}}=\rho^{2}P_{\rm{F}} ,\nonumber\\
&r^{\rm{DAKA}}_{\rm{B},\rm{F}}=\left(1-\rho^{2}\right)P_{\rm{F}} .
\end{align}

\subsubsection{$r^{\rm{DAKA}}_{\rm{J},\rm{F}}$ and $r^{\rm{DAKA}}_{\rm{J},\rm{B}}$}
\begin{figure}
\includegraphics[width=0.6\columnwidth]{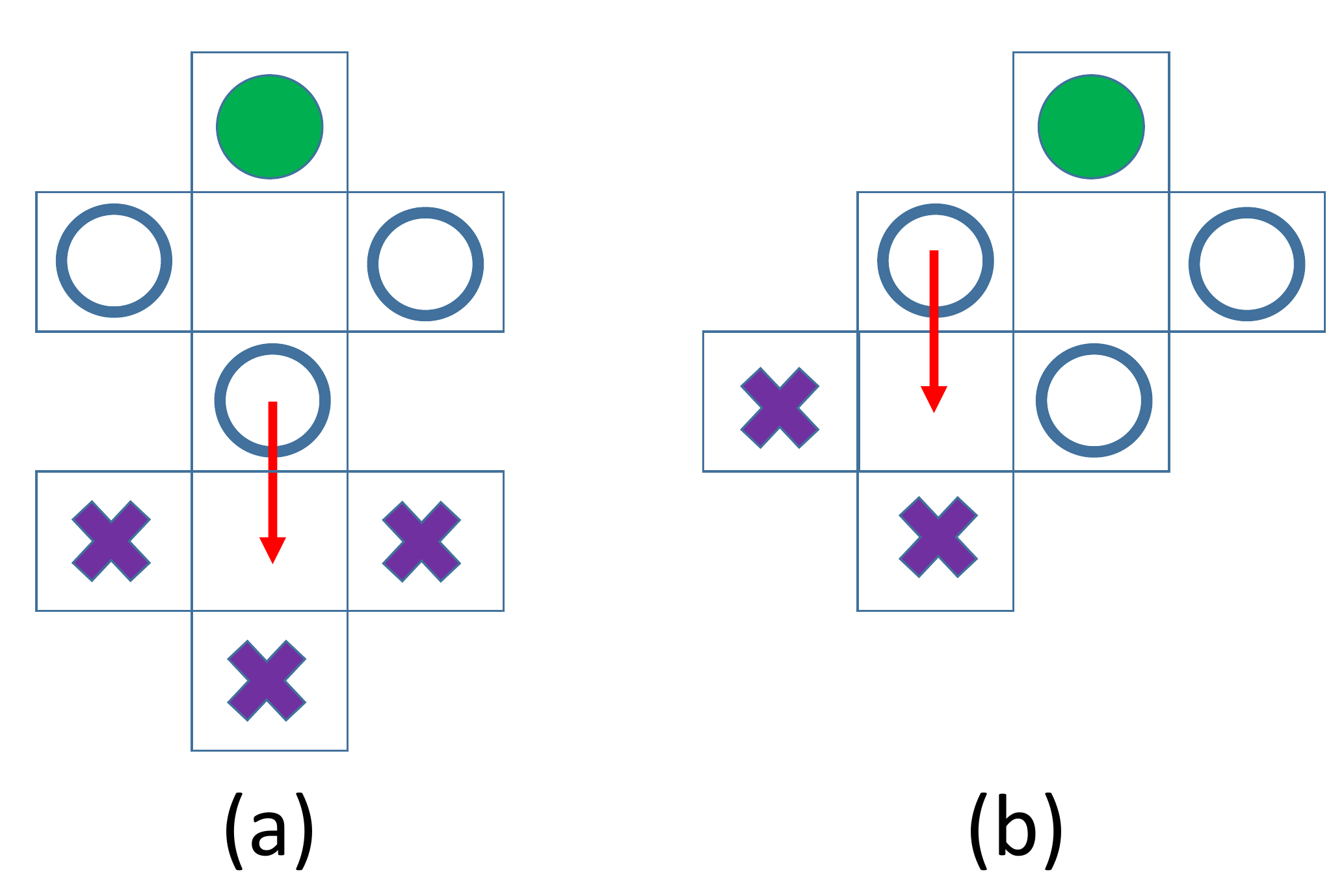}
\caption{An illustration of the transition ${\rm J\rightarrow F}$ in the DAKA model}
\label{rate_fig2}
\end{figure}

Before the move, the jammed configuration consists of the main particle (green circle in Fig. \ref{rate_fig2}) and the three jamming particles (empty circles). The state of the main particle can change only if one of the three jamming particles moves, and it can only change to a free state, not a blocked state, hence
\begin{align}
r^{\rm{DAKA}}_{\rm{J},\rm{B}}=0 .
\end{align}
The middle jamming particle (Fig. \ref{rate_fig2}a) can move with probability $P_{\rm{F}}$, while the jamming particles on the two sides (Fig. \ref{rate_fig2}b) can move with probability $P_{\rm{F}|\rho}$. Therefore, the rate $r^{\rm{DAKA}}_{\rm{J},\rm{F}}$ is
\begin{align}
r^{\rm{DAKA}}_{\rm{J},\rm{F}}=P_{\rm{F}}+2P_{\rm{F}|\rho}=P_{\rm{F}}+2\frac{1-\rho^{2}}{1-\rho^{3}}P_{\rm{F}} .
\end{align}

\subsubsection{$r^{\rm{DAKA}}_{\rm{F},\rm{B}}$}
\begin{figure}
\includegraphics[width=0.3\columnwidth]{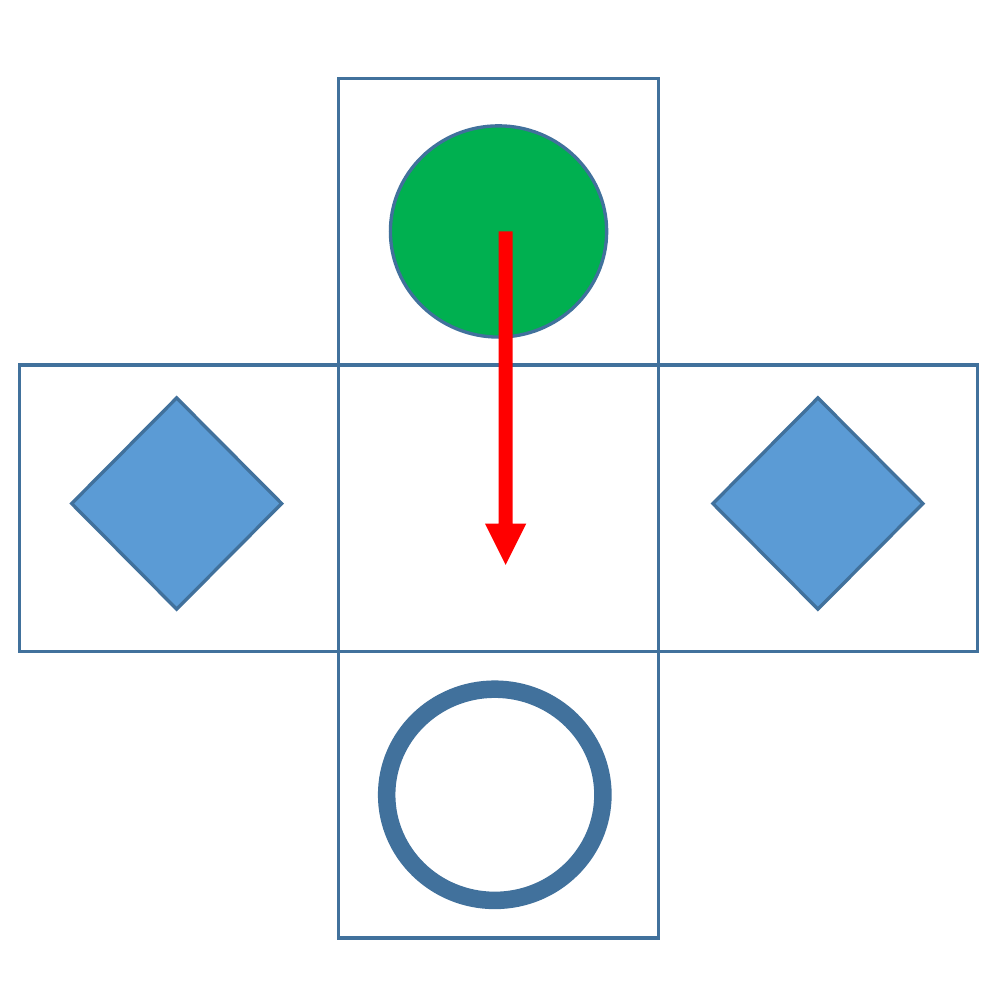}
\caption{An illustration of the transition ${\rm F\rightarrow B}$ in the DAKA model}
\label{rate_fig3}
\end{figure}

Before the move the configuration consists of the main free particle (green circle in Fig. \ref{rate_fig3}) and the particle which will block the main particle after the move (empty circle). The rate is given by the probability that the bottom site is occupied given that the main particle is free, $P_{\rho|{\rm F}}$, and thus
\begin{align}
r^{\rm{DAKA}}_{\rm{F},\rm{B}}=P_{\rho|{\rm F}}=\frac{\rho\left(1-\rho^{2}\right)}{1-\rho^{3}} .
\end{align}

\subsubsection{$r^{\rm{DAKA}}_{\rm{F},\rm{J}}$}
\begin{figure}
\includegraphics[width=0.6\columnwidth]{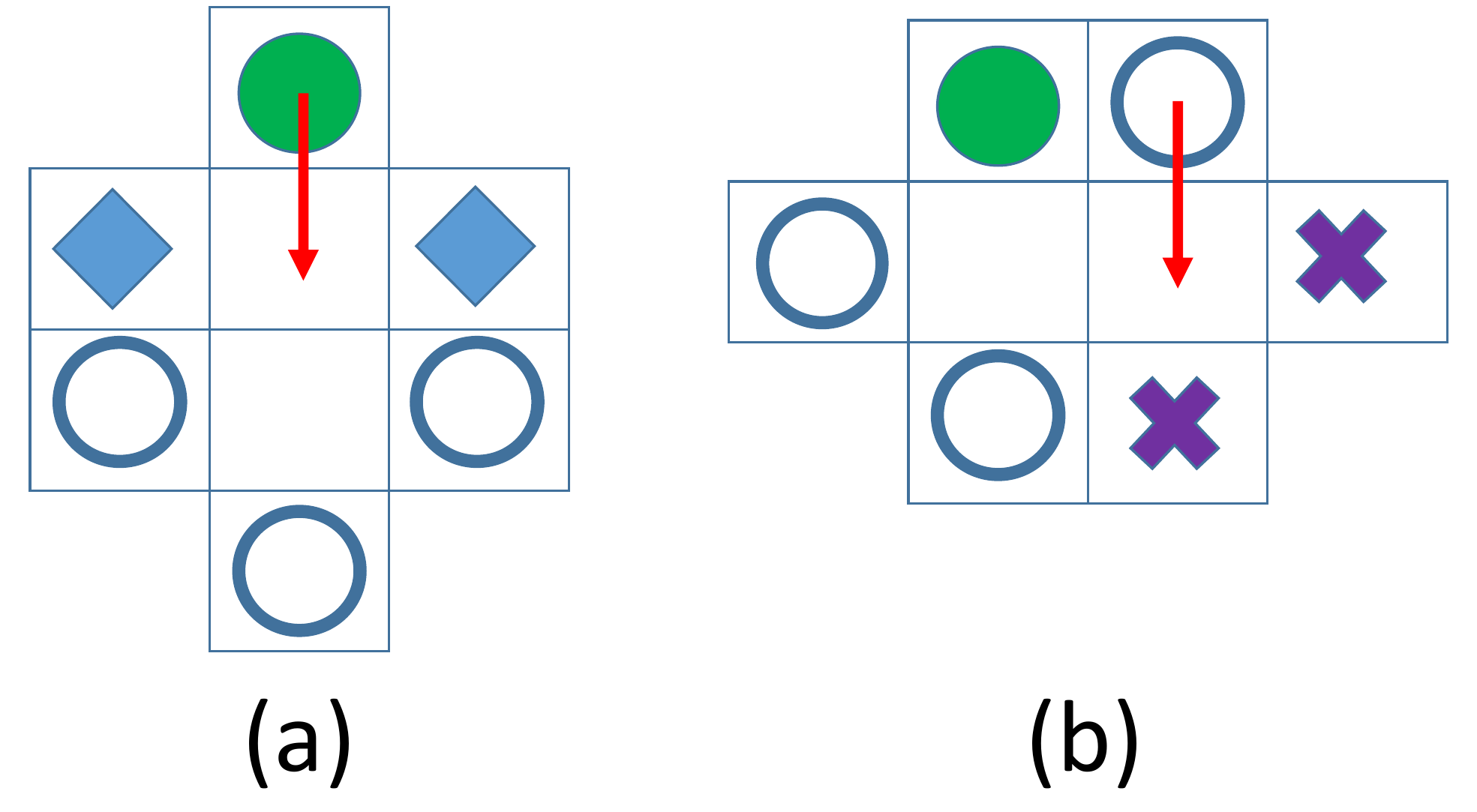}
\caption{An illustration of the transition ${\rm F\rightarrow J}$ in the DAKA model}
\label{rate_fig4}
\end{figure}

There are two ways in which a free particle can change into a jammed particle. In the first case, illustrated in Fig. \ref{rate_fig4}a, the main particle moves into a jammed configuration. This happens if the three jamming sites are occupied, which occur with probability $\rho^{3}$, and if the site between them is vacant, which occur with probability $P_{v|{\rm F}}$. Therefore, the rate for the first case is
\begin{align}
r^{DAKA,1}_{\rm{F},\rm{J}}=\rho^{3}P_{v|{\rm F}}=\frac{\rho^{3}\left(1-\rho\right)}{1-\rho^{3}} .
\end{align}

In the second case, illustrated in Fig. \ref{rate_fig4}b, a particle moves to occupy one of neighbors of the main particle. The rate for this case is
\begin{align}
r^{DAKA,2}_{\rm{F},\rm{J}}=2P_{\rho^{2}|{\rm F}}\rho P_{\rm{F}|vt}=\frac{2\rho^{3}}{\left(1-\rho^{3}\right)^{2}}P_{\rm{F}} .
\end{align}
The total rate $r^{\rm{DAKA}}_{\rm{F},\rm{J}}$ is given by
\begin{align}
r^{\rm{DAKA}}_{\rm{F},\rm{J}}=r^{DAKA,1}_{\rm{F},\rm{J}}+r^{DAKA,2}_{\rm{F},\rm{J}} .
\end{align}

\subsection{DBKA model}
\label{app_dbka}
\subsubsection{$r^{\rm{DBKA}}_{\rm{B},\rm{J}}$ and $r^{\rm{DBKA}}_{\rm{B},\rm{F}}$}
\begin{figure}
\includegraphics[width=0.3\columnwidth]{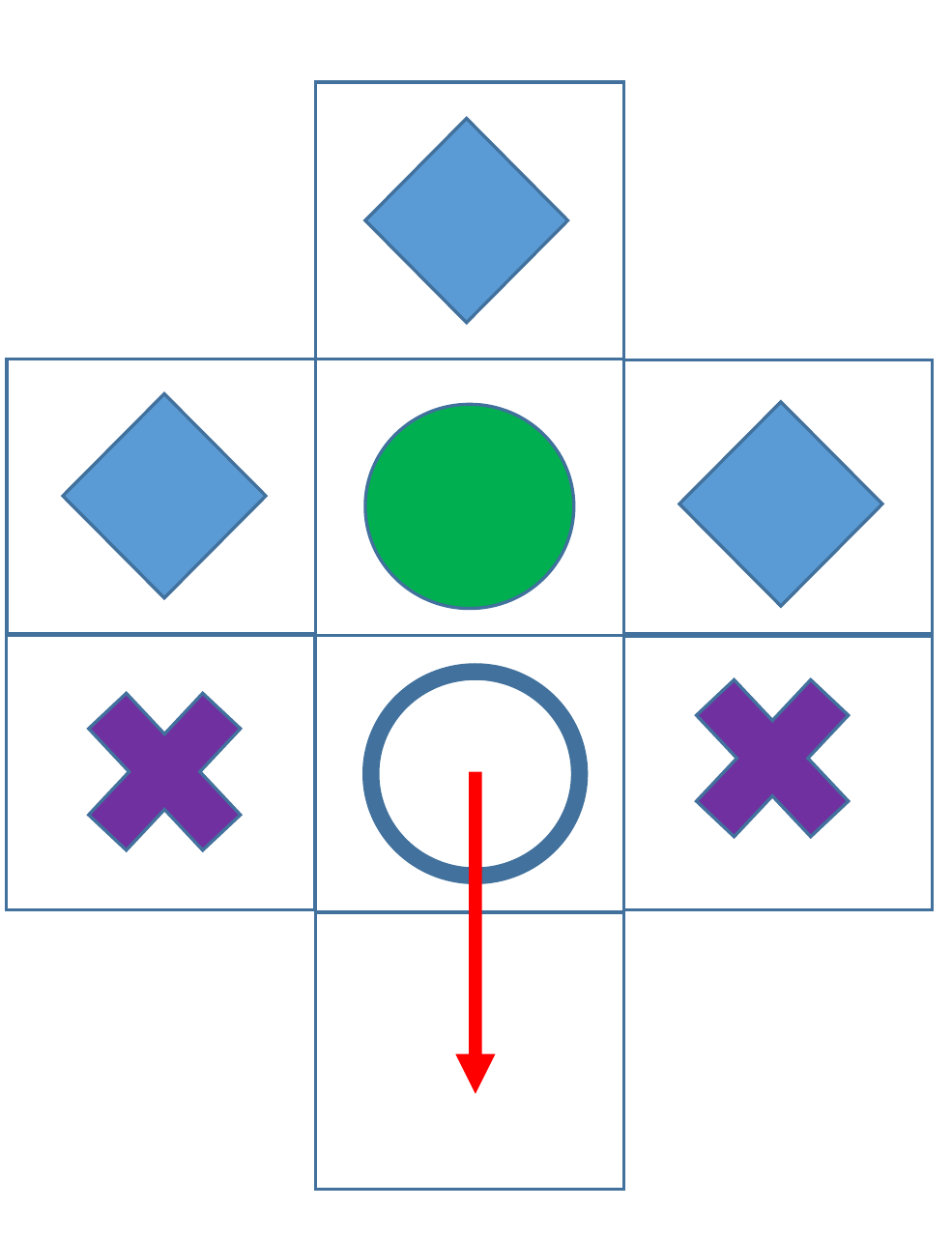}
\caption{An illustration of the transitions ${\rm B\rightarrow J}$ and ${\rm B\rightarrow F}$ in the DBKA model}
\label{rate_fig5}
\end{figure}

The main particle (green circle in Fig. \ref{rate_fig5}) can change its state only if the blocking particle (empty circle) moves. The rates are therefore
\begin{align}
&r^{\rm{DBKA}}_{\rm{B},\rm{J}}=\rho^{3}P_{\rm{F}|\rho}=\frac{\rho^{3}\left(1-\rho^{2}\right)}{1-\rho^{3}}P_{\rm{F}} ,\nonumber\\
&r^{\rm{DBKA}}_{\rm{B},\rm{F}}=\left(1-\rho^{3}\right)P_{\rm{F}|\rho}=\left(1-\rho^{2}\right)P_{\rm{F}} .
\end{align}

\subsubsection{$r^{\rm{DBKA}}_{\rm{J},\rm{B}}$ and $r^{\rm{DBKA}}_{\rm{J},\rm{F}}$}
\begin{figure}
\includegraphics[width=0.3\columnwidth]{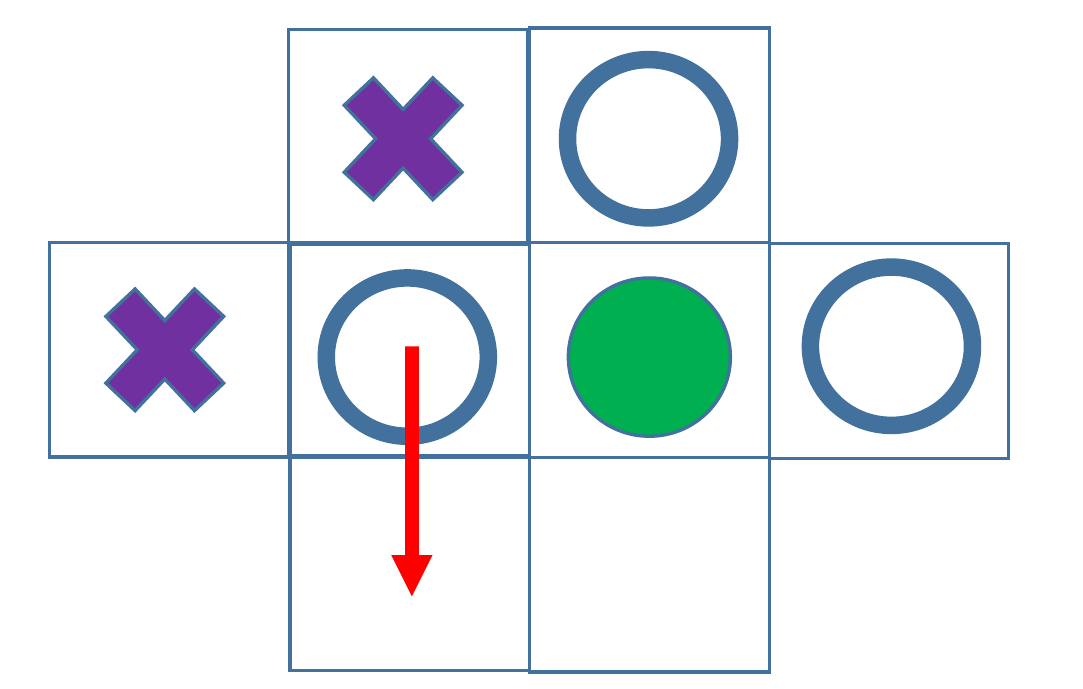}
\caption{An illustration of the transition ${\rm J\rightarrow F}$ in the DBKA model}
\label{rate_fig6}
\end{figure}

The main jammed particle (green circle in Fig. \ref{rate_fig6}) can change its state only if one of its two side neighbors move. The rates are therefore
\begin{align}
&r^{\rm{DBKA}}_{\rm{J},\rm{B}}=0 ,\nonumber\\
&r^{\rm{DBKA}}_{\rm{J},\rm{F}}=2P_{\rm{F}|\rho}=2\frac{1-\rho^{2}}{1-\rho^{3}}P_{\rm{F}} .
\end{align}

\subsubsection{$r^{\rm{DBKA}}_{\rm{F},\rm{B}}$ and $r^{\rm{DBKA}}_{\rm{F},\rm{J}}$}
\begin{figure}
\includegraphics[width=0.9\columnwidth]{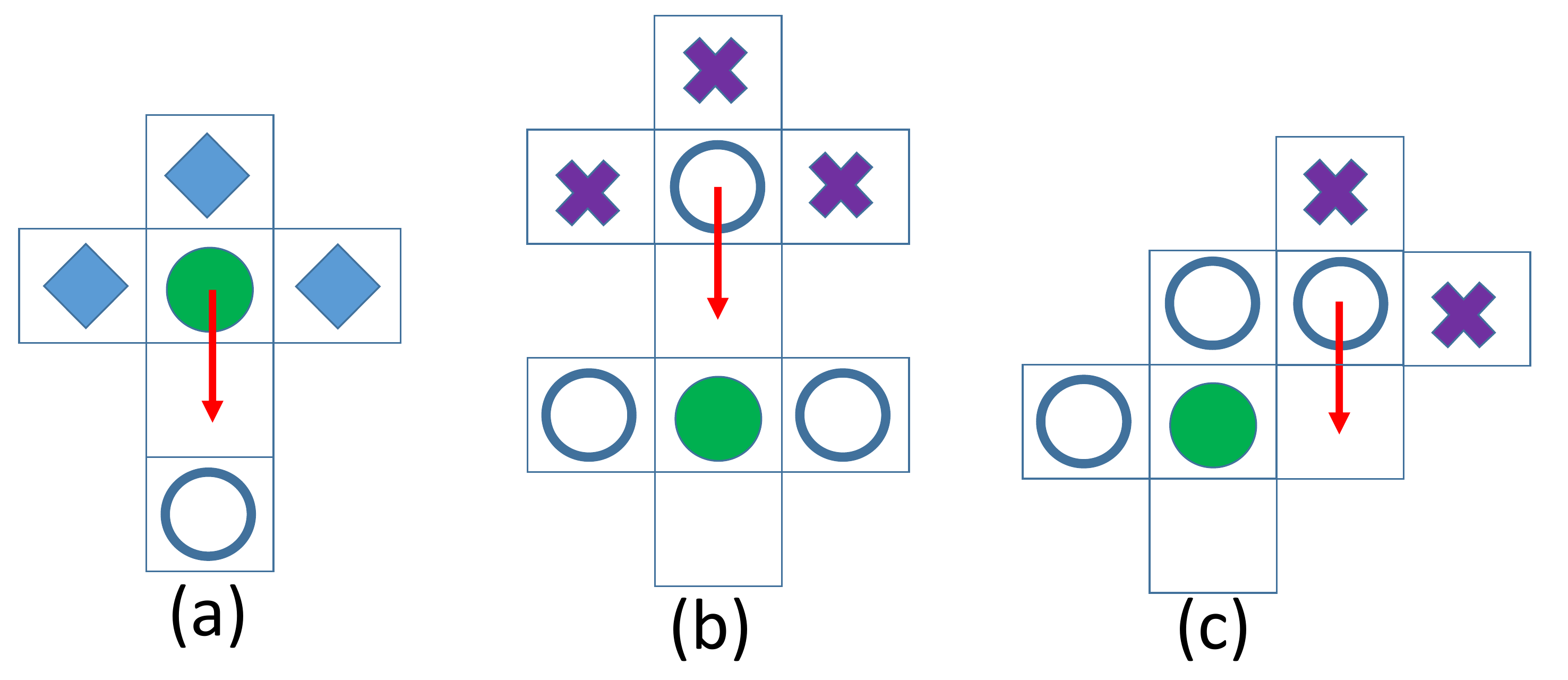}
\caption{An illustration of the transitions ${\rm F\rightarrow B}$ and ${\rm F\rightarrow J}$ in the DBKA model}
\label{rate_fig7}
\end{figure}

The transition ${\rm F\rightarrow B}$ is illustrated in Fig. \ref{rate_fig7}a, and its rate is
\begin{align}
r^{\rm{DBKA}}_{\rm{F},\rm{B}}=\rho .
\end{align}
The transition ${\rm F\rightarrow J}$ can occur in two different ways illustrated in Fig. \ref{rate_fig7}b and Fig. \ref{rate_fig7}c. The corresponding rates for the two cases are
\begin{align}
&r^{DBKA,1}_{\rm{F},\rm{J}}=P_{\rho^{2}|{\rm F}}\rho P_{\rm{F}|t}=\frac{\rho^{3}}{1-\rho^{3}}P_{\rm{F}} ,\nonumber\\
&r^{DBKA,2}_{\rm{F},\rm{J}}=2P_{\rho^{2}|{\rm F}}\rho P_{\rm{F}|\rho t}=\frac{2\rho^{3}\left(1-\rho^{2}\right)}{\left(1-\rho^{3}\right)^{2}}P_{\rm{F}} .
\end{align}
The total rate is therefore
\begin{align}
&r^{\rm{DBKA}}_{\rm{F},\rm{J}}=r^{DBKA,1}_{\rm{F},\rm{J}}+r^{DBKA,2}_{\rm{F},\rm{J}}=\nonumber\\
&=\frac{\rho^{3}}{1-\rho^{3}}\left(1+2\frac{1-\rho^{2}}{1-\rho^{3}}\right)P_{\rm{F}} .
\end{align}

\subsection{DKA model}
\label{app_dka}
\subsubsection{$r^{\rm{DKA}}_{\rm{B},\rm{J}}$ and $r^{\rm{DKA}}_{\rm{B},\rm{F}}$}
\begin{figure}
\includegraphics[width=0.3\columnwidth]{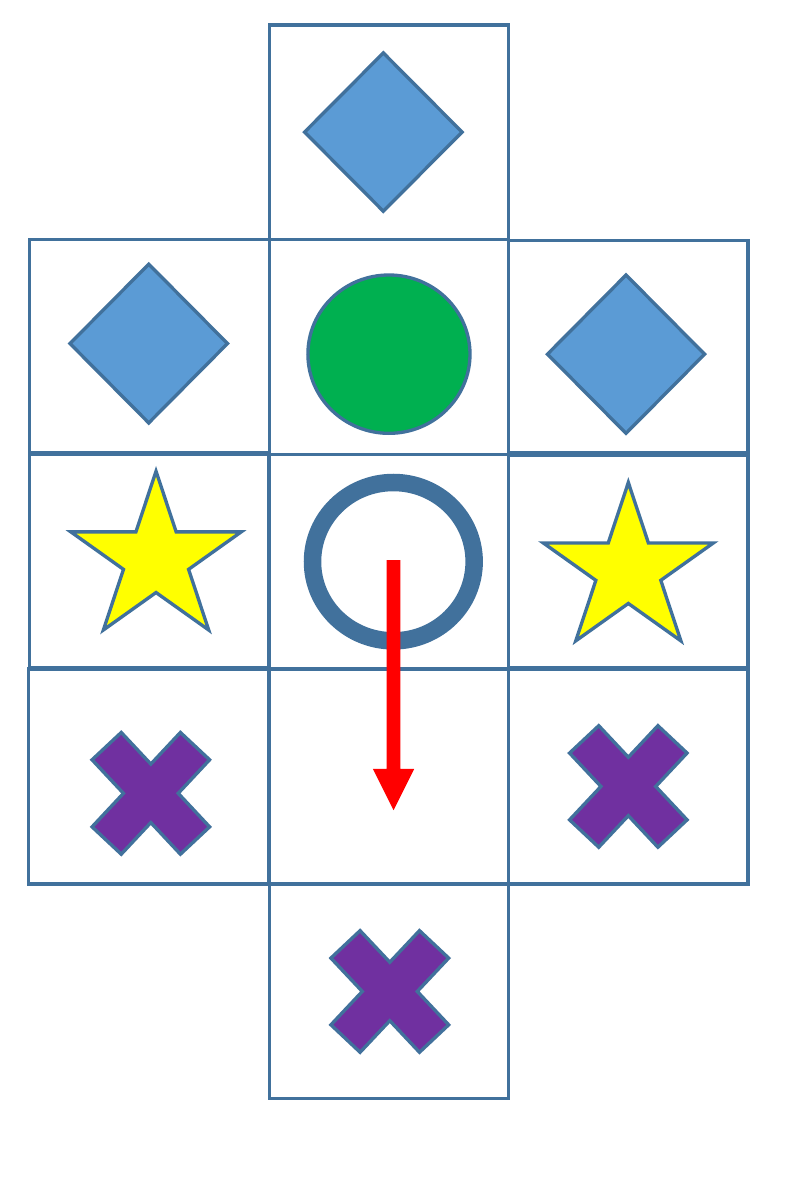}
\caption{An illustration of the transitions ${\rm B\rightarrow J}$ and ${\rm B\rightarrow F}$ in the DKA model}
\label{rate_fig8}
\end{figure}

Similarly to the corresponding rates in the DAKA and DBKA model, these transitions occur only when the blocking particle (empty circle in Fig. \ref{rate_fig8}) moves. In order for it to be free, it requires that at least one of its three neighbors after the move (purple $\times$) is vacant, and that at least one of its three neighbors before the move (the main particle depicted as a green circle and the two yellow stars) is vacant. Since the main particle already occupies one site, this transition is possible only if at least one of the two sites marked with a yellow star is vacant, which itself means that in order for the main particle to be jammed after the blocking particle moves, all three sites marked with a $\diamond$ must be occupied, otherwise it will be free after the blocking particle moves. Therefore, the rates are
\begin{align}
&r^{\rm{DKA}}_{\rm{B},\rm{J}}=\rho^{3}P_{\rm{F}|\rho}=\frac{\rho^{3}\left(1-\rho^{2}\right)}{1-\rho^{3}}P_{\rm{F}} ,\nonumber\\
&r^{\rm{DKA}}_{\rm{B},\rm{F}}=\left(1-\rho^{3}\right)P_{\rm{F}|\rho}=\left(1-\rho^{2}\right)P_{\rm{F}} .
\end{align}

\subsubsection{$r^{\rm{DKA}}_{\rm{J},\rm{B}}$ and $r^{\rm{DKA}}_{\rm{J},\rm{F}}$}
\begin{figure}
\includegraphics[width=0.9\columnwidth]{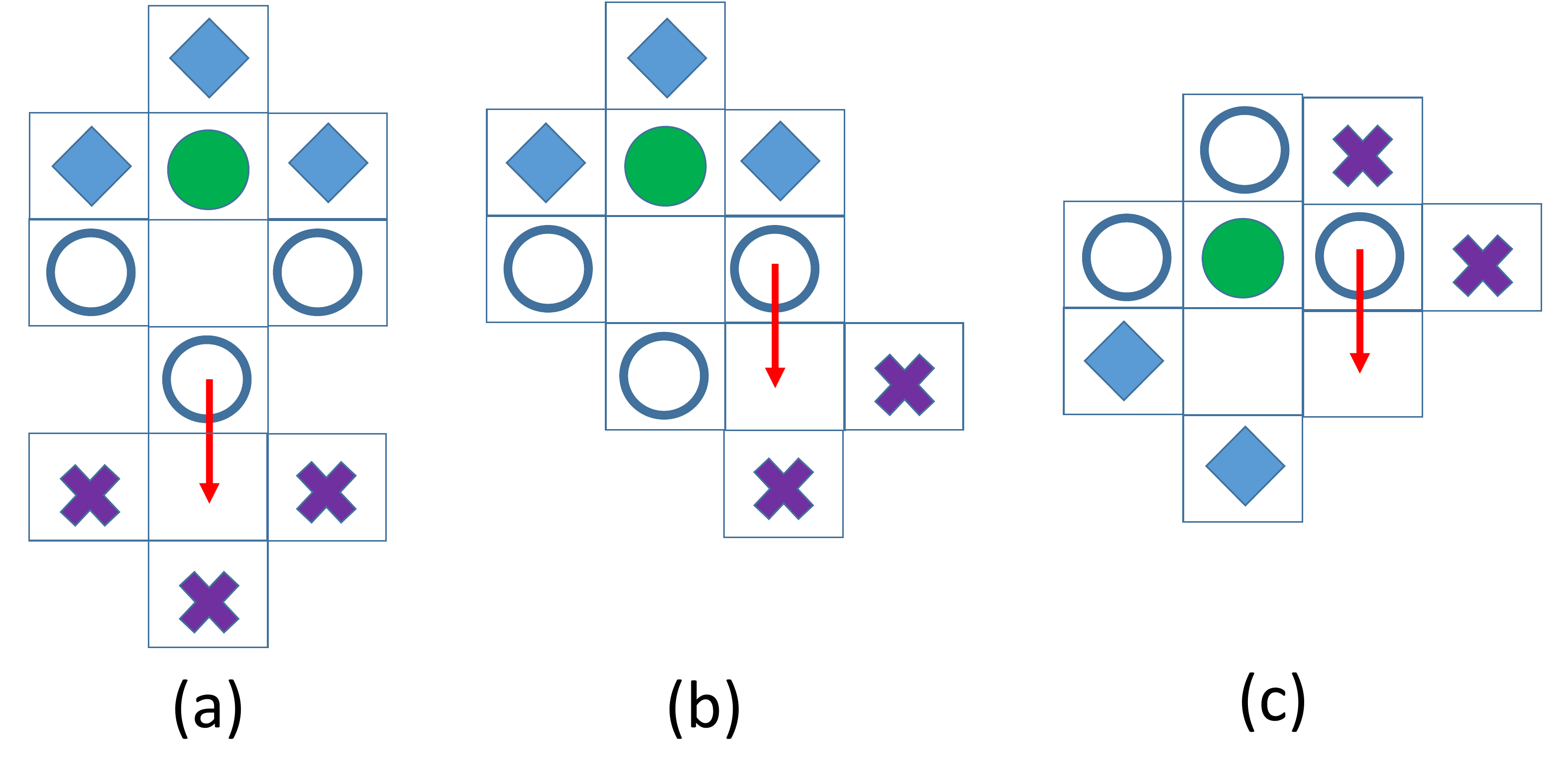}
\caption{An illustration of the transitions ${\rm J\rightarrow F}$ in the DKA model}
\label{rate_fig9}
\end{figure}

A jammed particle can change its state only if one of its neighbors moves, which causes it to become free. Therefore,
\begin{align}
r^{\rm{DKA}}_{\rm{J},\rm{B}}=0 .
\end{align}
The transition ${\rm J\rightarrow F}$ may occur in three different ways, illustrated in Fig. \ref{rate_fig9}. In the first case, shown in Fig. \ref{rate_fig9}a, at least one of the three sites marked $\diamond$ is vacant, and at least one of the sites marked $\times$ is vacant. The corresponding rate is
\begin{align}
r^{DKA,1}_{\rm{J},\rm{F}}=P_{\rho^{3}\left(1-\rho^3\right)|{\rm J}}P_{\rm{F}|v}=\frac{1}{2-\rho^{3}}P_{\rm{F}} .
\end{align}
In the second case, shown in Fig. \ref{rate_fig9}b, at least one of the three sites marked $\diamond$ is vacant, and at least one of the two sites marked $\times$ is vacant. The corresponding rate is
\begin{align}
r^{DKA,2}_{\rm{J},\rm{F}}=2P_{\rho^{3}\left(1-\rho^3\right)|{\rm J}}P_{\rm{F}|v,\rho}=2\frac{1-\rho^{2}}{\left(2-\rho^{3}\right)\left(1-\rho^{3}\right)}P_{\rm{F}} .
\end{align}
In the third case, shown in Fig. \ref{rate_fig9}c, at least one of the two sites marked $\diamond$ is vacant, and at least one of the two sites marked $\times$ is vacant. The corresponding rate is
\begin{align}
r^{DKA,3}_{\rm{J},\rm{F}}=2P_{\rho^{3}v\left(1-\rho^{2}\right)|{\rm J}}P_{\rm{F}|\rho,v,t}=2\frac{\left(1-\rho^{2}\right)^{2}}{\left(2-\rho^{3}\right)\left(1-\rho^{3}\right)^{2}}P_{\rm{F}} .
\end{align}

\subsubsection{$r^{\rm{DKA}}_{\rm{F},\rm{B}}$}
\begin{figure}
\includegraphics[width=0.3\columnwidth]{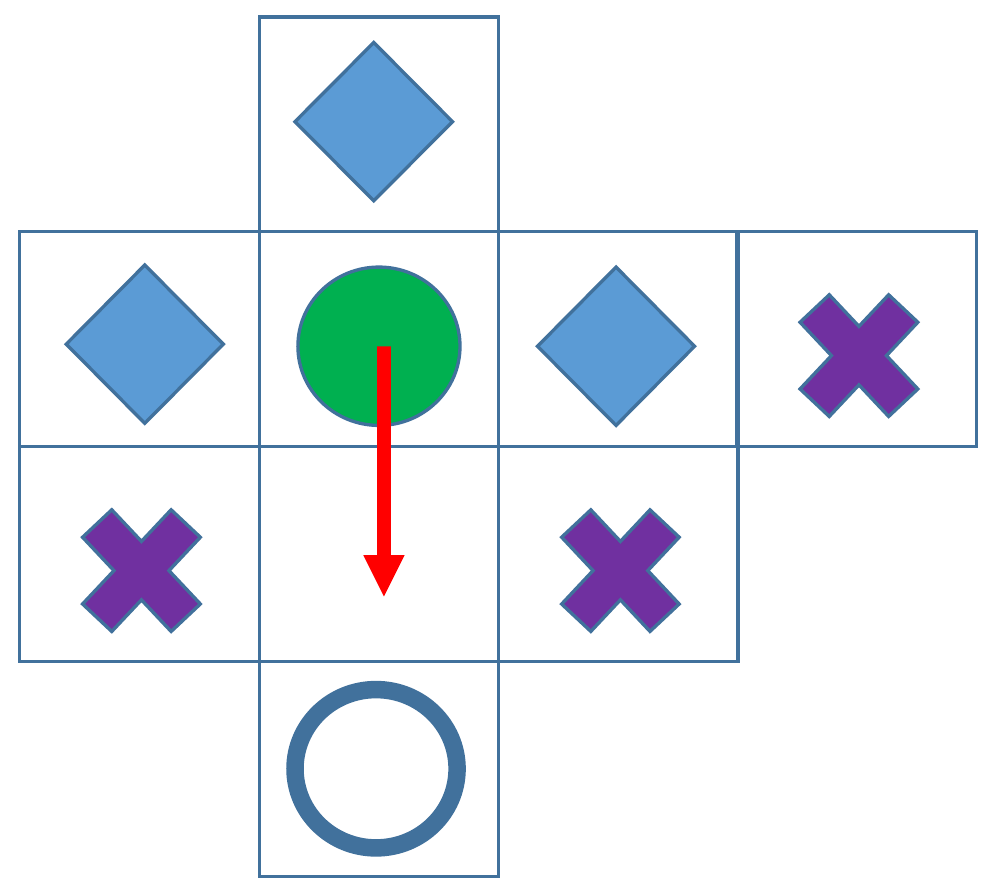}
\caption{An illustration of the transition ${\rm F\rightarrow B}$ in the DKA model}
\label{rate_fig10}
\end{figure}

This transition is shown in Fig. \ref{rate_fig10} and the rate is
\begin{align}
r^{\rm{DKA}}_{\rm{F},\rm{B}}=P_{\rho|{\rm F}}=\frac{\rho\left(1-\rho^{2}\right)}{1-\rho^{3}} .
\end{align}

\subsubsection{$r^{\rm{DKA}}_{\rm{F},\rm{J}}$}
\begin{figure}
\includegraphics[width=0.9\columnwidth]{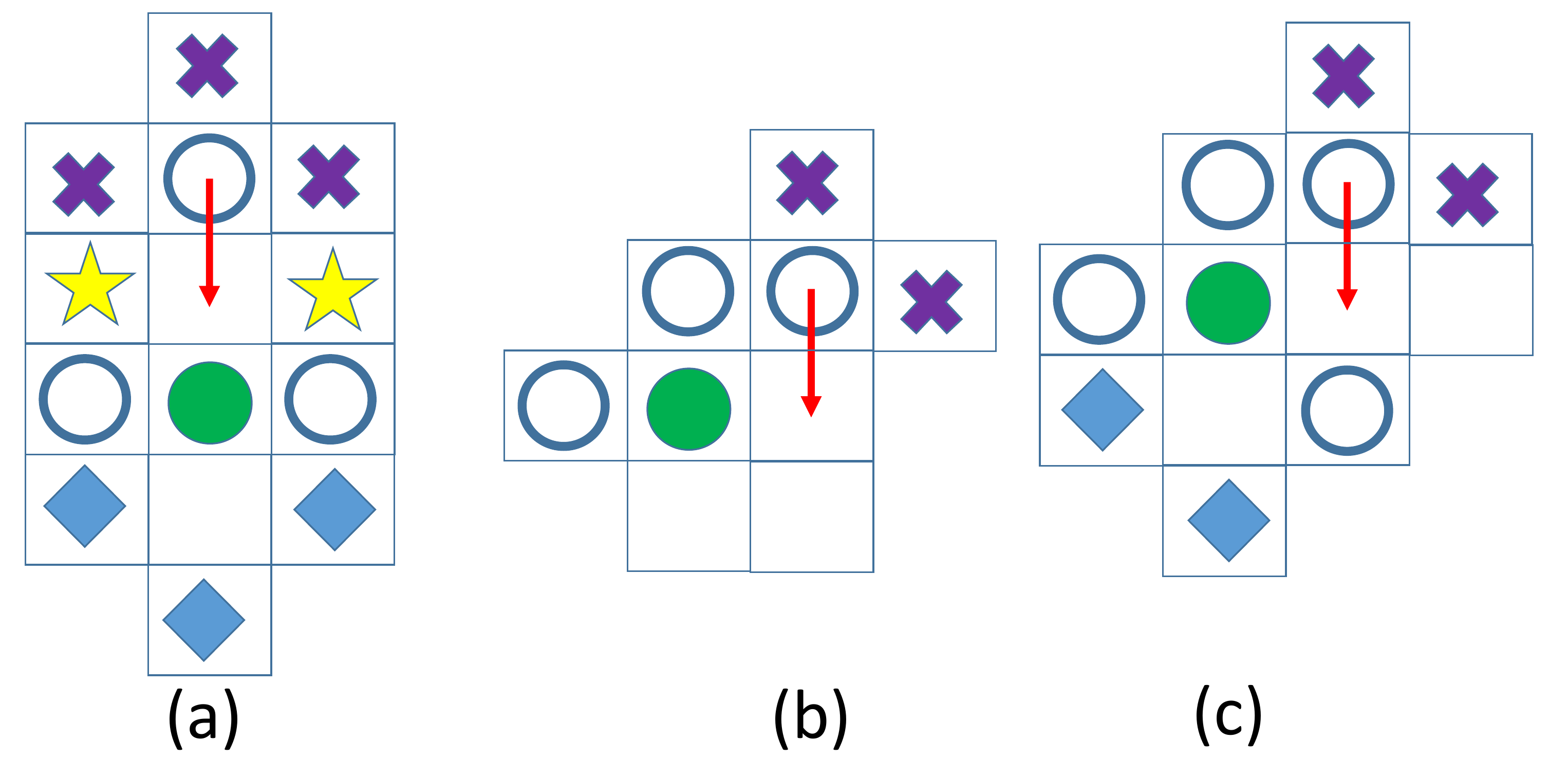}\\
\includegraphics[width=0.6\columnwidth]{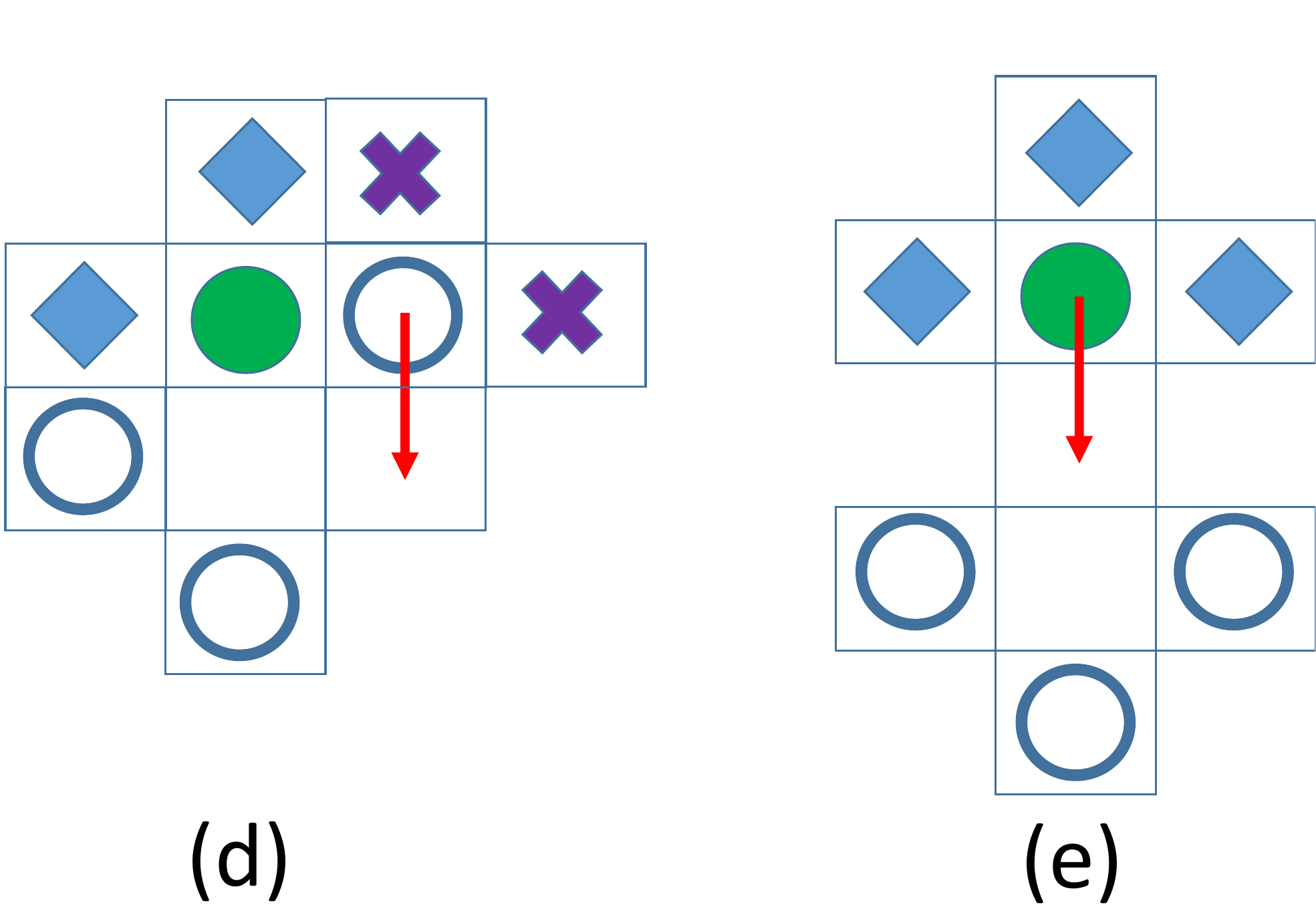}
\caption{An illustration of the transition ${\rm F\rightarrow J}$ in the DKA model}
\label{rate_fig11}
\end{figure}

There are five ways to transition from a free state to a jammed state. In the first case, shown in Fig. \ref{rate_fig11}a, at least one of the three $\diamond$ sites is vacant, at least one of the three $\times$ sites is vacant, and at least one of the two $\ast$ sites is vacant. The corresponding rate is
\begin{align}
r^{DKA,1}_{\rm{F},\rm{J}}=\rho P_{\rho^{2}|{\rm F}}P_{\rm{F}|\rho,t}=\frac{\rho^{3}\left(1-\rho^{2}\right)}{\left(1-\rho^{3}\right)^{2}}P_{\rm{F}} .
\end{align}
In the second case, shown in Fig. \ref{rate_fig11}b, at least one of the two $\times$ sites is vacant, and the corresponding rate is
\begin{align}
r^{DKA,2}_{\rm{F},\rm{J}}=2\rho P_{\rho^{2},v|{\rm F}}P_{\rm{F}|\rho,v\rho,t}=2\frac{\rho^{3}\left(1-\rho\right)\left(1-\rho^{2}\right)}{\left(1-\rho^{3}\right)^{4}}P_{\rm{F}} .
\end{align}
In the third case, shown in Fig. \ref{rate_fig11}c, at least one of the two $\diamond$ sites is vacant and at least one of the two $\times$ sites is vacant. The corresponding rate is
\begin{align}
&r^{DKA,3}_{\rm{F},\rm{J}}=2\rho v P_{\rho^{2},\rho|{\rm F}}P_{\rm{F}|\rho,\rho^{2}v,t}=\nonumber\\
&=2\frac{\rho^{4}\left(1-\rho\right)\left(1-\rho^{2}\right)^{2}}{\left(1-\rho^{3}\right)^{4}}P_{\rm{F}} .
\end{align}
In the fourth case, shown in Fig. \ref{rate_fig11}d, at least one of the two $\diamond$ sites and at least one of the two $\times$ sites are vacant. The corresponding rate is
\begin{align}
r^{DKA,4}_{\rm{F},\rm{J}}=2P_{\rho,\rho^{2}v|{\rm F}}P_{\rm{F}|\rho,v,t}=2\frac{\rho^{3}\left(1-\rho^{2}\right)^{2}}{\left(1-\rho^{3}\right)^{4}}P_{\rm{F}} .
\end{align}
In the fifth case, shown in Fig. \ref{rate_fig11}e, at least one of the three $\diamond$ sites is vacant and the corresponding rate is
\begin{align}
r^{DKA,5}_{\rm{F},\rm{J}}=\rho^{3}P_{v|{\rm F}}=\frac{\rho^{3}\left(1-\rho\right)}{1-\rho^{3}} .
\end{align}

\section{Stability of the stationary solution}
\label{app_stability}
In this Appendix we show that whenever the non-trivial stationary solution exists, it is also stable. Using Eq. (\ref{p3e1}), the evolution equations for $P_{\rm{B}}$ and $P_{\rm{F}}$ are explicitly
\begin{widetext}
\begin{align}
&\frac{\partial P_{\rm{B}}}{\partial t}=\left[\Omega_{\rm{F},\rm{B}}+\omega_{\rm{F},\rm{B}}P_{\rm{F}}+\omega_{\rm{J},\rm{B}}\left(1-P_{\rm{B}}-P_{\rm{F}}\right)-\left(\omega_{\rm{B},\rm{J}}+\omega_{\rm{B},\rm{F}}\right)P_{\rm{B}}\right]P_{\rm{F}} ,\nonumber\\
&\frac{\partial P_{\rm{F}}}{\partial t}=\left[\omega_{\rm{B},\rm{F}}P_{\rm{B}}+\omega_{\rm{J},\rm{F}}\left(1-P_{\rm{B}}-P_{\rm{F}}\right)-\left(\Omega_{\rm{F},\rm{B}}+\omega_{\rm{F},\rm{B}}P_{\rm{F}}+\Omega_{\rm{F},\rm{J}}+\omega_{\rm{F},\rm{J}}P_{\rm{F}}\right)\right]P_{\rm{F}} ,
\end{align}
\end{widetext}
which can be written in matrix form as
\begin{align}
\frac{\partial}{\partial t}\left(\begin{array}{c}P_{\rm{B}}\\P_{\rm{F}}\end{array}\right)=P_{\rm{F}}\left[{\cal M}\left(\begin{array}{c}P_{\rm{B}}\\P_{\rm{F}}\end{array}\right)+\left(\begin{array}{c}P^{0}_{\rm{B}}\\P^{0}_{\rm{F}}\end{array}\right)\right] ,\label{eveq2}
\end{align}
with the matrix ${\cal M}$ given explicitly by
\begin{align}
&{\cal M}=\nonumber\\
&-\left(\begin{array}{lr}\omega_{\rm{J},\rm{B}}+\omega_{\rm{B},\rm{J}}+\omega_{\rm{B},\rm{F}}&\omega_{\rm{J},\rm{B}}-\omega_{\rm{F},\rm{B}}\\\omega_{\rm{J},\rm{F}}-\omega_{\rm{B},\rm{F}}&\omega_{\rm{J},\rm{F}}+\omega_{\rm{F},\rm{B}}+\omega_{\rm{F},\rm{J}}\end{array}\right) ,
\end{align}
and the vector $\left(\begin{array}{c}P^{0}_{\rm{B}}\\P^{0}_{\rm{F}}\end{array}\right)$ by
\begin{align}
\left(\begin{array}{c}P^{0}_{\rm{B}}\\P^{0}_{\rm{F}}\end{array}\right)=\left(\begin{array}{c}\Omega_{\rm{F},\rm{B}}+\omega_{\rm{J},\rm{B}}\\\omega_{\rm{J},\rm{F}}-\Omega_{\rm{F},\rm{B}}-\Omega_{\rm{F},\rm{J}}\end{array}\right) .
\end{align}

The determinant of ${\cal M}$ is
\begin{align}
&\left|{\cal M}\right|=\left(\omega_{\rm{F},\rm{B}}+\omega_{\rm{F},\rm{J}}\right)\left(\omega_{\rm{B},\rm{J}}+\omega_{\rm{J},\rm{B}}\right)+\nonumber\\
&+\left(\omega_{\rm{B},\rm{J}}+\omega_{\rm{F},\rm{B}}\right)\omega_{\rm{J},\rm{F}}+\omega_{\rm{B},\rm{F}}\left(\omega_{\rm{F},\rm{J}}+\omega_{\rm{J},\rm{B}}+\omega_{\rm{J},\rm{F}}\right) ,
\end{align}
which is positive since for any set of kinetic constraints, including for models we do not consider here, $\omega_{\rm{J},\rm{F}}$ and at least one of $\omega_{\rm{B},\rm{F}}$ and $\omega_{\rm{B},\rm{J}}$ are positive.

We now look for stationary solutions of Eq. (\ref{eveq}) under the condition $0\leq P_{\rm{F}},P_{\rm{J}},P_{\rm{B}}\leq 1$. The solution $P_{\rm{F}}=0$ is always a stationary solution. We find that if there is another stationary solution with $P_{\rm{F}}>0$, then it is unique and given by
\begin{align}
&\left(\begin{array}{c}P_{\rm{B}}\\P_{\rm{F}}\end{array}\right)_{ss}=-{\cal M}^{-1}\left(\begin{array}{c}P^{0}_{\rm{B}}\\P^{0}_{\rm{F}}\end{array}\right) .\label{ss_sol}
\end{align}
In order to check its stability, we perturb $P_{\rm{B}}$ and $P_{\rm{F}}$ around it, such that $P_{\alpha}=P_{\alpha,ss}+\delta_{\alpha}$. Setting this perturbation in Eq. (\ref{eveq2}) yields
\begin{align}
&\frac{\partial}{\partial t}\left(\begin{array}{c}\delta_{\rm{B}}\\\delta_{\rm{F}}\end{array}\right)=\nonumber\\
&=\left(P_{\rm{F},ss}+\delta_{\rm{F}}\right)\left[{\cal M}\left(\begin{array}{c}P_{\rm{B}}\\P_{\rm{F}}\end{array}\right)+{\cal M}\left(\begin{array}{c}\delta_{\rm{B}}\\\delta_{\rm{F}}\end{array}\right)+\left(\begin{array}{c}P^{0}_{\rm{B}}\\P^{0}_{\rm{F}}\end{array}\right)\right]\nonumber\\
&=\delta_{\rm{F}}\left[{\cal M}\left(\begin{array}{c}P_{\rm{B}}\\P_{\rm{F}}\end{array}\right)+\left(\begin{array}{c}P^{0}_{\rm{B}}\\P^{0}_{\rm{F}}\end{array}\right)\right]+P_{\rm{F},ss}{\cal M}\left(\begin{array}{c}\delta_{\rm{B}}\\\delta_{\rm{F}}\end{array}\right) ,
\end{align}
where in the second step we kept terms linear in $\delta_{\alpha}$. Using Eq. (\ref{ss_sol}) yields
\begin{align}
\frac{\partial}{\partial t}\left(\begin{array}{c}\delta_{\rm{B}}\\\delta_{\rm{F}}\end{array}\right)=P_{\rm{F},ss}{\cal M}\left(\begin{array}{c}\delta_{\rm{B}}\\\delta_{\rm{F}}\end{array}\right) .
\end{align}
Since $P_{\rm{F},ss}>0$, the steady state solution is stable if the real part of both eigenvalues of ${\cal M}$ is negative. The two eigenvalues of ${\cal M}$ have the form
\begin{align}
\lambda_{\pm}=-\lambda_{0}\pm\lambda_{1} ,
\end{align}
where $\lambda_{0}>0$ and $\lambda_{1}$ can be negative, positive, or imaginary. Therefore, only $\lambda_{+}$ can have a positive real part, and that occurs only if $\lambda_{1}$ is positive. However, in that case $\lambda_{-}<0$, and since $\left|{\cal M}\right|>0$, and thus $\lambda_{+}$ must also be negative. Hence, if the non-trivial stationary solution exists, it is also stable.

\section{Stability of the $P_{\rm{F}}=0$ solution}
\label{app_stabpf0}
In this appendix we investigate the stability of the $P_{\rm{F}}=0$ state in the SMF approximation. We start from the driven models, and then consider the BKA model. 
\subsection{Driven models}
In order to check the stability of the $P_{\rm{F}}=0$ state in the driven models, we perturb $P_{\rm{B}}$ and $P_{\rm{F}}$ around it, such that $P_{\rm{F}}=\delta_{\rm{F}}$ and $P_{\rm{B}}=P_{\rm{B},ss}+\delta_{\rm{B}}$. Setting this perturbation in Eq. (\ref{eveq2}) and keeping only terms linear in $\delta_{\rm{B}}$ and $\delta_{\rm{F}}$ yields
\begin{align}
&\frac{\partial}{\partial t}\left(\begin{array}{c}\delta_{\rm{B}}\\\delta_{\rm{F}}\end{array}\right)=\delta_{\rm{F}}\left(\begin{array}{c}\omega_{\rm{B}}\\\omega_{\rm{F}}\end{array}\right) ,
\end{align}
with
\begin{align}
&\omega_{\rm{B}}=\Omega_{\rm{F},\rm{B}}+\omega_{\rm{J},\rm{B}}-\left(\omega_{\rm{J},\rm{B}}+\omega_{\rm{B},\rm{J}}+\omega_{\rm{B},\rm{F}}\right)P_{\rm{B},ss} \nonumber\\
&\omega_{\rm{F}}=\left(\omega_{\rm{B},\rm{F}}-\omega_{\rm{J},\rm{F}}\right)P_{\rm{B},ss}-\left(\Omega_{\rm{F},\rm{B}}+\Omega_{\rm{F},\rm{J}}-\omega_{\rm{J},\rm{F}}\right) .\label{pf0stab}
\end{align}
For any value of $\rho$ we find a critical $P_{\rm{B},c}$ above which $\omega_{\rm{F}}<0$ and thus the $P_{\rm{F}}=0$ solution is stable. This critical $P_{\rm{B},ss}$ is
\begin{align}
&P^{\rm{DAKA}}_{\rm{B},c}=\frac{3+2\rho-\rho^{3}}{2+2\rho+\rho^{2}+\rho^{3}+\rho^{4}} ,\nonumber\\
&P^{\rm{DBKA}}_{\rm{B},c}=\frac{2+\rho-\rho^{2}-\rho^{3}}{\left(1+\rho\right)\left(1+\rho^{3}\right)} ,\nonumber\\
&P^{\rm{DKA}}_{\rm{B},c}=\nonumber\\
&=\frac{5+8\rho+5\rho^2-2\rho^3-2\rho^4+3\rho^6+2\rho^7+\rho^8}{3+6\rho+5\rho^2+5\rho^{3}+7\rho^4+6\rho^5+2\rho^6-2\rho^7-2\rho^8-\rho^9} .
\end{align}
Note that $P_{\rm{B},c}>1$ for $\rho<0.618$ for all three models, which means that the $P_{\rm{F}}=0$ solution is unstable for $\rho<0.618$. The value of $0.618$ is the root of the polynomial $1-\rho^{2}\left(1+\rho\right)^{2}$.

In the BKA model, there are six states as outlined in Appendix \ref{app_bka}. In two of them, ${\rm B4}$ and ${\rm B3J}$, the particle cannot move and in the other four the particle can move in at least one direction. The trivial stationary solution in this model is $P_{\rm{B}3J}=1-P_{\rm{B}4}$ and $P_{\rm{B}2F2}=P_{\rm{B}FBF}=P_{\rm{B}F3}=P_{\rm{F}4}=0$. Perturbing the evolution equation around this solution to first order in the perturbation yields equations of the form
\begin{align}
&\frac{\partial\delta_{\alpha}}{\partial t}=\sum_{\beta}\left(\omega_{\beta,\alpha}+\Omega_{\beta,\alpha}P_{\rm{B}}\right)\delta_{\beta} ,
\end{align}
where $\omega_{\beta,\alpha}$ and $\Omega_{\beta,\alpha}$ depend only on the density, and $\alpha$ and $\beta$ are the states in which the particle can move in at least one direction. This may be written in matrix form as
\begin{align}
\frac{\partial}{\partial t}\delta=\left(\omega+P_{\rm{B}}\Omega\right)\delta .
\end{align}
The trivial stationary solution is stable if the real part of all the four eigenvalues of the matrix $\omega+P_{\rm{B}}\Omega$ is negative. Investigating this matrix numerically we find that all its eigenvalues are real, and that for $P_{\rm{B}}=1$ they are negative for all $\rho$. Therefore, the critical $P^{\rm{BKA}}_{\rm{B}4,c}$ is obtained when the determinant of the matrix equals zero, which yields
\begin{align}
P^{\rm{BKA}}_{\rm{B}4,c}=1-\frac{4\rho^{3}}{9-3\rho^{2}} .
\end{align}

\begin{figure}
\includegraphics[width=\columnwidth]{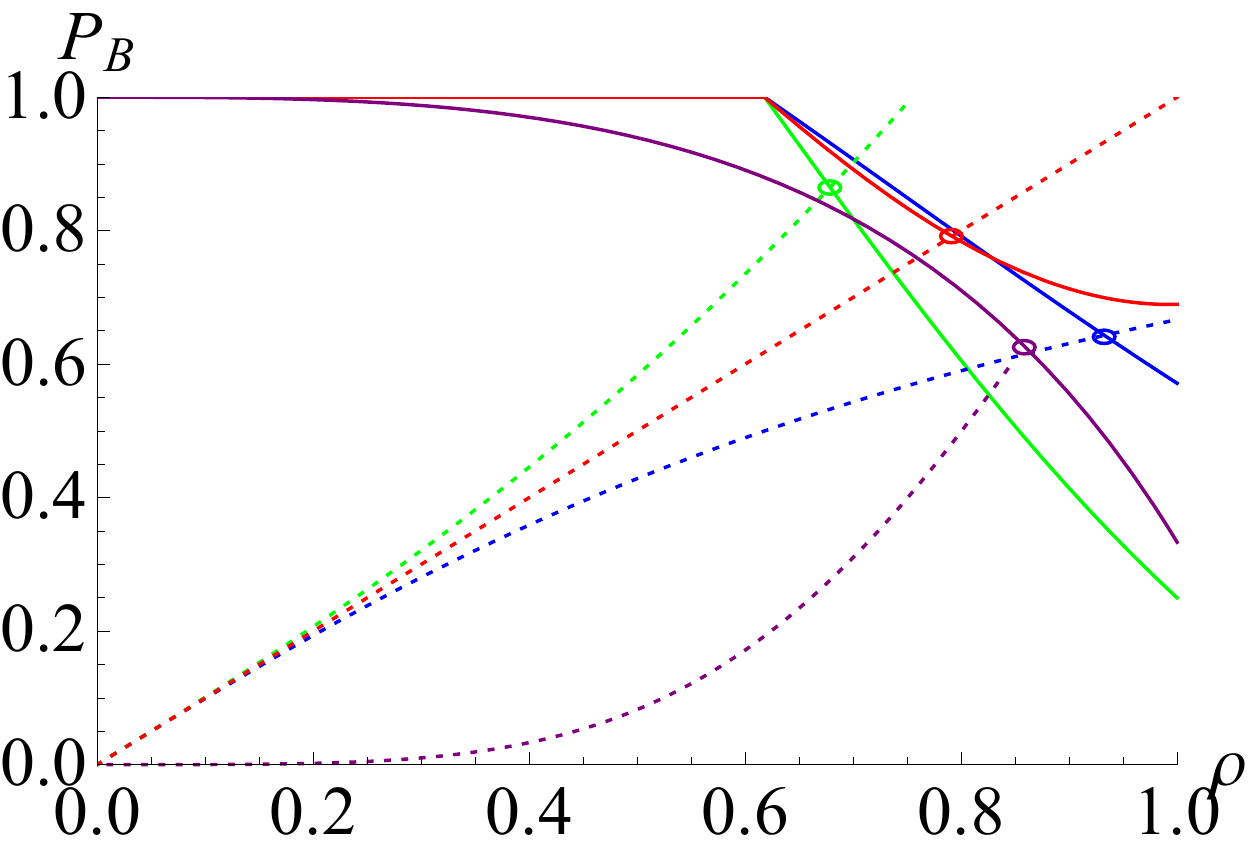}
\caption{The value of $P_{\rm{B},c}$ (continuous lines) and $P_{\rm{B},ss}$ (dotted lines) as a function of the density $\rho$ for the DKA model (red), DAKA model (blue), DBKA model (green), and BKA model (purple). The small circles are the points at which the non-trivial solution ceases to exist.}
\label{pf0stab_fig}
\end{figure}

For densities slightly below $\rho_{c}$, in which the non-trivial solution exists and is stable, the stationary value of $P_{\rm{B}}$ in the driven models is
\begin{align}
P_{\rm{B},ss}=\frac{\Omega_{\rm{F},\rm{B}}+\omega_{\rm{J},\rm{B}}}{\omega_{\rm{J},\rm{B}}+\omega_{\rm{B},\rm{J}}+\omega_{\rm{B},\rm{F}}} .
\end{align}
In the BKA model we find it numerically. Figure \ref{pf0stab_fig} shows $P_{\rm{B},c}$ and $P_{\rm{B},ss}$ as a function of $\rho$ for the four models DKA, DAKA, DBKA and BKA. Note that in the DKA and DBKA models, but not in the DAKA and BKA models, there is a range of densities for which $P_{\rm{B},c}<P_{\rm{B},ss}$.

\section{SMF approximation in the BKA Model}
\label{app_bka}
\begin{figure}
\includegraphics[width=0.9\columnwidth]{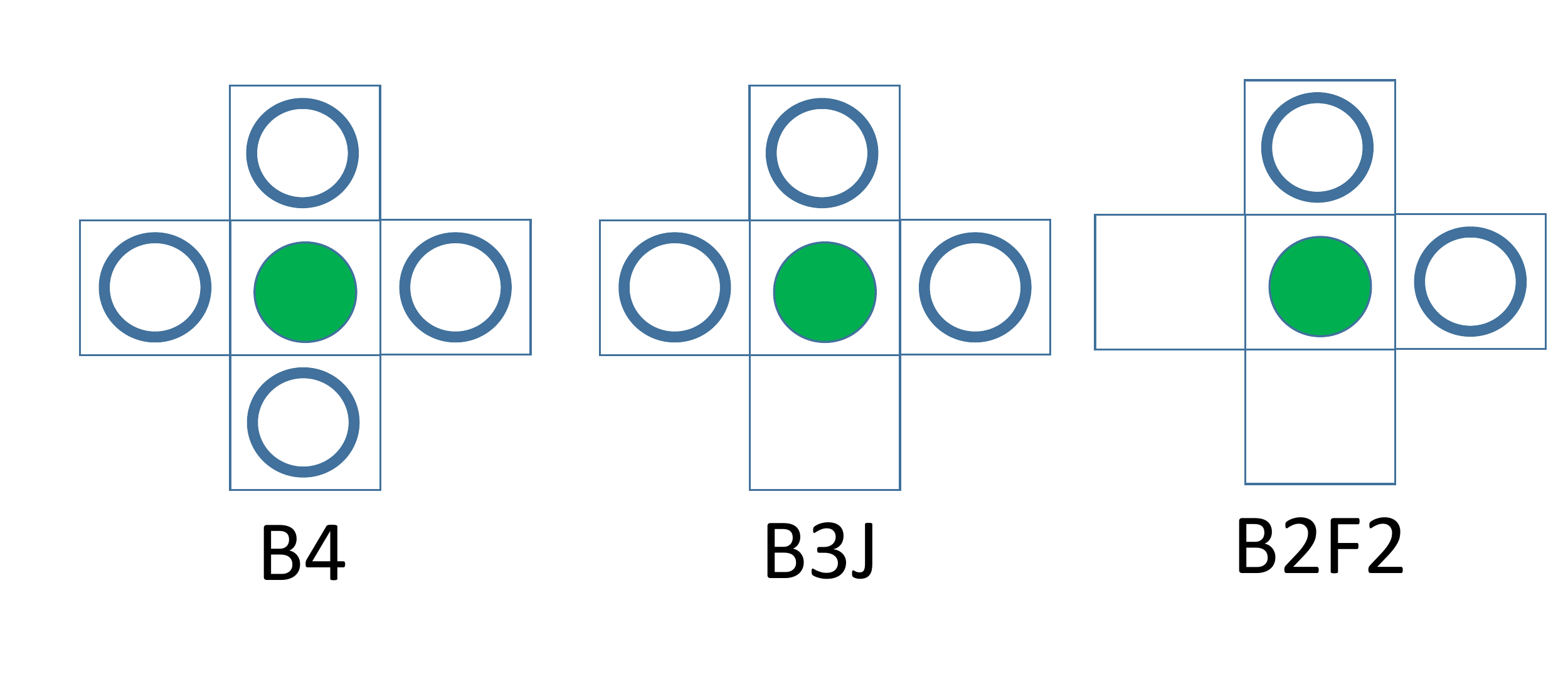}\\
\includegraphics[width=0.9\columnwidth]{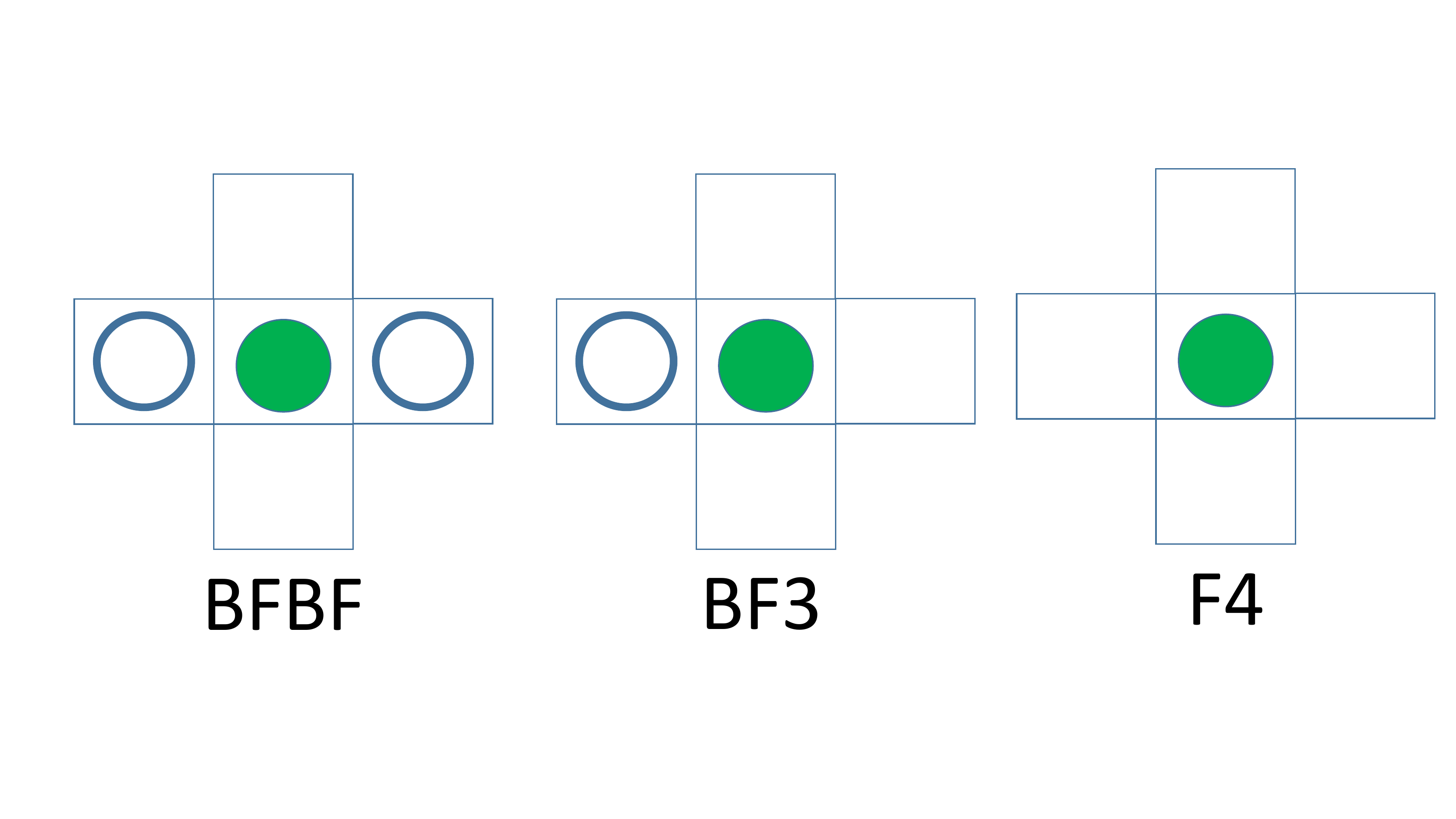}
\caption{An illustration of the six states of particles in the BKA model, up to symmetries.}
\label{bka_states}
\end{figure}

In the BKA model there are six states illustrated in Fig. \ref{bka_states}: ${\rm B4}$ in which all the four nearest neighbors of the main particle are occupied, ${\rm B3J}$ in which three of the four nearest neighbors are occupied and the fourth is vacant, ${\rm B2F2}$ in which two adjacent neighbors are occupied, ${\rm BFBF}$ in which two non-adjacent neighbors are occupied, ${\rm BF3}$ in which one neighbor is occupied, and ${\rm F4}$ in which all four neighbors are vacant. Since the sum of all six probabilities is $1$, there are five coupled non-linear equations. We numerically solve them in order to find the steady state.

\subsection{Conditional probabilities}
In the BKA model we are interested in the probability that a particle can move in a specific direction, $P_{{\rm M}}$, the probability that it can move in a specific direction given that its neighbor in the opposite direction is occupied, $P_{{\rm M}|o}$, the probability that it can move in a specific direction given that its neighbor in the direction perpendicular to its motion is occupied, $P_{{\rm M}|s}$, the probability that it can move in a specific direction given that its neighbor in the direction perpendicular to its motion is vacant, $P_{{\rm M}|v}$, the probability that it can move in a specific direction given that the target site is vacant, $P_{{\rm M}|t}$, the probability that it can move in a specific direction given that its neighbor in the direction perpendicular to its motion is occupied and that the target site is vacant, $P_{{\rm M}|st}$, and the probability that it can move in a specific direction given that both its neighbor in the direction perpendicular to its motion and the target site are vacant, $P_{{\rm M}|vt}$. The probability $P_{{\rm M}}$ is
\begin{align}
P_{{\rm M}}=\frac{1}{2}\left(P_{\rm{B}2F2}+P_{\rm{B}FBF}\right)+\frac{3}{4}P_{\rm{B}F3}+P_{\rm{F}4} .
\end{align}
The conditional probability $P_{{\rm M}|o}$ is
\begin{widetext}
\begin{align}
P_{{\rm M}|o}=\frac{P_{{\rm M}\cap o}}{\rho}=\rho^{-1}\left(P_{\rm{B}2F2\cap o}+P_{\rm{B}F3\cap o}\right)=\rho^{-1}\left(\frac{1}{2}P_{\rm{B}2F2}+\frac{1}{4}P_{\rm{B}F3}\right) .
\end{align}
The conditional probability $P_{{\rm M}|s}$ is
\begin{align}
P_{{\rm M}|s}=\frac{P_{{\rm M}\cap s}}{\rho}=\rho^{-1}\left(P_{\rm{B}2F2\cap s}+P_{\rm{B}FBF\cap s}+P_{\rm{B}F3\cap s}\right)=\rho^{-1}\left(\frac{1}{4}P_{\rm{B}2F2}+\frac{1}{2}P_{\rm{B}FBF}+\frac{1}{4}P_{\rm{B}F3}\right) .
\end{align}
The conditional probability $P_{{\rm M}|v}$ is
\begin{align}
P_{{\rm M}|v}=\frac{P_{{\rm M}\cap v}}{1-\rho}=\left(1-\rho\right)^{-1}\left(P_{\rm{B}2F2\cap v}+P_{\rm{B}F3\cap v}+P_{\rm{F}4\cap v}\right)=\left(1-\rho\right)^{-1}\left(\frac{1}{4}P_{\rm{B}2F2}+\frac{1}{2}P_{\rm{B}F3}+P_{\rm{F}4}\right) .
\end{align}
The conditional probability $P_{{\rm M}|t}$ is
\begin{align}
P_{{\rm M}|t}=\frac{P_{{\rm M}\cap t}}{1-\rho}=\frac{P_{{\rm M}}}{1-\rho} .
\end{align}
The conditional probabilities $P_{{\rm M}|vt}$ and $P_{{\rm M}|st}$ are
\begin{align}
&P_{{\rm M}|vt}=\frac{P_{{\rm M}\cap v\cap t}}{\left(1-\rho\right)^{2}}=\frac{P_{{\rm M}\cap v}}{\left(1-\rho\right)^{2}}=\frac{P_{{\rm M}|v}}{1-\rho} ,\nonumber\\
&P_{{\rm M}|st}=\frac{P_{{\rm M}\cap s\cap t}}{\rho\left(1-\rho\right)}=\frac{P_{{\rm M}\cap s}}{\rho\left(1-\rho\right)}=\frac{P_{{\rm M}|s}}{1-\rho} .
\end{align}
\end{widetext}

\subsection{$r^{\rm{BKA}}_{\rm{B}4,B3J}$}
\begin{figure}
\includegraphics[width=0.6\columnwidth]{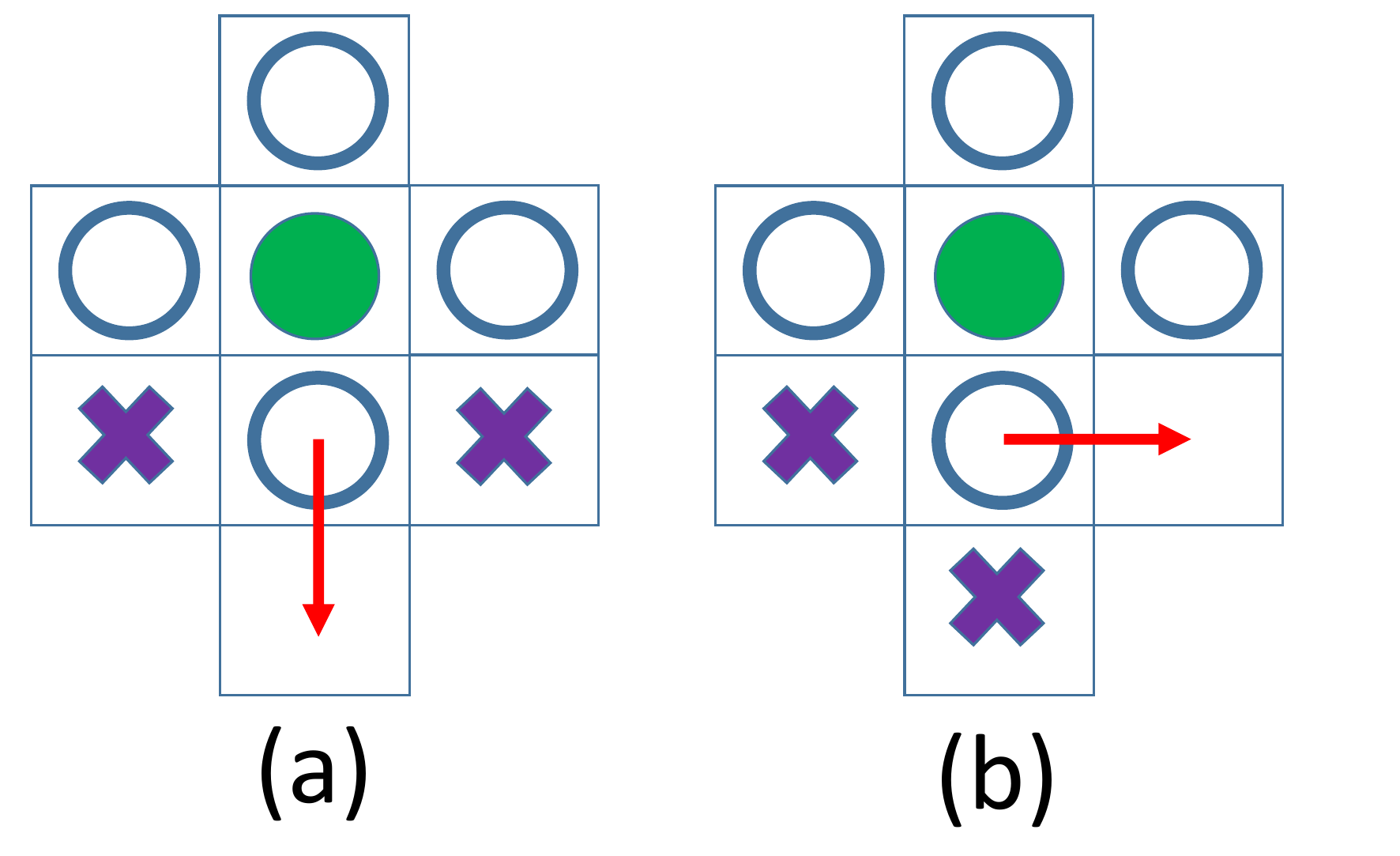}
\caption{An illustration of the transition ${\rm B4\rightarrow B3J}$}
\label{rates_bka_1}
\end{figure}

This transition is illustrated in Fig. \ref{rates_bka_1}. At least one of the $\times$ sites is vacant. The rate is
\begin{align}
r^{\rm{BKA}}_{\rm{B}4,B3J}=P_{{\rm M}|o}+2P_{{\rm M}|s}
\end{align}

\subsection{Outgoing rates from ${\rm B3J}$}
\begin{figure}
\includegraphics[width=0.6\columnwidth]{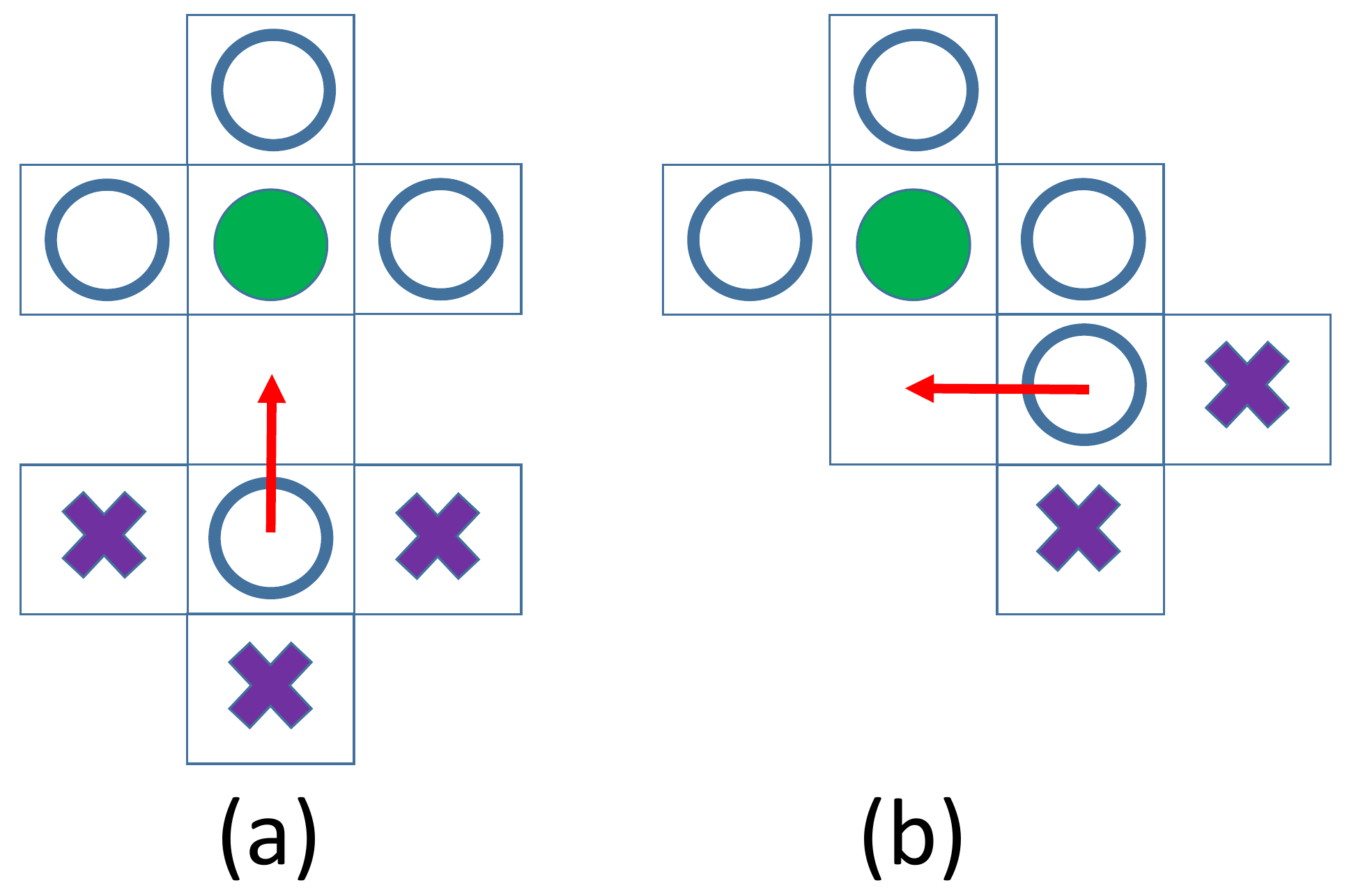}\\
\includegraphics[width=0.6\columnwidth]{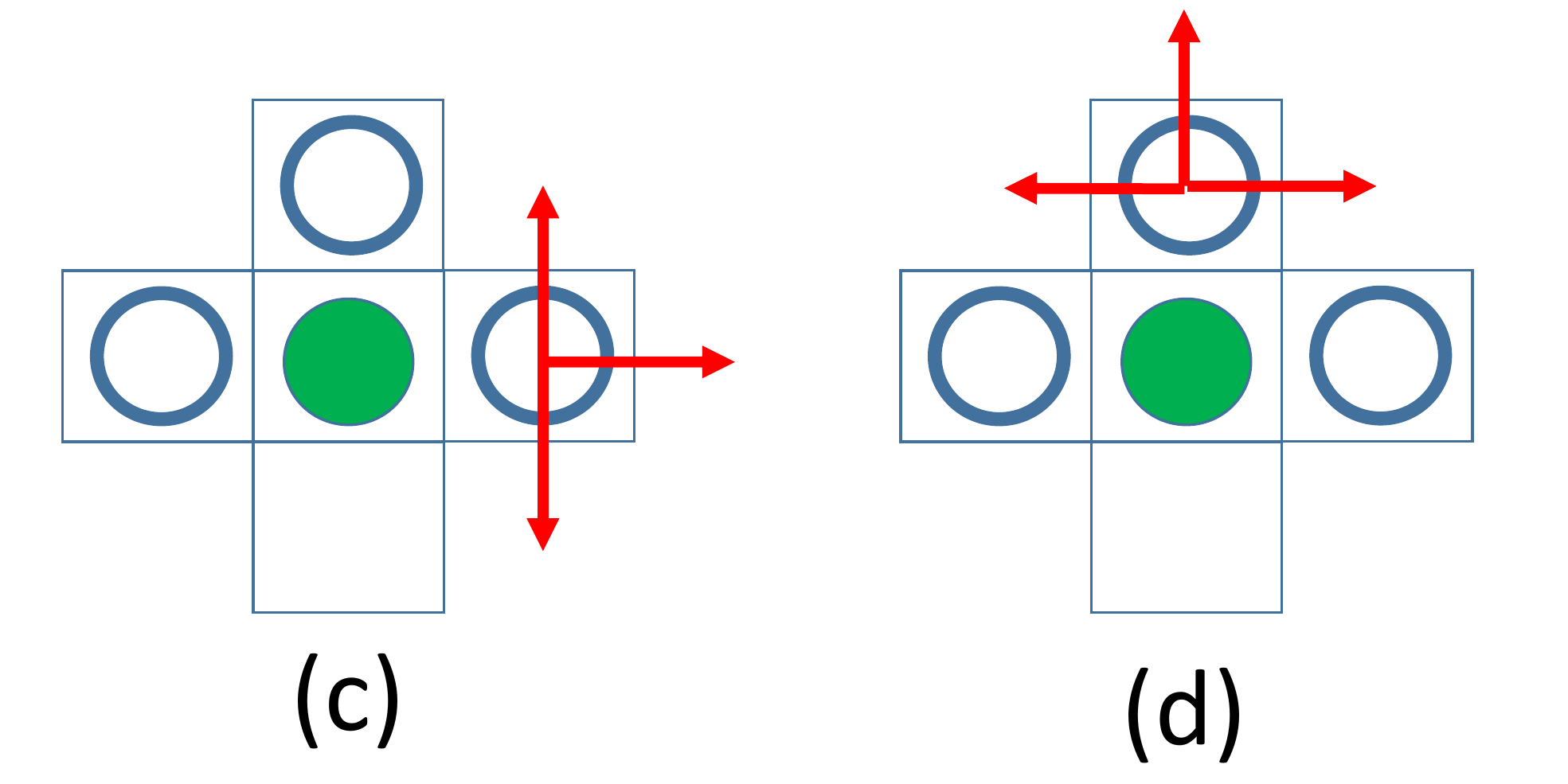}
\caption{An illustration of the outgoing transitions from ${\rm B3J}$}
\label{rates_bka_2}
\end{figure}

The outgoing rates from state ${\rm B3J}$ are illustrated in Fig. \ref{rates_bka_2}. In Figs. \ref{rates_bka_2}a-b at least one of the $\times$ sites is vacant, and in Fig. \ref{rates_bka_2}c-d the transition occur if the particle moves in one of the three directions. The rates are
\begin{align}
&r^{\rm{BKA}}_{\rm{B}3J,B4}=\frac{1}{4}\left(\rho P_{{\rm M}|t}+2\rho P_{{\rm M}|st}\right) ,\nonumber\\
&r^{\rm{BKA}}_{\rm{B}3J,B2F2}=\frac{1}{4}\left(4P_{{\rm M}|s}+2P_{{\rm M}|o}\right) ,\nonumber\\
&r^{\rm{BKA}}_{\rm{B}3J,BFBF}=\frac{1}{4}\left(2P_{{\rm M}|s}+P_{{\rm M}|o}\right) .
\end{align}

\subsection{Outgoing rates from ${\rm B2F2}$}
\begin{figure}
\includegraphics[width=0.9\columnwidth]{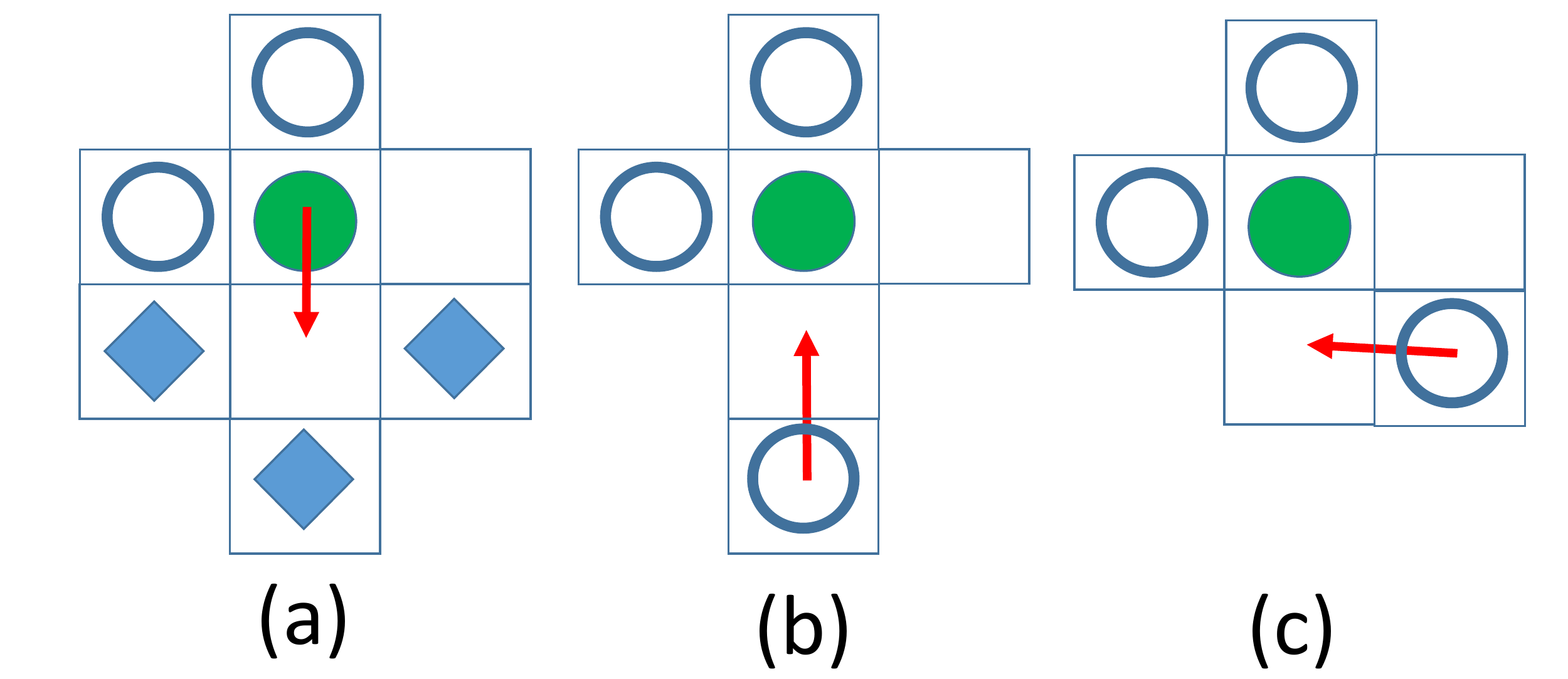}\\
\includegraphics[width=0.6\columnwidth]{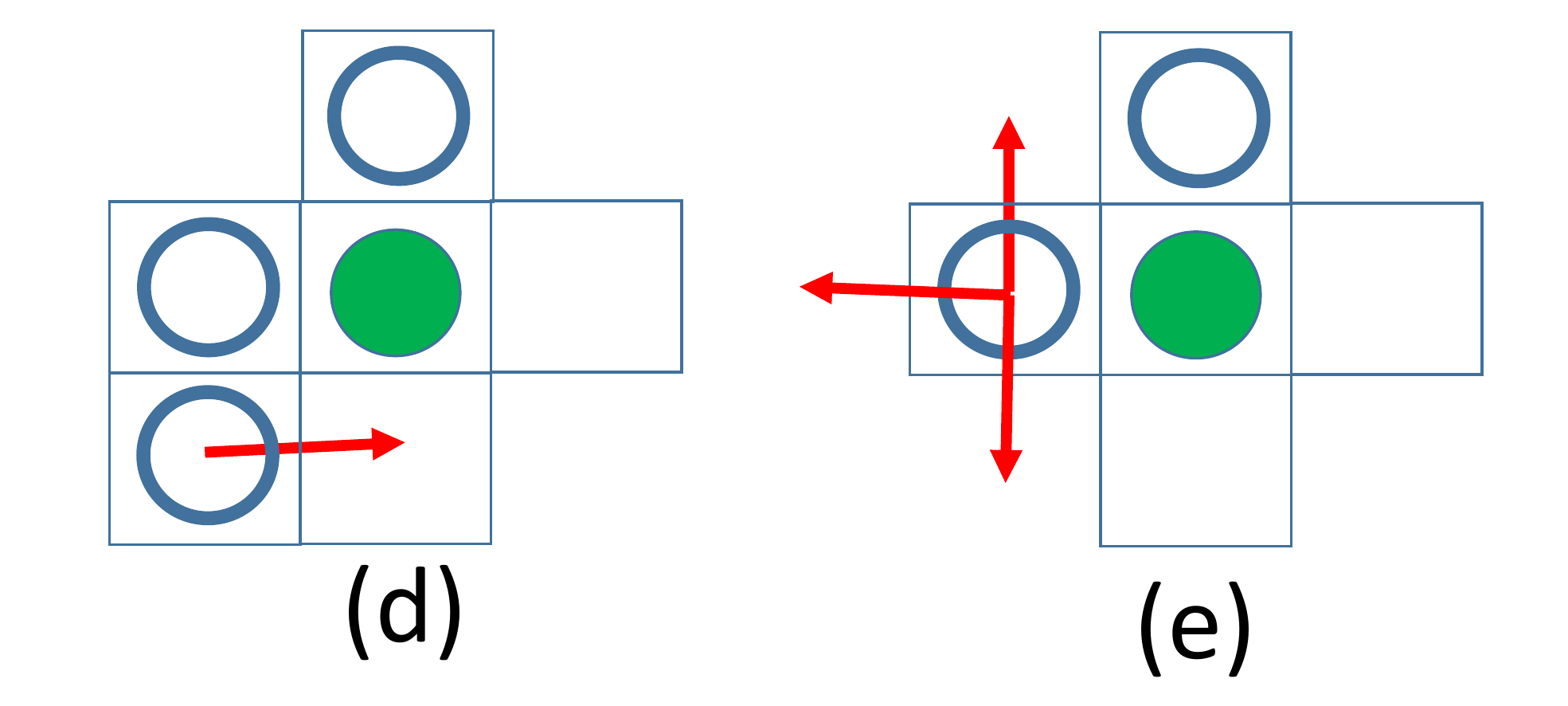}
\caption{An illustration of the outgoing transitions from ${\rm B2F2}$}
\label{rates_bka_3}
\end{figure}

The outgoing rates from state ${\rm B2F2}$ are illustrated in Fig. \ref{rates_bka_3}. Figure \ref{rates_bka_3}a, shows the transition induced by the main particle moving. The state after the move depends on the three sites marked $\diamond$. Figures \ref{rates_bka_3}b-d shows the transition to ${\rm B3J}$ induced by an incoming particle. Figure \ref{rates_bka_3}e shows the transition to ${\rm BF3}$ induced by the movement of one of the blocking particles. The rates are
\begin{align}
&r^{\rm{BKA}}_{\rm{B}2F2,B3J}=\frac{1}{2}\left[\rho^{3}+\rho P_{{\rm M}|t}+\rho P_{{\rm M}|vt}+\rho P_{{\rm M}|st}\right] ,\nonumber\\
&r^{\rm{BKA}}_{\rm{B}2F2,BFBF}=\frac{1}{2}\rho^{2}\left(1-\rho\right) ,\nonumber\\
&r^{\rm{BKA}}_{\rm{B}2F2,BF3}=\frac{1}{2}\left[3\rho\left(1-\rho\right)^{2}+2P_{{\rm M}|s}+P_{{\rm M}|o}\right] ,\nonumber\\
&r^{\rm{BKA}}_{\rm{B}2F2,F4}=\frac{1}{2}\left(1-\rho\right)^{3} .
\end{align}

\subsection{Outgoing rates from ${\rm BFBF}$}
\begin{figure}
\includegraphics[width=\columnwidth]{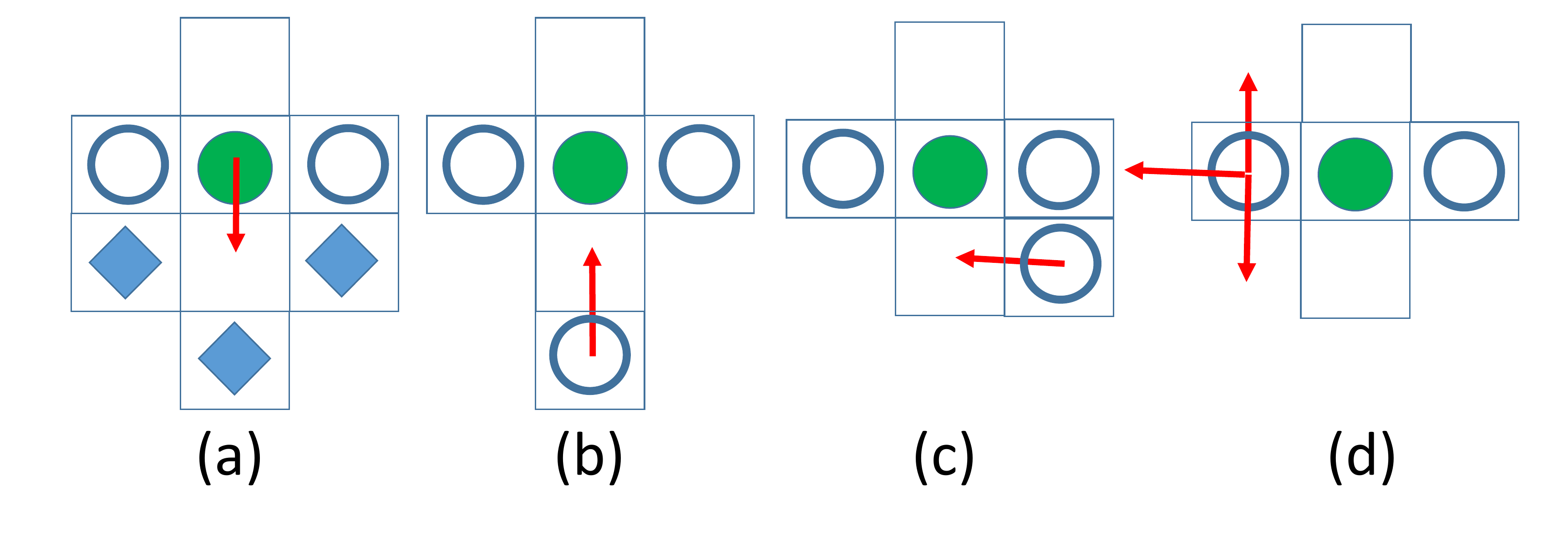}
\caption{An illustration of the outgoing transitions from ${\rm BFBF}$}
\label{rates_bka_4}
\end{figure}

The outgoing rates from state ${\rm BFBF}$ are illustrated in Fig. \ref{rates_bka_4}. Figure \ref{rates_bka_4}a, shows the transition induced by the main particle moving. The state after the move depends on the three sites marked $\diamond$. Figures \ref{rates_bka_4}b-c shows the transition to ${\rm B3J}$ induced by an incoming particle. Figure \ref{rates_bka_4}d shows the transition to ${\rm BF3}$ induced by the movement of one of the blocking particles. The rates are
\begin{align}
&r^{\rm{BKA}}_{\rm{B}FBF,B3J}=\frac{1}{2}\left(\rho^{3}+\rho P_{{\rm M}|t}+2\rho P_{{\rm M}|st}\right) ,\nonumber\\
&r^{\rm{BKA}}_{\rm{B}FBF,B2F2}=\rho^{2}\left(1-\rho\right) ,\nonumber\\
&r^{\rm{BKA}}_{\rm{B}FBF,BF3}=\frac{1}{2}\left[3\rho\left(1-\rho\right)^{2}+2P_{{\rm M}|s}+P_{{\rm M}|o}\right] ,\nonumber\\
&r^{\rm{BKA}}_{\rm{B}FBF,F4}=\frac{1}{2}\left(1-\rho\right)^{3} .
\end{align}

\subsection{Outgoing rates from ${\rm BF3}$}
\begin{figure}
\includegraphics[width=\columnwidth]{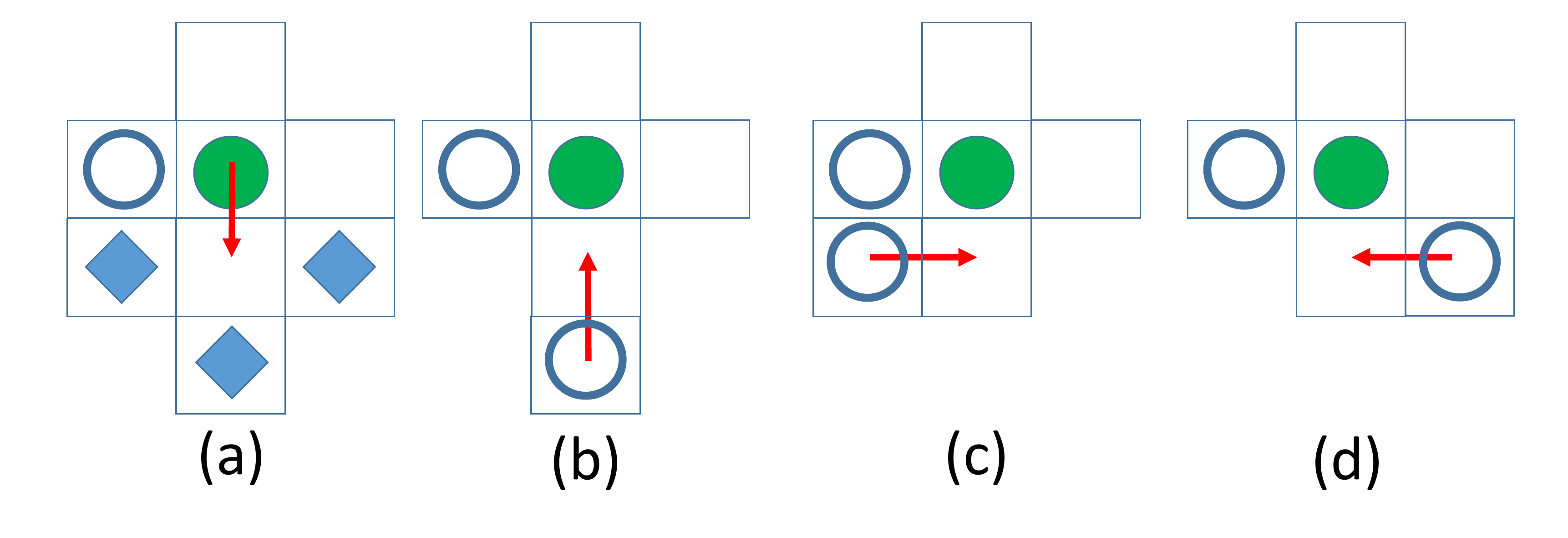}\\
\includegraphics[width=0.9\columnwidth]{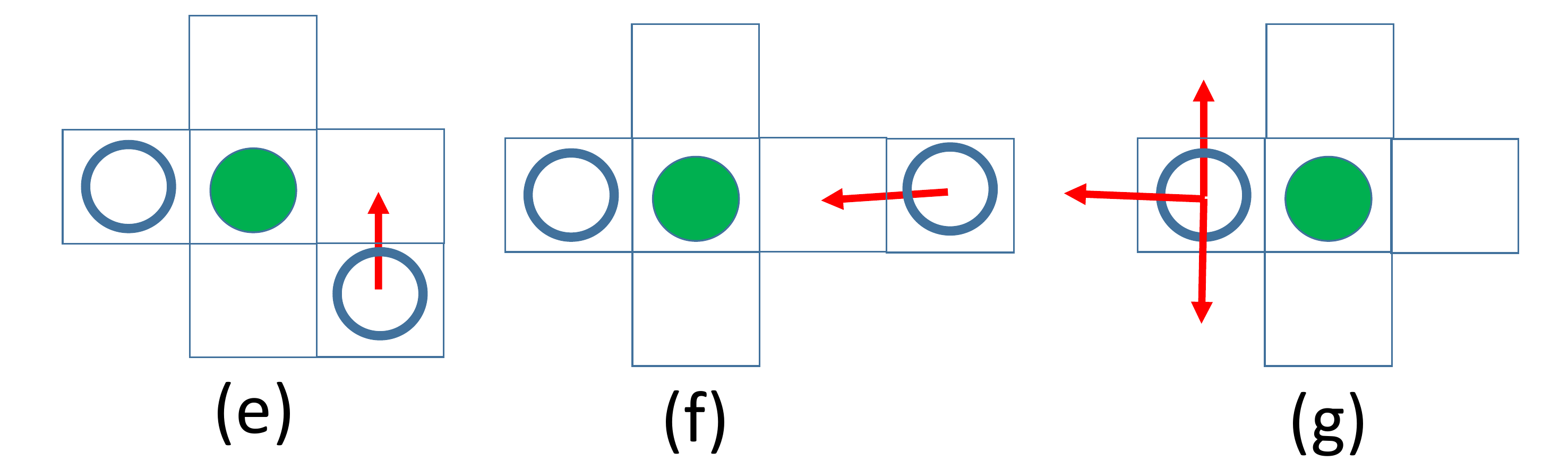}
\caption{An illustration of the outgoing transitions from ${\rm BF3}$}
\label{rates_bka_5}
\end{figure}

The outgoing rates from state ${\rm BF3}$ are illustrated in Fig. \ref{rates_bka_5}. Figure \ref{rates_bka_5}a, shows the transition induced by the main particle moving in either of the three directions. The state after the move depends on the three sites marked $\diamond$. Figures \ref{rates_bka_5}b-d shows the transition to ${\rm B2F2}$ induced by an incoming particle. Figures \ref{rates_bka_5}e-f shows the transition to ${\rm B2F2}$ induced by an incoming particle. Figure \ref{rates_bka_5}g shows the transition to ${\rm F4}$ induced by the movement of one of the blocking particles. The rates are
\begin{align}
&r^{\rm{BKA}}_{\rm{B}F3,B3J}=\frac{3}{4}\rho^{3} ,\nonumber\\
&r^{\rm{BKA}}_{\rm{B}F3,B2F2}=\frac{1}{2}\left[3\rho^{2}\left(1-\rho\right)+\rho P_{{\rm M}|t}+\rho P_{{\rm M}|st}+\rho P_{{\rm M}|vt}\right] ,\nonumber\\
&r^{\rm{BKA}}_{\rm{B}F3,BFBF}=\frac{1}{4}\left[3\rho^{2}\left(1-\rho\right)+2\rho P_{{\rm M}|vt}+\rho P_{{\rm M}|t}\right] ,\nonumber\\
&r^{\rm{BKA}}_{\rm{B}F3,F4}=\frac{1}{4}\left[3\left(1-\rho\right)^{3}+2P_{{\rm M}|s}+P_{{\rm M}|o}\right] .
\end{align}

\subsection{Outgoing rates from ${\rm F4}$}
\begin{figure}
\includegraphics[width=0.9\columnwidth]{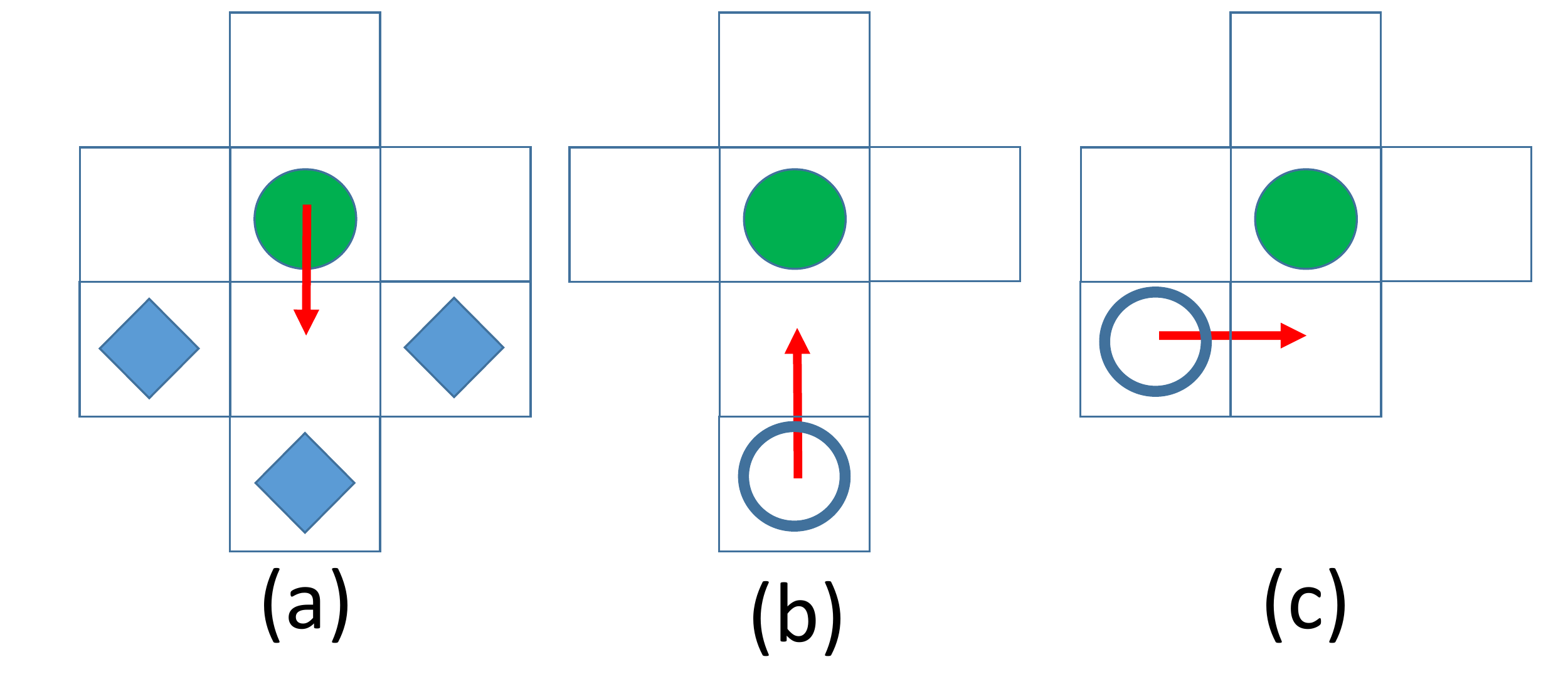}
\caption{An illustration of the outgoing transitions from ${\rm F4}$}
\label{rates_bka_6}
\end{figure}

The outgoing rates from state ${\rm F4}$ are illustrated in Fig. \ref{rates_bka_6}. Figure \ref{rates_bka_6}a, shows the transition induced by the main particle moving in either of the four directions. The state after the move depends on the three sites marked $\diamond$. Figures \ref{rates_bka_6}b-c shows the transition to ${\rm BF3}$ induced by an incoming particle. The rates are
\begin{align}
&r^{\rm{BKA}}_{\rm{F}4,B3J}=\rho^{3} ,\nonumber\\
&r^{\rm{BKA}}_{\rm{F}4,B2F2}=2\rho^{2}\left(1-\rho\right) ,\nonumber\\
&r^{\rm{BKA}}_{\rm{F}4,BFBF}=\rho^{2}\left(1-\rho\right) ,\nonumber\\
&r^{\rm{BKA}}_{\rm{F}4,BF3}=3\rho\left(1-\rho\right)^{2}+\rho P_{{\rm M}|t}+2\rho P_{{\rm M}|vt} .
\end{align}

\end{document}